\documentclass[12pt]{article}

\usepackage[
    a4paper,
    top=2.5cm,
    bottom=2.5cm,
    left=2.5cm,
    right=2.5cm
]{geometry}

\usepackage{graphicx}
\usepackage[numbers]{natbib}
\usepackage{bm}
\usepackage{amsmath}
\usepackage{amssymb}
\usepackage{booktabs}

\usepackage{amsthm}
\usepackage{algorithm}
\usepackage{algpseudocode}
\usepackage{float}
\usepackage{xurl}

\usepackage[font=footnotesize, labelfont=bf, labelsep=space]{caption}

\usepackage{xcolor}

\begin{document}

\title{Equal-response probing with replicate calibration identifies scaling relations from noisy observations of complex systems}

\author{Dingbang Yan$^{1,*}$, Zhaoyue Xu$^{2}$, Hua-Dong Yao$^{1}$}

\date{}

\maketitle

\noindent
$^{1}$Department of Mechanical Engineering, Chalmers University of Technology, Gothenburg, Sweden.\\
$^{2}$Department of Animal Sciences, Wageningen University, Wageningen, The Netherlands.\\
$^{*}$Correspondence to: dingbang@chalmers.se

\begin{abstract}
    Dimensionless descriptions of complex physical systems can sometimes be reduced further when multiple input groups combine into a single composite scaling coordinate. Identifying such structure from noisy response evaluations is difficult because apparent departures may reflect either genuine response structure or evaluation noise. Here we develop equal-response probing with replicate calibration to assess whether two dimensionless inputs support a one-group representation under prescribed directional and collapse tolerances. Equal-response geometry provides a candidate coordinate without direct numerical differentiation, while a second stage evaluates the accuracy of the resulting one-dimensional collapse. Replicate evaluations estimate the noise contribution to both stages, enabling reducible, not-red and unresolved outcomes. We evaluate the framework across eight model-based benchmark regimes spanning rough-pipe flow, cell-monolayer wound closure, microfluidic droplet generation and column buckling. Six regimes satisfy both criteria and two fail both, while known reduced coordinates are recovered in exact-reduction benchmarks. Controlled synthetic tests assess false-alarm behaviour, sensitivity to structural departures and noise calibration. These results show how replicate-calibrated equal-response probing distinguishes regime-specific scaling structure from variation attributable to evaluation noise while retaining an unresolved outcome when evaluation noise precludes a definitive verdict at the prescribed tolerances.
\end{abstract}

\section*{Introduction}

Dimensional analysis is ubiquitous across the physical sciences and engineering, where complex relations involving many physical quantities are routinely expressed in terms of a smaller set of dimensionless groups. By eliminating dependence on the particular choice of units, these groups provide scale-invariant descriptions of physical behaviour and form the basis of similarity analysis across systems of different sizes and operating conditions~\citep{Buckingham1914,Bridgman1922,Barenblatt1996ScalingSA}. Dimensionless groups such as the Reynolds, Péclet and capillary numbers have become standard descriptors in fluid mechanics, transport phenomena and multiphase flows, while analogous groups are widely used in solid mechanics, thermal sciences and biological systems. Dimensional analysis therefore provides a common language for organising multivariable physical problems and revealing relations that extend beyond a particular experimental or computational configuration.

The value of dimensional analysis extends beyond dimensional consistency. By replacing dimensional variables with a smaller number of dimensionless groups, it can substantially reduce the parameter space that must be explored experimentally or computationally~\citep{Buckingham1914,Bridgman1922,Xie2022}. The scale invariance of these groups also underpins similitude, allowing observations obtained under one set of scales or operating conditions to inform physically similar systems at another~\citep{Barenblatt1996ScalingSA,sedov1960similarity,kline2012similitude}. Moreover, dimensionless groups often compare competing physical mechanisms, making them useful not only as compact coordinates but also as interpretable indicators of the processes governing a response~\citep{Bridgman1922,Xie2022}. These properties have made dimensional analysis an effective route for simplifying complex physical systems while retaining their essential structure.

Despite these advantages, classical dimensional analysis does not generally determine the lowest-dimensional representation of a response. The Buckingham $\Pi$ theorem identifies admissible sets of dimensionless groups, but these groups are not unique and the theorem alone does not determine the functional relation among them~\citep{Buckingham1914,Xie2022}. More importantly for the present problem, even after the relevant dimensionless inputs and a physical regime of interest have been specified, dimensional analysis alone does not establish how many of those inputs are actually required to represent the response over that regime~\citep{constantine2017data,jofre2020data,bakarji2022dimensionally,Yuan2025}. A response expressed in terms of two dimensionless inputs may genuinely require both coordinates, or its variation over the prescribed regime may be captured by a single composite dimensionless group. In the latter case, the two-input response admits a one-dimensional relation between the response and the composite coordinate. Determining whether such a further reduction is justified therefore requires information about the response itself beyond that supplied by dimensional analysis.

Several data-driven approaches have been developed to extract lower-dimensional structure beyond that provided directly by classical dimensional analysis. Early semi-empirical methods combined dimensional constraints with regression to infer scaling relations from experimental data~\citep{mendez2005scaling}. Data-driven dimensional analysis subsequently connected the identification of unique and relevant dimensionless groups with active-subspace and ridge-function representations~\citep{constantine2017data,constantine2015active,constantine2014active,constantine2016many,seshadri2019dimension}. More recent dimensionless-learning approaches incorporate dimensional invariance into regression, sparse modelling and machine learning to discover dimensionless groups and low-dimensional scaling relations directly from data~\citep{Xie2022,bakarji2022dimensionally}, with successful applications to scarce and noisy experimental measurements~\citep{Xie2022}. Related methods have been developed to identify dominant dimensionless quantities in systems containing multiple physical regimes~\citep{ZHANG2024116728}, and information-theoretic criteria have been introduced to quantify the relevance of candidate dimensionless quantities without relying on derivative-based sensitivity measures~\citep{zhang2025mutual}. Most recently, IT-$\Pi$ used an information-theoretic irreducible-error bound to rank dimensionless variables by predictive information and to determine the minimum number of dimensionless inputs required to maximize predictability ~\citep{Yuan2025}. These developments substantially extend classical dimensional analysis by identifying which dimensionless coordinates are most informative and how many are required to represent a response.

A distinct challenge arises when response evaluations are noisy. An imperfect one-dimensional collapse may reflect genuine two-input structure, but it may also be induced by evaluation noise. Demonstrating that a reduced coordinate can be recovered robustly from noisy data does not by itself distinguish between these two contributions. Identifying whether such a reduced representation is supported therefore requires separating noise-induced variation from structural departure, comparing the latter with a prescribed tolerance, and recognising when the noise level is too large to support a definite conclusion.

Here we develop an equal-response probing framework with replicate calibration to assess whether two dimensionless inputs support a one-group representation under prescribed directional and collapse tolerances. In logarithmic coordinates, exact one-group structure has a distinctive geometry: its regular connected level-set branches are parallel straight lines. An equal-response crossing on the branch through an anchor therefore determines the axial response direction without requiring direct numerical differentiation. For responses that depart from exact one-group structure, the resulting finite-scale read-out provides an approximate direction, with the associated chord and interpolation errors quantified separately. The framework combines two complementary assessments. The first stage tests whether finite-scale response directions remain consistent across the prescribed regime, with replicate evaluations providing a calibrated reference for the directional variation attributable to evaluation noise. The second stage tests the accuracy of the resulting one-dimensional collapse and quantifies the noise contribution to the observed residual. Combining these two assessments with prescribed directional and collapse tolerances gives three possible outcomes: reducible, not-red or unresolved, the latter retaining uncertainty when evaluation noise prevents a definite structural conclusion. We evaluate the framework across eight regimes drawn from rough-pipe flow, cell-monolayer wound closure, microfluidic droplet generation and column buckling, together with controlled synthetic response fields. The physical examples test whether known and approximate composite coordinates can be identified across distinct governing relations and whether genuinely two-input regimes are distinguished, while the synthetic tests quantify false-alarm behaviour, sensitivity to structural departures, noise calibration and the effect of the prescribed tolerances. The resulting framework provides a regime-specific, noise-resolved basis for determining when noisy dimensionless responses support reduction to a single composite coordinate.

\section*{Results}

\subsection*{A noise-resolved test for further dimensionless reduction}

\begin{figure}[htbp]
    \centering
    \includegraphics[width=0.8\linewidth]{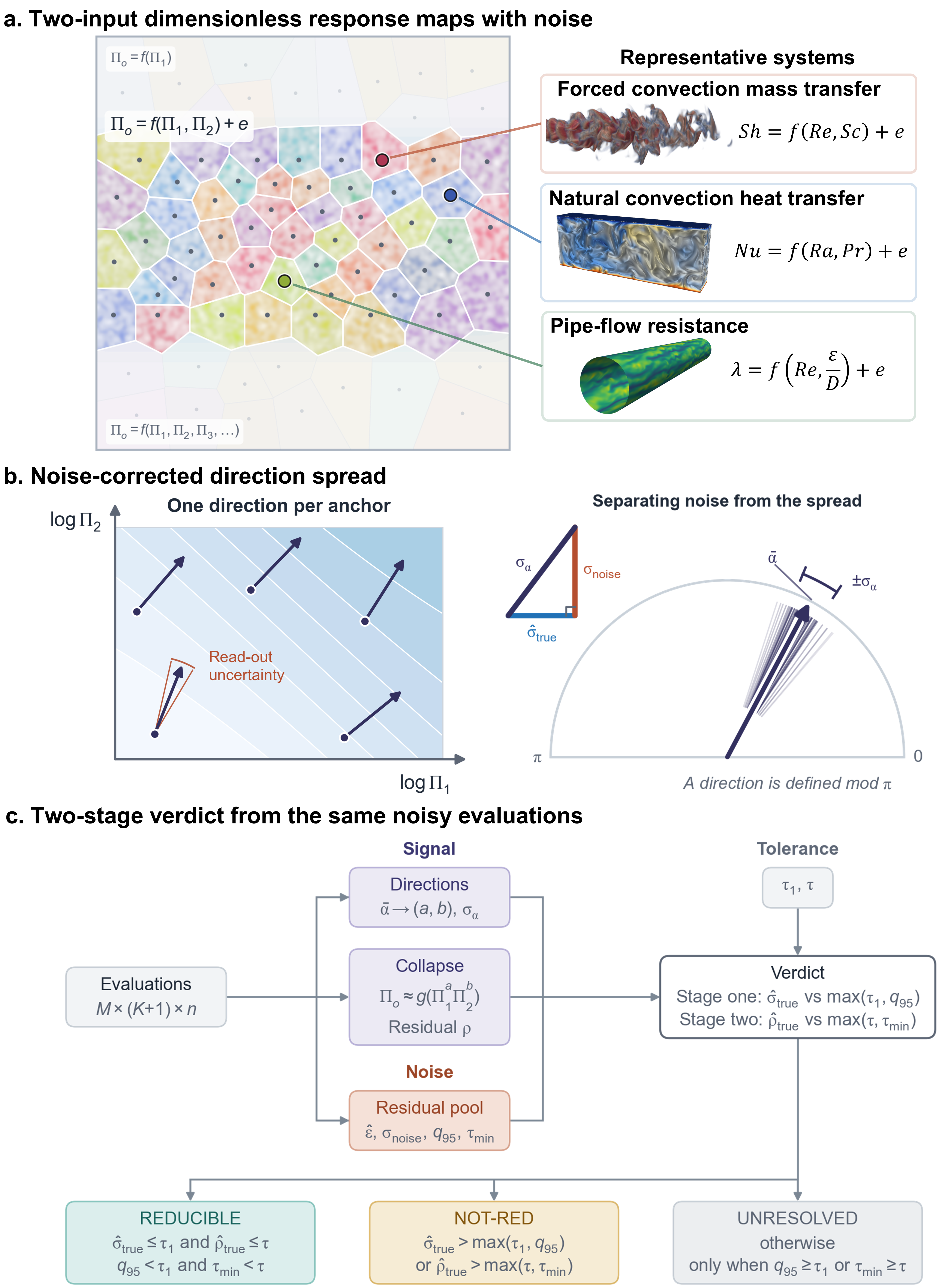}
    \caption{\textbf{Noise-resolved two-stage framework for further dimensionless reduction.}
    \textbf{a}, Examples of two-input dimensionless responses in forced-convection mass transfer, natural-convection heat transfer and pipe-flow resistance.
    \textbf{b}, Directional test. Finite-scale probes recover local response directions at $M$ anchors in logarithmic coordinates. Their axial mean $\bar\alpha$ defines the candidate reduced coordinate, while their spread $\sigma_\alpha$ measures directional variation; repeated evaluations quantify the contribution of evaluation noise.
    \textbf{c}, Two-stage decision framework. Stage one tests directional consistency and stage two tests the accuracy of the resulting one-dimensional collapse. Each stage compares a noise-corrected departure with a prescribed tolerance and a noise-based reference, yielding reducible, not-red or unresolved. The two stage-wise verdicts are combined into the final verdict.}
    \label{fig:methodology}
\end{figure}

Dimensional analysis reduces a physical relation to a dependence among dimensionless groups, but does not determine how many of those groups are required to describe the response over a given regime~\citep{Buckingham1914,constantine2017data,Yuan2025}. For the two-input maps that arise widely in transport and mechanics---for example, $Sh=f(Re,Sc)$ in forced-convection mass transfer, $Nu=f(Ra,Pr)$ in natural-convection heat transfer and $\lambda=f(Re,\epsilon/D)$ in pipe flow (Fig.~\ref{fig:methodology}a)---the relevant question is whether the response can be reduced to a single power-law coordinate,
\begin{align}
    \hat u=\Pi_1^{a}\Pi_2^{b},
\end{align}
such that the noise-free dimensionless response can be represented by a univariate relation $\Pi_o=G(\hat u)$. When this reduction is valid, a two-dimensional response surface collapses onto a one-dimensional curve; when it is not, enforcing such a collapse removes physically relevant variation. In practice, each evaluation is observed with evaluation noise,
\begin{align}
    y^{\rm obs} = \Pi_o + e = f(\Pi_1,\Pi_2) + e,
\end{align}
so exact one-dimensionality cannot be inferred directly from the observed map. The practical question is therefore how to determine whether the response is reducible within a prescribed tolerance when the observed response is contaminated by evaluation noise.

We address this problem with a two-stage test applied to the same set of noisy evaluations (Fig.~\ref{fig:methodology}b,c). The first stage exploits a geometric consequence of one-group reducibility. In logarithmic coordinates, $x_1=\ln\Pi_1$ and $x_2=\ln\Pi_2$, a one-group response depends only on the coordinate
\begin{align}
    u=ax_1+bx_2.
\end{align}
Its non-zero gradient is therefore everywhere parallel to the fixed direction $(a,b)$. For an exactly one-group response, the connected level-set branch through an anchor is a straight line in logarithmic coordinates. In two dimensions, the anchor and a single equal-response crossing therefore determine a contour direction whose normal gives the axial response direction, without requiring direct numerical differentiation. For responses that depart from exact one-group structure, the contour may be curved and the resulting chord gives a finite-scale approximation to the local direction; the associated chord and angular-interpolation errors are quantified in Supplementary Fig.~\ref{fig:s6}. We probe this property at $M$ anchors distributed across the regime. Around each anchor, responses are evaluated repeatedly at the anchor and at $K$ points on a finite-scale half-ellipse. The equal-response crossing of the probe identifies a finite-scale contour direction, whose normal provides an estimate of the local gradient direction (Fig.~\ref{fig:methodology}b). If the response is reducible to a single group, these directions should remain consistent throughout the regime. Their axial mean $\bar\alpha$ defines the recovered coordinate direction,
\begin{align}
    (a,b)=(\cos\bar\alpha,\sin\bar\alpha),
\end{align}
whereas their axial spread $\sigma_\alpha$ measures the observed directional variation.

Evaluation noise also perturbs each directional read-out and therefore contributes to $\sigma_\alpha$. Repeated evaluations are used to estimate this contribution, yielding a noise-induced spread $\sigma_{\rm noise}$. We then define the noise-corrected directional spread
\begin{align}
    \hat\sigma_{\rm true} = \sqrt{ \max\left( \sigma_\alpha^2-\sigma_{\rm noise}^2, 0 \right) }.
\end{align}
To determine whether a non-zero $\hat\sigma_{\rm true}$ can arise from noise alone, we construct a synthetic direction field in which all anchors share the same noise-free read-out direction while retaining their anchor-specific noise perturbations. This defines the equal-read-out-direction null used to calibrate $q_{95}$. Repeating the procedure gives the distribution of $\hat\sigma_{\rm true}$, generated by noise alone, whose 95th percentile is denoted by $q_{95}$. The detailed resampling procedure and noise assumptions are given in Methods.

Directional consistency is necessary but not sufficient to establish a useful one-dimensional reduction. The angular statistic quantifies variation in local directions but does not directly measure the accuracy of the resulting one-dimensional collapse. We therefore test the collapse directly in a second stage. The anchors are projected onto candidate directions near $\bar\alpha$, and their averaged responses are fitted by a univariate curve. The smallest resulting normalised residual $\rho$ quantifies the fitting error of the one-dimensional representation selected by this direction-search and polynomial-fitting procedure. Visually similar collapse plots can nevertheless correspond to substantially different residuals, so the collapse is assessed quantitatively rather than by visual inspection alone (Supplementary Fig.~\ref{fig:s4}).

Even an exactly reducible response has a non-zero observed collapse residual in the presence of evaluation noise. We therefore estimate the noise contribution from the repeated evaluations. The corresponding analytical noise scale is
\begin{align}
    \tau_{\min} = \frac{\hat\varepsilon}{\sqrt{n}\,s_y},
\end{align}
where $\hat\varepsilon$ is the estimated single-evaluation noise scale and $s_y$ is the standard deviation of the averaged anchor responses. Removing this contribution in quadrature gives the noise-corrected collapse residual
\begin{align}
    \hat\rho_{\rm true} = \sqrt{ \max\left( \rho^2-\tau_{\min}^2,\, 0 \right) },
\end{align}
which represents the collapse residual remaining after removal of the estimated noise contribution. The derivation is given in Methods.

The two stages quantify complementary departures from one-group behaviour. Stage one compares the noise-corrected directional spread $\hat\sigma_{\rm true}$ with a prescribed angular tolerance $\tau_1$, while $q_{95}$ is the 95th percentile under the equal-read-out-direction null. Stage two compares the noise-corrected collapse residual $\hat\rho_{\rm true}$ with a prescribed collapse tolerance $\tau$, while $\tau_{\min}$ gives the analytical noise scale of the collapse residual. Together, these quantities define reducible, not-red and unresolved outcomes; the exact decision rules are given in Methods. The overall result is reducible only when both stages return reducible, not-red when either stage returns not-red, and unresolved otherwise (Fig.~\ref{fig:methodology}c). An unresolved result indicates that evaluation noise prevents a definite verdict at the prescribed tolerance, rather than implying irreducibility.

The two stages test different aspects of the same reduction. Stage one assesses whether the local directions are consistent across the regime, whereas stage two directly quantifies the accuracy of the resulting one-dimensional collapse. Directional consistency alone therefore does not determine the quality of the collapse, motivating the complementary second-stage test.

\subsection*{Regime-dependent reducibility in rough-pipe flow}

\begin{figure}[htbp]
    \centering
    \includegraphics[width=\linewidth]{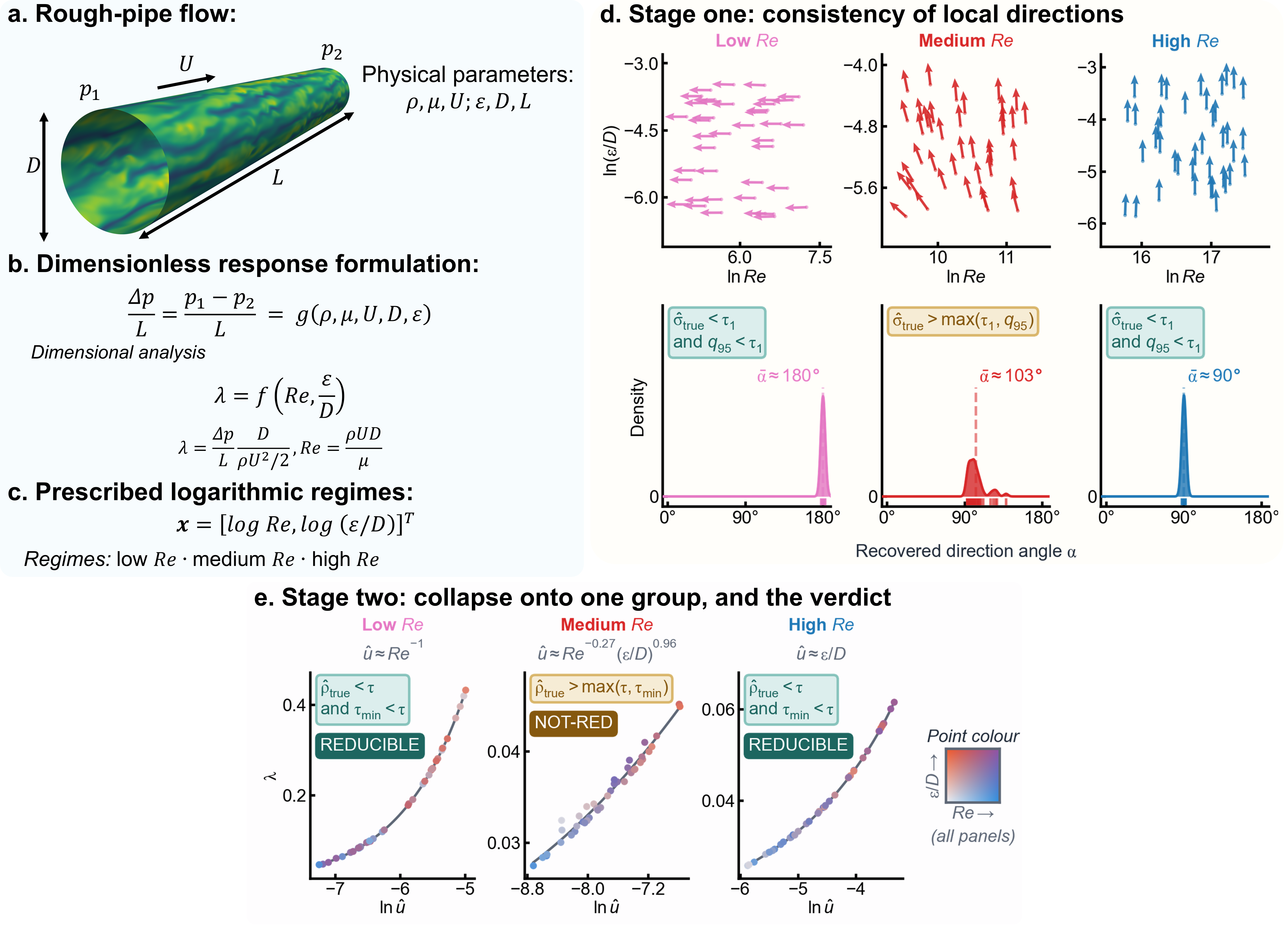}
    \caption{\textbf{Regime-dependent reducibility in rough-pipe flow.}
    \textbf{a}, Pressure drop across a rough pipe of diameter $D$ and length $L$.
    \textbf{b}, Dimensional analysis gives the two-input relation $\lambda=f(Re,\epsilon/D)$.
    \textbf{c}, Three prescribed logarithmic regimes spanning low, medium and high Reynolds numbers.
    \textbf{d}, Stage-one directional analysis. Top: local directions recovered at $M=40$ anchors. Bottom: distributions of the axial direction angle $\alpha$, with dashed lines indicating the axial mean $\bar\alpha$.
    \textbf{e}, Stage-two collapse onto the recovered one-dimensional coordinate $\hat u$. Low- and high-$Re$ regimes collapse onto a single curve, whereas the medium-$Re$ regime retains systematic two-input variation. Point colour indicates position in the $(Re,\epsilon/D)$ plane. Numerical values are given in Table~\ref{tab:pipe_verdicts}.}
    \label{fig:flow_resistance}
\end{figure}

We first test the framework on rough-pipe flow, for which the same physical relation exhibits different effective dimensionality across the $(Re,\epsilon/D)$ plane. For steady flow through a rough pipe of diameter $D$ and length $L$ (Fig.~\ref{fig:flow_resistance}a), dimensional analysis reduces
\begin{align}
    \Delta p/L=g(\rho,\mu,U,D,\epsilon)
\end{align}
to the two-input relation
\begin{align}
    \lambda=f(Re,\epsilon/D)
\end{align}
(Fig.~\ref{fig:flow_resistance}b)~\citep{moody1944friction,colebrook1939correspondence}.

We consider three logarithmic regimes (Fig.~\ref{fig:flow_resistance}c). At low $Re$, with $Re\in[10^2,2\times10^3]$ and $\epsilon/D\in[10^{-3},5\times10^{-2}]$, the laminar relation $\lambda=64/Re$ is independent of roughness~\citep{White2016}. At medium $Re$, with $Re\in[10^4,10^5]$ and $\epsilon/D\in[2\times10^{-3},2\times10^{-2}]$, both Reynolds number and relative roughness contribute appreciably across the regime, and their relative importance varies across the domain. At high $Re$, with $Re\in[5\times10^6,5\times10^7]$ and $\epsilon/D\in[2\times10^{-3},5\times10^{-2}]$, the viscous term is at least about two orders of magnitude smaller than the roughness term and is typically several orders smaller, so the response approaches the fully rough dependence on $\epsilon/D$~\citep{nikuradse1950laws,colebrook1939correspondence}. The same noisy-evaluation protocol is applied in all three regimes (Methods), with $\tau_1=4^\circ$ and $\tau=5\%$. The resulting stage-one and stage-two statistics are summarised in Table~\ref{tab:pipe_verdicts}.

Stage one recovers the change in directional structure across the three rough-pipe regimes (Fig.~\ref{fig:flow_resistance}d). In the low-$Re$ regime, the recovered directions cluster close to the $\ln Re$ axis. The observed spread is $\sigma_\alpha=0.83^\circ$, while the estimated noise-induced spread is $\sigma_{\rm noise}=0.95^\circ$, giving $\hat\sigma_{\rm true}=0$. With $q_{95}=0.78^\circ<\tau_1=4^\circ$, stage one therefore returns reducible. In the medium-$Re$ regime, the directions vary substantially across the domain: $\sigma_\alpha=10.12^\circ$ and $\sigma_{\rm noise}=0.72^\circ$ give $\hat\sigma_{\rm true}=10.10^\circ$, which exceeds both $\tau_1=4^\circ$ and $q_{95}=0.56^\circ$. Stage one therefore returns not-red. At high $Re$, the directions cluster close to the $\ln(\epsilon/D)$ axis. Here $\sigma_\alpha=0.75^\circ$ and $\sigma_{\rm noise}=1.36^\circ$ give $\hat\sigma_{\rm true}=0$, while $q_{95}=1.37^\circ<\tau_1$, so stage one again returns reducible.

Stage two gives the same regime separation (Fig.~\ref{fig:flow_resistance}e). In the low-$Re$ regime, the recovered coordinate is 
\begin{align}
    \hat u\approx Re^{-1.00},
\end{align}
consistent with the laminar dependence $\lambda=64/Re$. The observed collapse residual is $\rho=0.0080$, below the analytical noise scale $\tau_{\min}=0.0090$, so $\hat\rho_{\rm true}=0$ and stage two returns reducible. In the high-$Re$ regime, the recovered coordinate is
\begin{align}
    \hat u\approx(\epsilon/D)^{1.00},
\end{align}
with $\rho=0.0086<\tau_{\min}=0.0088$ and hence $\hat\rho_{\rm true}=0$, again giving reducible. By contrast, in the medium-$Re$ regime the best recovered coordinate is
\begin{align}
    \hat u\approx Re^{-0.27}(\epsilon/D)^{0.96},
\end{align}
but the collapse remains broad, with $\rho=0.149$ and $\hat\rho_{\rm true}=0.149$, compared with $\tau_{\min}=0.0101$ and $\tau=0.05$. Stage two therefore returns not-red.

The two stages therefore identify the same regime dependence. The low-$Re$ response reduces to a Reynolds-number-controlled group, the medium-$Re$ response retains two-input dependence, and the high-$Re$ response reduces to a roughness-controlled group. Thus, the same physical law is effectively one-dimensional at both ends of the Reynolds-number range but retains genuine two-input dependence in the medium-$Re$ regime. Importantly, the corresponding exponents are recovered from the noisy response data rather than supplied from the known limiting relations.

\begin{table}[htbp]
    \centering
    \caption{\textbf{Two-stage results for the three rough-pipe-flow regimes.} Values are shown for one noise realisation per regime with $\tau_1=4^\circ$ and $\tau=5\%$. Angles are in degrees. The reported $\hat u$ corresponds to the direction selected in stage two; $\hat u$ and $1/\hat u$ represent the same one-dimensional coordinate. Exact decision rules are given in Methods.}
    \label{tab:pipe_verdicts}
    \footnotesize
    \setlength{\tabcolsep}{2.5pt}
    \renewcommand{\arraystretch}{1.05}
    \begin{tabular}{@{}lrrrrlrrrl@{}}
        \toprule
        & \multicolumn{4}{c}{Stage one} & & \multicolumn{3}{c}{Stage two} & \\
        \cmidrule(lr){2-5} \cmidrule(lr){7-9}
        Regime & $\sigma_\alpha$ & $\sigma_{\rm noise}$ & $q_{95}$ & $\hat\sigma_{\rm true}$ & $\hat u$ & $\rho$ & $\tau_{\min}$ & $\hat\rho_{\rm true}$ & Verdict \\
        \midrule
        Pipe, low $Re$       & 0.83  & 0.95 & 0.78 & 0.00  & $Re^{-1.00}$                          & 0.0080 & 0.0090 & 0     & reducible \\
        Pipe, medium $Re$    & 10.12 & 0.72 & 0.56 & 10.10 & $Re^{-0.27}(\epsilon/D)^{0.96}$       & 0.149  & 0.0101 & 0.149 & not-red \\
        Pipe, high $Re$      & 0.75  & 1.36 & 1.37 & 0.00  & $(\epsilon/D)^{1.00}$                 & 0.0086 & 0.0088 & 0     & reducible \\
        \bottomrule
    \end{tabular}
\end{table}

\subsection*{Regime-dependent reducibility in wound closure}

\begin{figure}[htbp]
    \centering
    \includegraphics[width=0.8\linewidth]{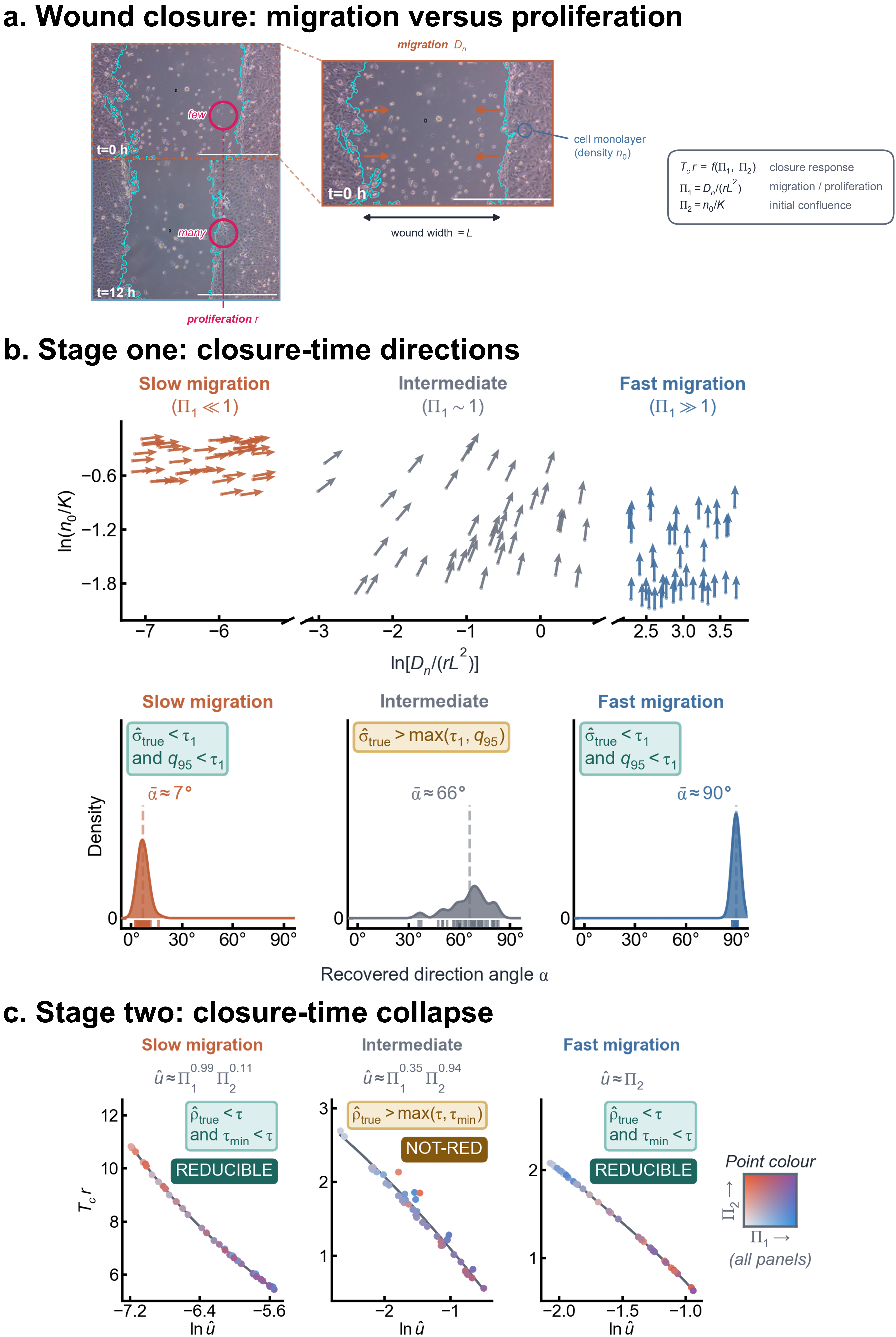}
    \caption{\textbf{Regime-dependent reducibility in wound closure.}
    \textbf{a}, Scratch-wound closure driven by cell migration ($D_n$) and proliferation ($r$), with dimensionless response $T_c r=f(\Pi_1,\Pi_2)$, where $\Pi_1=D_n/(rL^2)$ and $\Pi_2=n_0/K$.
    \textbf{b}, Stage-one directional analysis. Top: local directions recovered at $M=40$ anchors in the slow, intermediate and fast migration regimes. Bottom: distributions of the axial direction angle $\alpha$, with dashed lines indicating the axial mean $\bar\alpha$.
    \textbf{c}, Stage-two collapse onto the recovered one-dimensional coordinate $\hat u$. Slow- and fast-migration regimes collapse onto single curves, whereas the intermediate regime retains systematic two-input variation. Point colour indicates position in the $(\Pi_1,\Pi_2)$ plane. Numerical values are given in Table~\ref{tab:wound_verdicts}.}
    \label{fig:wound_closure}
\end{figure}

We next consider wound closure, for which the response map is obtained numerically rather than from a closed-form relation. In a scratch-wound assay, a gap of width $L$ in a cell monolayer closes through cell migration into the gap and proliferation behind the advancing front (Fig.~\ref{fig:wound_closure}a)~\citep{liang2007vitro,trepat2009physical}. A Fisher--KPP description of cell-sheet wound closure~\citep{habbal2014assessing,johnston2015estimating} involves the migration coefficient $D_n$, proliferation rate $r$, carrying capacity $K$ and initial density $n_0$. Dimensional analysis gives the closure response
\begin{align}
    T_c r=f(\Pi_1,\Pi_2),
\end{align}
with
\begin{align}
    \Pi_1=\frac{D_n}{rL^2}, \quad \Pi_2=\frac{n_0}{K},
\end{align}
where $\Pi_1$ compares migration and proliferation time scales and $\Pi_2$ represents the initial confluence. Each evaluation of the response is obtained by numerically solving the governing equation (Methods).

\begin{table}[htbp]
    \centering
    \caption{\textbf{Two-stage results for the three wound-closure regimes.} Values are shown for one noise realisation per regime with $\tau_1=4^\circ$ and $\tau=5\%$. Angles are in degrees. The reported $\hat u$ corresponds to the direction selected in stage two. Exact decision rules are given in Methods.}
    \label{tab:wound_verdicts}
    \footnotesize
    \setlength{\tabcolsep}{2.5pt}
    \renewcommand{\arraystretch}{1.05}
    \begin{tabular}{@{}lrrrrlrrrl@{}}
        \toprule
        & \multicolumn{4}{c}{Stage one} & & \multicolumn{3}{c}{Stage two} & \\
        \cmidrule(lr){2-5} \cmidrule(lr){7-9}
        Regime & $\sigma_\alpha$ & $\sigma_{\rm noise}$ & $q_{95}$ & $\hat\sigma_{\rm true}$ & $\hat u$ & $\rho$ & $\tau_{\min}$ & $\hat\rho_{\rm true}$ & Verdict \\
        \midrule
        Slow         & 2.57  & 2.05 & 1.54 & 1.55  & $\Pi_1^{0.99}\Pi_2^{0.11}$ & 0.0142 & 0.0077 & 0.0119 & reducible \\
        Intermediate & 11.10 & 0.46 & 0.35 & 11.10 & $\Pi_1^{0.35}\Pi_2^{0.94}$ & 0.184  & 0.0092 & 0.184  & not-red \\
        Fast         & 0.62  & 0.60 & 0.48 & 0.17  & $\Pi_2^{1.00}$             & 0.0075 & 0.0080 & 0      & reducible \\
        \bottomrule
    \end{tabular}
\end{table}

We consider three regimes (Fig.~\ref{fig:wound_closure}b): slow migration, with $\Pi_1\in[6\times10^{-4},5\times10^{-3}]$ and $\Pi_2\in[0.40,0.90]$; an intermediate regime, with $\Pi_1\in[0.03,3]$ and $\Pi_2\in[0.12,0.90]$; and fast migration, with $\Pi_1\in[8,50]$ and $\Pi_2\in[0.10,0.45]$. In the slow-migration limit, closure is governed primarily by the migration--proliferation balance represented by $\Pi_1$, whereas in the fast-migration limit the density rapidly homogenises across the gap and the subsequent closure is governed primarily by the initial confluence $\Pi_2$. In the intermediate regime, both groups contribute. The resulting two-stage statistics are summarised in Table~\ref{tab:wound_verdicts}.

Stage one recovers these changes in directional structure (Fig.~\ref{fig:wound_closure}b). In the slow-migration regime, the local directions cluster close to the $\ln\Pi_1$ axis, with $\bar\alpha\approx7^\circ$, whereas in the fast-migration regime they cluster close to the $\ln\Pi_2$ axis, with $\bar\alpha\approx90^\circ$. Their noise-corrected directional spreads are $\hat\sigma_{\rm true}=1.55^\circ$ and $0.17^\circ$, respectively, both below the prescribed tolerance $\tau_1=4^\circ$; the corresponding $q_{95}$ values, $1.54^\circ$ and $0.48^\circ$, also lie below $\tau_1$. By contrast, the directions vary strongly across the intermediate regime, producing a broad distribution centred at $\bar\alpha\approx66^\circ$. Here $\hat\sigma_{\rm true}=11.10^\circ$, well above both $\tau_1=4^\circ$ and $q_{95}=0.35^\circ$. Stage one therefore returns reducible, not-red and reducible for the slow, intermediate and fast regimes, respectively.

Stage two gives the same regime separation (Fig.~\ref{fig:wound_closure}c). In the slow-migration regime, the recovered coordinate is
\begin{align}
    \hat u\approx\Pi_1^{0.99}\Pi_2^{0.11},
\end{align}
and the observed residual is $\rho=0.0142$. After removal of the analytical noise contribution, $\hat\rho_{\rm true}=0.0119$, well below $\tau=0.05$, so stage two returns reducible. In the fast-migration regime, the recovered coordinate approaches
\begin{align}
    \hat u\approx\Pi_2,
\end{align}
with $\rho=0.0075<\tau_{\min}=0.0080$ and hence $\hat\rho_{\rm true}=0$, again giving reducible. In the intermediate regime, the best one-dimensional coordinate,
\begin{align}
    \hat u\approx\Pi_1^{0.35}\Pi_2^{0.94},
\end{align}
leaves a much larger residual, with $\rho=0.184$ and $\hat\rho_{\rm true}=0.184$, compared with $\tau_{\min}=0.0092$ and $\tau=0.05$. The remaining colour variation at similar values of $\hat u$ shows that the second input continues to affect the closure response after projection, giving a not-red verdict.

The slow-migration regime therefore represents an approximate rather than exact reduction: small directional and collapse departures remain, but both are well within the prescribed tolerances, so the response is classified as reducible. By contrast, the fast-migration regime approaches a nearly pure $\Pi_2$ dependence. Thus, as in rough-pipe flow, the same governing model changes from effectively one-dimensional to genuinely two-dimensional and back again as the physical regime changes.

\subsection*{Exact reductions across domains: droplet generation and column buckling}

\begin{figure}[htbp]
    \centering
    \includegraphics[width=\linewidth]{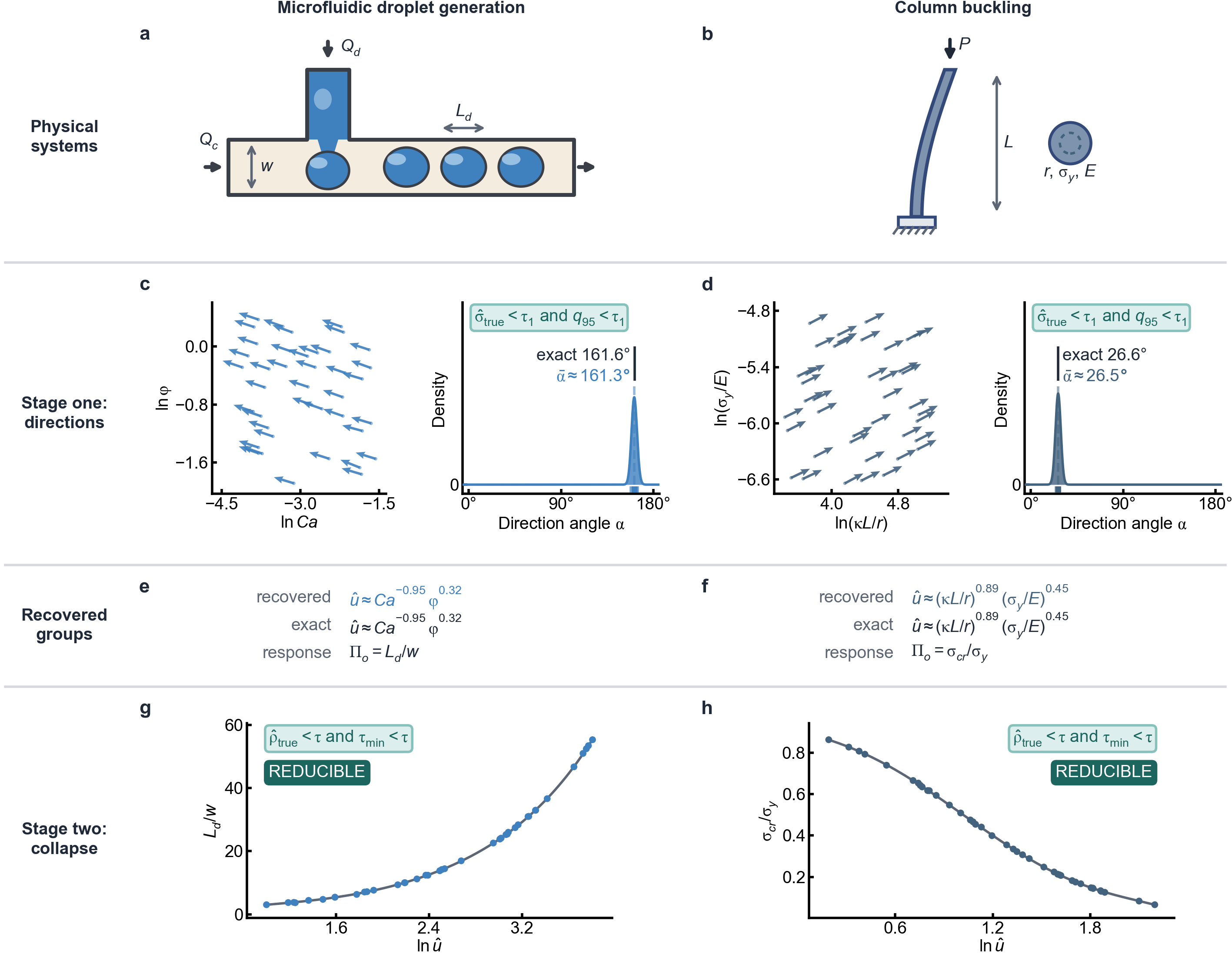}
    \caption{\textbf{Recovery of exact one-group coordinates across two physical domains.}
    \textbf{a}, Microfluidic droplet generation controlled by capillary number $Ca$ and flow-rate ratio $\phi=Q_d/Q_c$.
    \textbf{b}, Column buckling governed by slenderness $\kappa L/r$ and material ratio $\sigma_y/E$.
    \textbf{c}, \textbf{d}, Stage-one directional analysis showing the recovered local directions and their axial distributions; black marks indicate the exact directions.
    \textbf{e}, \textbf{f}, Recovered one-group coordinates compared with their exact values.
    \textbf{g}, \textbf{h}, Stage-two collapse of the responses onto the recovered coordinates. Both systems return reducible. Numerical values are given in Table~\ref{tab:cross_domain}.}
    \label{fig:cross_domain}
\end{figure}

The previous systems test how reducibility changes across physical regimes. We next consider two systems for which the response depends exactly on a known one-group coordinate, but with different physical origins and functional forms (Fig.~\ref{fig:cross_domain}).

\begin{table}[htbp]
    \centering
    \caption{\textbf{Two-stage results for microfluidic droplet generation and column buckling.} Values are shown for one noise realisation per system with $\tau_1=4^\circ$ and $\tau=5\%$. Angles are in degrees. The reported $\hat u$ corresponds to the direction selected in stage two. Exact decision rules are given in Methods.}
    \label{tab:cross_domain}
    \footnotesize
    \setlength{\tabcolsep}{2.5pt}
    \renewcommand{\arraystretch}{1.05}
    \begin{tabular}{@{}lrrrrlrrrl@{}}
        \toprule
        & \multicolumn{4}{c}{Stage one} & & \multicolumn{3}{c}{Stage two} & \\
        \cmidrule(lr){2-5} \cmidrule(lr){7-9}
        System & $\sigma_\alpha$ & $\sigma_{\rm noise}$ & $q_{95}$ & $\hat\sigma_{\rm true}$ & $\hat u$ & $\rho$ & $\tau_{\min}$ & $\hat\rho_{\rm true}$ & Verdict \\
        \midrule
        Microfluidic    & 1.12 & 1.41 & 1.34 & 0.00 & $Ca^{-0.95}\phi^{0.32}$                   & 0.0067 & 0.0073 & 0 & reducible \\
        Column buckling & 0.76 & 1.02 & 0.87 & 0.00 & $(\kappa L/r)^{0.89}(\sigma_y/E)^{0.45}$ & 0.0070 & 0.0089 & 0 & reducible \\
        \bottomrule
    \end{tabular}
\end{table}

In a microfluidic T-junction, droplet generation depends on the capillary number $Ca$ and the flow-rate ratio $\phi=Q_d/Q_c$ (Fig.~\ref{fig:cross_domain}a)~\citep{thorsen2001dynamic,garstecki2006formation}. We use the constructed benchmark
\begin{align}
    \frac{L_d}{w}=Ca^{-1}\phi^{1/3},
\end{align}
over $Ca\in[10^{-2},3\times10^{-1}]$ and $\phi\in[0.1,2]$. The response therefore depends exactly on the single group $Ca^{-1}\phi^{1/3}$. Under the unit-norm convention used for the recovered direction, the equivalent coordinate is
\begin{align}
    \hat u=Ca^{-0.95}\phi^{0.32},
\end{align}
corresponding to an axial direction of $161.6^\circ$. The dependence on $Ca$ and $\phi$ is motivated by microfluidic droplet formation, whereas the specific exponents are chosen for this benchmark rather than taken as a published correlation.

For column buckling, the Rankine--Gordon relation
\begin{align}
    \frac{\sigma_{cr}}{\sigma_y} = \frac{\pi^2}{\pi^2+(\kappa L/r)^2\sigma_y/E}
\end{align}
depends on the slenderness $\kappa L/r$ and material ratio $\sigma_y/E$ only through the product
\begin{align}
    (\kappa L/r)^2(\sigma_y/E)
\end{align}
(Fig.~\ref{fig:cross_domain}b)~\citep{ross1999strength,timoshenko2012theory}. Over $\kappa L/r\in[20,200]$ and $\sigma_y/E\in[10^{-3},10^{-2}]$, the corresponding unit-normalised coordinate is
\begin{align}
    \hat u=(\kappa L/r)^{0.89}(\sigma_y/E)^{0.45},
\end{align}
with axial direction $26.6^\circ$. Neither the exact group nor the functional dependence of the response on that group is supplied to the framework. The resulting two-stage statistics are summarised in Table~\ref{tab:cross_domain}.

Stage one recovers both known directions closely (Fig.~\ref{fig:cross_domain}c,d). For droplet generation, the recovered mean direction is $\bar\alpha\approx161.3^\circ$, compared with the exact $161.6^\circ$. The observed spread, $\sigma_\alpha=1.12^\circ$, is smaller than the estimated noise-induced spread of $1.41^\circ$, giving $\hat\sigma_{\rm true}=0$. For column buckling, $\bar\alpha\approx26.5^\circ$ compared with the exact $26.6^\circ$, while $\sigma_\alpha=0.76^\circ$ and $\sigma_{\rm noise}=1.02^\circ$ give $\hat\sigma_{\rm true}=0$. In both cases the noise-corrected spread lies below $\tau_1=4^\circ$, with $q_{95}=1.34^\circ$ and $0.87^\circ$, respectively, and stage one returns reducible.

Stage two also recovers the exact one-group coordinates (Fig.~\ref{fig:cross_domain}e,f). For droplet generation, the recovered coordinate is $\hat u\approx Ca^{-0.95}\phi^{0.32}$, while for column buckling it is $\hat u\approx(\kappa L/r)^{0.89}(\sigma_y/E)^{0.45}$. In both systems, the responses collapse onto single curves (Fig.~\ref{fig:cross_domain}g,h). The observed residuals, $\rho=0.0067$ and $0.0070$, lie below the corresponding analytical noise scales, $\tau_{\min}=0.0073$ and $0.0089$, giving $\hat\rho_{\rm true}=0$ in both cases. Stage two therefore returns reducible for both systems.

Across these two domains, the framework therefore recovers the known exact group and identifies no collapse departure beyond that attributable to evaluation noise. This recovery is achieved without supplying either the group direction or the functional form of the one-dimensional response.

\subsection*{Eight regimes on a common decision map}

\begin{figure}[htbp]
    \centering
    \includegraphics[width=\linewidth]{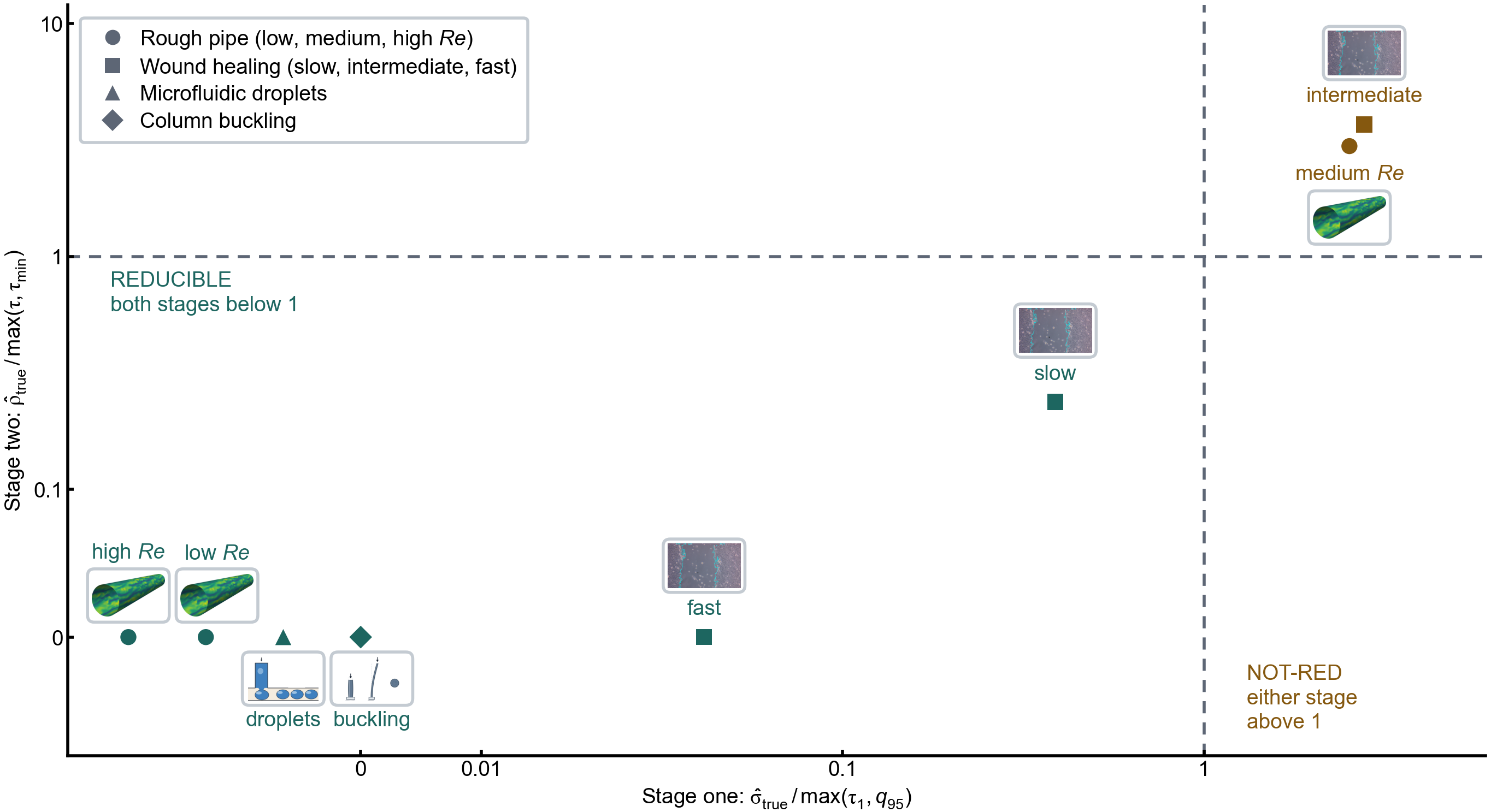}
    \caption{\textbf{Eight benchmark regimes on a common decision map.} Each regime is positioned by the stage-one ratio $\hat\sigma_{\rm true}/\max(\tau_1,q_{95})$ and the stage-two ratio $\hat\rho_{\rm true}/\max(\tau,\tau_{\min})$, with $\tau_1=4^\circ$ and $\tau=5\%$. Dashed lines mark unity on each axis. For all eight benchmarks, $q_{95}<\tau_1$ and $\tau_{\min}<\tau$, so no case is unresolved; the lower-left region therefore corresponds to reducible, whereas a ratio above unity on either axis gives not-red. Marker shape identifies the physical system and colour the verdict. The axes use a linear region near zero and logarithmic scaling at larger values to accommodate cases with zero noise-corrected departure. Numerical values are reported in Tables~\ref{tab:pipe_verdicts}, \ref{tab:wound_verdicts} and \ref{tab:cross_domain}.}
    \label{fig:decision_map}
\end{figure}

We finally place all eight benchmark regimes on a common decision map (Fig.~\ref{fig:decision_map}). The horizontal axis shows the stage-one departure,
\begin{align}
    \frac{\hat\sigma_{\rm true}}{\max(\tau_1,q_{95})},
\end{align}
and the vertical axis the stage-two departure,
\begin{align}
    \frac{\hat\rho_{\rm true}}{\max(\tau,\tau_{\min})}.
\end{align}
A value above one therefore indicates a resolved departure at the corresponding stage.

For all eight benchmarks, the noise-based references lie below the prescribed tolerances, with $q_{95}\le1.54^\circ<\tau_1=4^\circ$ and $\tau_{\min}\le0.0101<\tau=0.05$. None of the benchmark cases is therefore unresolved, and the two axes reduce to the estimated departures normalised by their respective tolerances. All six reducible regimes lie well inside the lower-left region: their stage-one ratios are at most $0.39$ and their stage-two ratios at most $0.24$. By contrast, the medium-$Re$ pipe-flow and intermediate wound-closure regimes lie beyond unity on both axes, with stage-one ratios of approximately $2.5$ and $2.8$ and stage-two ratios of approximately $3.0$ and $3.7$, respectively. The two stages therefore give consistent classifications across all eight regimes.

The benchmark cases are therefore well separated from the thresholds used in the two-stage decision rule. They test whether the framework recovers the expected group structure and distinguishes reducible from non-reducible regimes when the ground truth is known. Cases closer to the classification thresholds are examined separately using synthetic response fields, including directional departures comparable to $q_{95}$, collapse residuals comparable to $\tau_{\min}$, and variations in the prescribed tolerances (Supplementary Figs.~\ref{fig:s1}--\ref{fig:s3} and \ref{fig:s5}).

\section*{Discussion}

Scaling relations provide compact descriptions of physical responses, but identifying them from noisy observations requires distinguishing genuine response structure from variation introduced by evaluation noise. In the present setting, this question takes the form of whether two dimensionless inputs can be combined into a single composite coordinate that preserves the response over a specified regime. We address this problem using equal-response probing with replicate calibration. Across eight regimes drawn from four physical systems, the framework recovered the expected one-group structure in the reducible cases and identified the two regimes that retained genuine two-input dependence. In rough-pipe flow and wound closure, the same governing relation changed from effectively one-dimensional to two-input and back again as the physical regime changed. In the droplet-generation and column-buckling benchmarks, the recovered directions agreed with the known directions to within a few tenths of a degree and the collapse residuals did not exceed the analytical noise scale. The slow-migration wound-closure regime further showed that the framework can identify an approximate reduction: small departures remained, but both stayed within the prescribed tolerances. These results emphasise that the existence of a reduced scaling representation is a property of the response over a specified regime rather than of a physical law as a whole, consistent with approaches that seek regime-dependent or local low-dimensional structure rather than a single global reduction~\citep{ZHANG2024116728,romor2024local}.

The two stages address complementary signatures of one-group structure. For a differentiable response of the form
\begin{align}
    y = g(a x_1 + b x_2),
\end{align}
all non-zero gradients are parallel to the same direction. In the present two-input setting, this gives exact one-group reducibility a particularly simple geometric signature: the connected equal-response branches are parallel in logarithmic coordinates, and their normals define a common response direction. Equal-response probing therefore provides a finite-scale route to the candidate composite coordinate without requiring direct numerical differentiation. Directional consistency is a necessary signature of exact reducibility, but it does not by itself quantify the accuracy of the resulting one-dimensional representation. Response fields with identical gradient-vector statistics can exhibit substantially different collapse residuals (Supplementary Note~\ref{sec:sn1_direction_collapse}). Active-subspace methods use gradient information to identify important response directions and provide approximation-error bounds under suitable assumptions~\citep{constantine2014active,constantine2015active}. In the present framework, stage two directly evaluates the residual of the fitted one-dimensional representation, while replicate evaluations estimate the contribution of evaluation noise. Together with the equal-response directional read-out, this provides a quantitative assessment of the candidate reduction at prescribed directional and collapse tolerances. Obtaining reliable gradients can also be difficult when response evaluations are expensive or noisy, motivating surrogate-based and gradient-free alternatives for subspace identification~\citep{wycoff2021sequential,gautier2022fully,tripathy2016gaussian}. Conversely, a small collapse residual does not imply an exactly common local direction; it establishes only that the remaining departure from a one-dimensional representation is sufficiently small at the prescribed tolerance. Mathematically, the one-group representation is also related to single-index models and sufficient dimension reduction methods, which seek low-dimensional linear projections that retain response-relevant information in multivariate predictors~\citep{li1991sliced}. The framework is therefore complementary to approaches that infer dimensionless forms or global surrogates from existing data~\citep{bakarji2022dimensionally,XU2022111145,villar2023dimensionless,hang2023novel,murari2023combining,shi2025hybrid}. When such a surrogate is available, its gradients can provide directional information, while the collapse criterion can still be used to assess whether the resulting reduction is sufficiently accurate.

The prescribed tolerances and the replicate-based noise references play different roles in distinguishing structural departure from variation attributable to evaluation noise. The tolerances $\tau_1$ and $\tau$ specify how much directional and collapse departure is acceptable for the intended reduction, whereas $q_{95}$ and $\tau_{\min}$ quantify what can be distinguished in the presence of evaluation noise. The latter are calibrated from replicate evaluations rather than prescribed externally. No parametric distribution is supplied to the framework: the noise magnitude is estimated from the replicate evaluations and the stage-one null distribution is generated by residual resampling, subject to the assumed additive evaluation errors that are independent and identically distributed, have a common mean that is constant across evaluated points, and have finite variance. Synthetic tests characterise the empirical calibration and its limits. For $800$ randomly generated exactly reducible fields, the stage-one rejection rate at $q_{95}$ is $5.00\%$ at the calibration noise and $1.00\%$ at four times that noise, indicating conservative behaviour in the higher-noise ensemble (Supplementary Fig.~\ref{fig:s2}). However, within the calibration-noise ensemble, the rejection rate is $8.0\%$ ($19$ of $237$ runs) when $q_{95}$ is less than three times the noise-free read-out spread. Thus, aggregate agreement with the nominal rate does not imply uniform calibration across fields, particularly when deterministic read-out errors become comparable to the null threshold. For stage two, the pooled median ratio $\rho/\tau_{\min}$ is $1.04$, $0.98$ and $0.97$ at $0.25$, $1$ and $4$ times the calibration noise, respectively (Supplementary Fig.~\ref{fig:s3}a). These values support the analytical noise scale over the tested ensembles, while profile-dependent direction-search and polynomial-approximation errors become visible at the lowest noise level (Supplementary Note~\ref{sec:sn2_stage2_low_noise}). Stage-one detection reaches $80\%$ when the noise-free gradient-direction spread is $1.41$--$1.62$ times $q_{95}$, and all tested runs with $\sigma_{\rm grad}\ge2q_{95}$ are detected (Supplementary Fig.~\ref{fig:s1}). Varying $\tau_1$ on the same synthetic runs further shows how the prescribed tolerance changes the balance between unresolved and incorrect resolved verdicts (Supplementary Fig.~\ref{fig:s5} and Supplementary Table~\ref{tab:tau1}). These results clarify the role of the unresolved outcome: when evaluation noise is too large relative to the prescribed tolerance to distinguish a structural departure, the framework does not force either a reducible or not-red classification. The tolerance therefore remains an application-dependent choice, while replicate calibration provides an independent noise-based reference against which that choice can be interpreted.

Several aspects define the present scope. The current framework is developed for reduction from two dimensionless inputs to one composite group. This setting is special in two related ways. Geometrically, one equal-response direction uniquely determines its axial normal in two dimensions, allowing a finite-scale response direction to be recovered from a single equal-response crossing. Statistically, the resulting directions can be represented by a single axial angle, which enables the doubled-angle directional statistics and the associated noise decomposition used here. Neither feature extends unchanged to higher dimensions. Even for reduction from $d$ inputs to one group, recovering the response direction from level-set geometry requires multiple independent tangent directions, and the directional-dispersion and noise-calibration statistics would need to be reformulated. More generally, higher-dimensional reduction need not terminate at one group: a response in $d$ inputs may reduce to an $r$-dimensional representation with $r>1$, in which case the relevant object is a common response subspace rather than a single common direction. Extending the present framework would therefore require both a higher-dimensional geometric read-out and a corresponding noise-resolved treatment of subspace variation, rather than a direct extension of the current axial statistics.

The noise model also defines the present scope. Evaluation errors are assumed to be additive, independent and identically distributed, to have a common mean that is constant across evaluated points, and to have finite variance. Spatially varying bias, correlated errors, temporal drift and heteroscedasticity are not considered here. The geometric read-out additionally requires a crossing on the level-set branch through the anchor. Branch ambiguity in non-monotone responses and directional non-identifiability in locally constant regions are not resolved by the present crossing-selection rule. The finite-scale directional read-out and finite anchor sampling introduce additional approximations, although the read-out errors remain small for the benchmark settings used here (Supplementary Fig.~\ref{fig:s6}). The present validation further relies on benchmark and synthetic response fields for which the noise-free response is available, with synthetic evaluation noise added to the evaluations. This controlled setting provides the reference quantities needed to assess the recovered directions, collapse residuals and final verdicts directly. Extending the framework to measured responses and more complex noise structures therefore represents a natural next step.

Within the scope considered here, the results show how noisy dimensionless evaluations can be used to assess a candidate one-group scaling representation over a specified regime. Equal-response probing supplies a candidate coordinate and tests directional consistency, while replicate calibration estimates the variation attributable to evaluation noise. The collapse test then assesses the residual error of the fitted one-dimensional representation. The joint verdict combines these complementary criteria at prescribed tolerances, retaining an unresolved outcome when the noise-based resolution criteria are not met. A reducible verdict supports the candidate representation under both criteria; a not-red verdict identifies failure of at least one criterion under the implemented protocol. These outcomes provide a regime-specific assessment of reduced scaling structure, rather than an unconditional assignment of dimensionality to an entire physical relation.

\section*{Methods}

\subsection*{Notation and the two statistics}

The noise-free dimensionless response is $\Pi_o\equiv y=f(\Pi_1,\Pi_2)$, evaluated over a rectangular regime $\Omega$ in the logarithmic coordinates $x_1=\ln\Pi_1$ and $x_2=\ln\Pi_2$. Each observed evaluation is
\begin{align}
    y^{\rm obs} = y + e.
\end{align}
A reduction to one group means
\begin{align}
    y = g \bigl(a x_1 + b x_2 \bigr),\quad \hat u = \Pi_1^{\,a}\Pi_2^{\,b}, \quad a^2+b^2=1,
    \label{eq:ridge}
\end{align}
with $g$ an unknown univariate profile~\citep{FriedmanTukey1974,pinkus2015ridge,constantine2016many}; the exponent vector of $\hat u$ lies in the null space of the dimension matrix, so $\hat u$ remains dimensionless for any $(a,b)$ (Supplementary Methods~\ref{sec:sm1_nullspace})~\citep{Buckingham1914,Bridgman1922}. The direction $(a,b)$ is axial and is represented by $\alpha=\operatorname{atan2}(b,a)$ modulo $\pi$. If $g$ is differentiable, $\nabla y=g'(a x_1+b x_2)\,(a,b)$ is parallel to one direction everywhere in $\Omega$; conversely, a gradient that is everywhere parallel to one direction on a convex $\Omega$ makes $y$ a function of $a x_1+b x_2$ alone. Writing $\alpha_{\rm grad}(\bm x)$ for the axial angle of the analytic gradient wherever $\nabla y\neq0$, the response is exactly reducible on $\Omega$ if and only if all non-zero gradients are parallel to a common direction, equivalently if the axial spread of $\alpha_{\rm grad}$, denoted $\sigma_{\rm grad}$, is zero. The second statistic, the collapse residual $\rho$, measures the departure of the averaged anchor responses from the univariate polynomial fit selected by the stage-two procedure, with degrees-of-freedom correction and normalisation by the sample standard deviation of those responses. It assesses the accuracy of the fitted representation on the sampled anchors and is not determined by the direction statistics, since response fields with identical sets of sampled gradients can exhibit markedly different collapse residuals (Supplementary Note~\ref{sec:sn1_direction_collapse}).

\paragraph{Noise model.}
Each observed evaluation is $y^{\rm obs}=y+e$, where the errors of distinct evaluations are independent and identically distributed, have a common mean $m_e$ that is constant across evaluated points, and have finite common variance $\varepsilon^2$. Neither the distribution of $e$, the common mean $m_e$, nor $\varepsilon$ is assumed known. The common mean cancels from probe--anchor differences in stage one and is absorbed by the intercept of the stage-two fit; spatially varying mean errors lie outside the present noise model.

\subsection*{Evaluation protocol}

The fixed protocol and analysis settings are summarised in Supplementary Table~\ref{tab:protocol}. A run consists of $M$ anchors, $K$ probe points per anchor plus the anchor itself, and $n$ repeated evaluations of every point, $M(K+1)n$ evaluations in all (Fig.~\ref{fig:methodology}b). All settings below were fixed before any benchmark was run and are identical for every system.

\paragraph{Anchors.}
Let $\Omega=[x_1^{-},x_1^{+}]\times[x_2^{-},x_2^{+}]$ have half-widths $L_1$ and $L_2$. We sample $M$ anchors $\bm c_i$ independently and uniformly from the interior region obtained by excluding margins of $\eta L_1$ and $\eta L_2$ from the corresponding boundaries of $\Omega$. The parameter $\eta$ sets the minimum distance between an anchor and the boundary, ensuring that a finite-scale probe can be placed around each anchor. The same anchors are subsequently used to evaluate the collapse residual. Unless otherwise stated, we use $M=40$ and $\eta=0.20$ in all benchmark calculations.

\paragraph{Probes.}
Around each anchor, we define a half-ellipse
\begin{align}
    \bm p_i(t) = \bm c_i + \bigl(h_{i1}\cos t,\,h_{i2}\sin t\bigr), \quad t\in[0,\pi],
    \label{eq:probe}
\end{align}
with semi-axes chosen as large as possible while the corresponding centred ellipse remains within $\Omega$,
\begin{align}
    h_{i1} &= \min\bigl(c_{i1}-x_1^{-},\,x_1^{+}-c_{i1}\bigr), \\
    h_{i2} &= \min\bigl(c_{i2}-x_2^{-},\,x_2^{+}-c_{i2}\bigr).
\end{align}
The anchor margins therefore ensure that $h_{i1}\ge\eta L_1$ and $h_{i2}\ge\eta L_2$. Because $\alpha$ and $\alpha+\pi$ represent the same axial direction, only one half of the ellipse is required.

At each anchor, the probe is evaluated at $K$ angles equally spaced over $[0,\pi]$, including both endpoints,
\begin{align}
    t_j = \frac{j\pi}{K-1}, \quad j=0,\ldots,K-1.
\end{align}
The parameter $K$ controls the angular discretisation of the probe and is fixed before the responses are evaluated. We use $K=15$ throughout the benchmarks reported here, corresponding to an angular spacing of approximately $12.9^\circ$. The query plan is fixed a priori and does not depend on the observed responses.

\paragraph{Repeats.}
Each probe point and anchor is evaluated $n$ times. The averaged responses are used in both stages, while the within-repeat residuals are used to estimate the evaluation noise. The total evaluation budget is therefore
\begin{align}
    M(K+1)n.
\end{align}
Unless otherwise stated, we use $M=40$, $K=15$ and $n=6$, corresponding to $3840$ evaluations per run. Each run of the framework comprises $3840$ response evaluations. This count excludes the preliminary noise-free evaluations used to set the injected noise level $\varepsilon$ for each benchmark ($200$ preliminary anchors), as well as additional noise-free evaluations used to compute reference gradient directions and collapse residuals.

\subsection*{Stage one: direction read-out and noise-corrected spread}

\paragraph{One direction per anchor.}
For anchor $i$, let $\bar y_{ij}^{\rm p}$ denote the mean of the $n$ evaluations at queried probe angle $t_j$, and let $\bar y_i^{\rm a}$ denote the corresponding mean at the anchor. We define the signed profile
\begin{align}
    s_{ij}=\bar y_{ij}^{\rm p}-\bar y_{i}^{\rm a},\quad j=0,\ldots,K-1,
    \label{eq:profile}
\end{align}
which vanishes where the probe intersects the equal-response contour through the anchor. The crossing angle $\hat t_i$ is obtained by linear interpolation between adjacent queried angles that bracket a sign change. If more than one sign change occurs, we select the pair with the smallest $|s_{ij}|+|s_{i,j+1}|$; if no sign change occurs, we take the queried angle with the smallest $|s_{ij}|$. The displacement $\bm p_i(\hat t_i)-\bm c_i$ approximates the local contour tangent, and its normal defines the local response direction,
\begin{align}
    \bm n_i=\bigl(h_{i2}\sin\hat t_i,\;-h_{i1}\cos\hat t_i\bigr),\qquad
    \alpha_i=\arg\bm n_i\bmod\pi.
    \label{eq:readout}
\end{align}
Here $\arg\bm n_i$ is the polar angle of $\bm n_i$; the direction is defined modulo $\pi$ because the sign of the normal is arbitrary. For a circular probe, $h_{i1}=h_{i2}$, Eq.~\eqref{eq:readout} reduces to $\alpha_i=\hat t_i-\pi/2\bmod\pi$. For an exact one-group response, an equal-response crossing on the straight level-set branch through the anchor recovers the axial response direction independently of the probe size and aspect ratio. A sufficient condition is that the ridge profile $g$ be injective over the range of projected coordinates covered by the anchor and its probe: equal responses then imply equal projected coordinates. For a non-monotone profile, equal responses may instead occur on distinct level-set branches, and the crossing-selection rule above does not guarantee selection of the branch through the anchor. For a general response, a crossing on the same curved branch gives a chord whose normal provides a finite-scale approximation to the local gradient direction. The corresponding chord and angular-interpolation errors are quantified in Supplementary Fig.~\ref{fig:s6}. When no sign change is found, the minimum-magnitude fallback supplies a numerical read-out but does not carry the same geometric guarantee. 

\paragraph{Axial statistics.}
The $M$ directions are summarised by their axial mean and spread~\citep{mardia2009directional},
\begin{align}
    z=\frac1M\sum_{i=1}^{M}e^{2\mathrm i\alpha_i}, \quad \bar\alpha=\tfrac12\arg z, \quad \sigma_\alpha=\tfrac12\sqrt{-2\ln|z|},
    \label{eq:axial}
\end{align}
where all angular quantities entering the doubled-angle expressions and the analytical bounds below are evaluated in radians and are converted to degrees only for reporting and for comparison with $\tau_1$. The spread $\sigma_\alpha$ is reported in degrees. It vanishes if and only if all directions coincide modulo $\pi$, and for tightly clustered directions it approaches the ordinary standard deviation of the angles. The recovered direction is $(a,b)=(\cos\bar\alpha,\sin\bar\alpha)$, giving $\hat u=\Pi_1^{\cos\bar\alpha}\Pi_2^{\sin\bar\alpha}$. Because the direction is axial, $(a,b)$ and $(-a,-b)$ are equivalent and correspond to reciprocal group coordinates.

\paragraph{Noise-induced directional spread from the residual pool.}
Noise alone produces a non-zero directional spread, so the observed $\sigma_\alpha$ cannot be interpreted directly as structural variation (Fig.~\ref{fig:methodology}b). We estimate the noise-induced spread by re-running the complete direction read-out after removing the spatial structure of the response. For each of the $M(K+1)$ evaluated points, the within-repeat residuals are multiplied by $\sqrt{n/(n-1)}$ to correct for the reduction in variance introduced by centring. Pooling these residuals gives $M(K+1)n$ values, whose standard deviation provides an estimate $\hat\varepsilon$ of the single-evaluation error scale.

For each anchor, we use a Monte Carlo residual-resampling procedure to propagate the estimated evaluation noise through the direction read-out. The resampling follows the general bootstrap principle of approximating sampling variation from an empirical distribution~\citep{efron1992bootstrap,davison1997bootstrap}. Specifically, $N_{\rm MC}$ perturbed profiles are generated. For each Monte Carlo replicate at anchor $i$, one sample of size $n$ is drawn with replacement from the residual pool and averaged to generate a perturbation of the anchor response. Independently, a separate sample of size $n$ is drawn and averaged for each probe point. The perturbed probe--anchor differences are formed using the same sampled anchor perturbation for all probe angles at that anchor. This preserves the shared anchor-noise contribution, and hence the covariance structure among the probe--anchor differences within each perturbed profile. Each perturbed profile is passed through the same read-out procedure as the observed profile. If $\delta_{ir}$ denotes the axial deviation of the $r$th perturbed direction from the baseline read-out direction obtained from the original averaged profile at anchor $i$, wrapped to $(-\pi/2,\pi/2]$, we define
\begin{align}
    \sigma_{\rm noise}^2 =-\tfrac12\ln\Bigl|\frac1M\sum_{i=1}^{M}\varphi_i\Bigr|, \quad \varphi_i =\frac{1}{N_{\rm MC}}\sum_{r=1}^{N_{\rm MC}}e^{2\mathrm i\delta_{ir}} .
    \label{eq:sigma_noise}
\end{align}
This quantity is the axial spread expected when all noise-free read-out directions are equal and the observed variation arises solely from anchor-specific read-out uncertainty. Under this null model, the common direction factors out of $\mathbb E[z]$, yielding the population counterpart of Eq.~\eqref{eq:sigma_noise} without requiring the read-out uncertainties to have the same distribution at every anchor. Because the perturbed profiles pass through the same interpolation and crossing-selection procedure as the observations, these numerical effects are represented consistently in both.

\paragraph{Noise-corrected spread and null threshold.}
For a fixed underlying response field, let $\alpha_i^{0}$ denote the direction returned by the noise-free read-out at anchor $i$, and let $\delta_i$ denote the random perturbation induced by evaluation noise, so that
\begin{align}
    \alpha_i=\alpha_i^{0}+\delta_i \pmod{\pi}.
\end{align}
All expectations below are conditional on the fixed response field. Writing
\begin{align}
    \varphi_i=\mathbb E \left[e^{2\mathrm i\delta_i}\right]
\end{align}
for the anchor-specific characteristic value of the read-out perturbation, the expected axial resultant in Eq.~\eqref{eq:axial} is
\begin{align}
    \mathbb E[z] = \frac1M \sum_{i=1}^{M} e^{2\mathrm i\alpha_i^{0}} \varphi_i .
    \label{eq:Ez}
\end{align}

If the read-out perturbations have the same distribution at every anchor, $\varphi_i\equiv\varphi$, Eq.~\eqref{eq:Ez} factorises as
\begin{align}
    \mathbb E[z] = \varphi z^{0}, \quad z^{0} = \frac1M \sum_{i=1}^{M} e^{2\mathrm i\alpha_i^{0}} .
\end{align}
Defining
\begin{align}
    \sigma_{\rm true}^{2} = -\tfrac12\ln|z^{0}|, \quad \sigma_{\rm noise}^{2} = -\tfrac12\ln|\varphi|,
\end{align}
gives
\begin{align}
    -\tfrac12\ln|\mathbb E[z]| = \sigma_{\rm true}^{2} + \sigma_{\rm noise}^{2}.
    \label{eq:additivity}
\end{align}
Thus, at the level of the expected axial resultant, the structural and noise-induced spreads add exactly in quadrature when the anchor-specific read-out perturbations share a common distribution. Here $\sigma_{\rm true}$ denotes the structural spread of the noise-free read-out directions; deterministic discrepancies between the finite-scale read-out and the analytic gradient direction are assessed separately (Supplementary Fig.~\ref{fig:s6}).

This decomposition motivates the finite-sample estimator
\begin{align}
    \hat\sigma_{\rm true} = \sqrt{\max\bigl(\sigma_\alpha^2-\sigma_{\rm noise}^2,0 \bigr)},
    \label{eq:quadrature}
\end{align}
where $\sigma_\alpha$ is obtained from the observed directions and $\sigma_{\rm noise}$ from the anchor-specific Monte Carlo perturbations through Eq.~\eqref{eq:sigma_noise}. The truncation at zero accounts for finite-sample fluctuations that can give $\sigma_\alpha<\sigma_{\rm noise}$. Accordingly, $\hat\sigma_{\rm true}=0$ means that the observed directional spread does not exceed the estimated noise contribution. Because Eq.~\eqref{eq:quadrature} is applied to the resultant of one finite run rather than to $|\mathbb E[z]|$, its finite-$M$ behaviour is examined in Supplementary Note~\ref{sec:sn3_axial_quadrature}.

Identical read-out distributions are not required for calibration under the equal-read-out-direction null. Let
\begin{align}
    \bar\varphi=\frac1M\sum_{i=1}^M\varphi_i.
\end{align}
For $\bar\varphi\neq0$, write
\begin{align}
    \varphi_i=\bar\varphi(1+\epsilon_i), \quad \sum_{i=1}^M\epsilon_i=0.
\end{align}
Equation~\eqref{eq:Ez} can then be written as
\begin{align}
    \mathbb E[z] = \bar\varphi \left( z^{0}+r \right), \quad r = \frac1M \sum_{i=1}^{M} \epsilon_i e^{2\mathrm i\alpha_i^{0}} .
    \label{eq:heterogeneous}
\end{align}
Under the equal-read-out-direction null used to calibrate $q_{95}$, all $\alpha_i^{0}$ coincide modulo $\pi$, and therefore $r=0$ exactly, irrespective of the anchor-to-anchor variation in $\varphi_i$. Hence
\begin{align}
    -\tfrac12\ln|\mathbb E[z]| = -\tfrac12\ln|\bar\varphi|
\end{align}
under the null, which is precisely the population quantity estimated by Eq.~\eqref{eq:sigma_noise}. The noise-induced directional spread can therefore be calibrated without assuming identical read-out uncertainties across anchors.

Away from the null, the term $r$ quantifies the coupling between structural variation in the noise-free directions and anchor-to-anchor heterogeneity in the read-out uncertainty. Supplementary Note~\ref{sec:sn3_axial_quadrature} shows that the remainder in the quadrature decomposition vanishes as the structural spread tends to zero and derives an explicit bound in terms of the heterogeneity of the $\varphi_i$. Supplementary Table~\ref{tab:kappa} reports plug-in evaluations of this bound expression for the benchmark cases and a subset of the synthetic fields. End-to-end tests on $902$ synthetic fields further show that, when $\sigma_{\alpha,\mathrm{clean}}>2\sigma_{\rm noise}$, the median ratio $\hat\sigma_{\rm true}/\sigma_{\alpha,\mathrm{clean}}$ is $1.00$ at the calibration noise and $0.98$ at four times that noise. As the departure approaches $\sigma_{\rm noise}$, the estimator tends to read low, making stage one less likely to declare not-red (Supplementary Note~\ref{sec:sn3_axial_quadrature}).

The finite-sample null distribution of $\hat\sigma_{\rm true}$ is constructed by imposing the equal-read-out-direction null on the data using the anchor-specific perturbation distributions estimated from the current run. The anchor-specific perturbation distributions already remove the structural differences between anchors because each deviation is measured relative to the baseline read-out direction obtained from the original averaged profile before Monte Carlo perturbation. Drawing one deviation from each anchor-specific distribution produces a synthetic null realisation. The resulting directional spread represents the variation expected from noise alone if all noise-free read-out directions were identical, while preserving the anchor-specific distributions of noise-induced directional error.

For each of $N_{\rm F}$ synthetic null runs, one deviation is drawn independently from the perturbation distribution of every anchor, consistent with the independent-evaluation noise model. The axial spread of the resulting deviations is computed using Eq.~\eqref{eq:axial}, and the same correction in Eq.~\eqref{eq:quadrature} is applied. We define $q_{95}$ as the 95th percentile of the resulting null distribution. This construction does not presuppose that the observed response is reducible. The threshold $q_{95}$ is the estimated 95th percentile of the resampled equal-read-out-direction null distribution. Its nominal $5\%$ exceedance level refers to that estimated distribution; the false-alarm rate of the complete procedure additionally depends on how accurately the resampled perturbations represent the actual read-out errors. An observed $\hat\sigma_{\rm true}>q_{95}$ is therefore treated as evidence against the null under this calibration, whereas $\hat\sigma_{\rm true}\le q_{95}$ indicates that the statistic does not exceed the calibrated threshold. The false-alarm calibration of this threshold is evaluated on exactly reducible synthetic fields in Supplementary Fig.~\ref{fig:s2}, while its detection power for non-zero directional departures is quantified in Supplementary Fig.~\ref{fig:s1}. Unless otherwise stated, we use $N_{\rm MC}=1200$ perturbations per anchor and $N_{\rm F}=6400$ synthetic null runs.

\paragraph{Stage-one verdict.}
The user specifies an angular-spread tolerance $\tau_1$, defined as the largest structural axial spread acceptable for the intended use. The stage-one verdict is
\begin{align}
    \text{stage one}=
    \begin{cases}
        \text{not-red} & \hat\sigma_{\rm true}>\max(\tau_1,q_{95}),\\[1pt]
        \text{reducible} & \hat\sigma_{\rm true}\le\tau_1\ \text{and}\ q_{95}<\tau_1,\\[1pt]
        \text{unresolved}& \text{otherwise}.
    \end{cases}
    \label{eq:v1}
\end{align}
A structural departure is declared only when the estimated spread exceeds both the user-defined tolerance and the 95th percentile of the noise-only null distribution. Conversely, a reduction is declared only when the estimated structural spread lies within the tolerance and the null threshold itself lies below that tolerance. The unresolved branch therefore arises only when $q_{95}\ge\tau_1$ and the observed spread does not exceed the null threshold: in this regime, the 95th-percentile noise-only fluctuation exceeds the user-defined tolerance, so the noise-only threshold reaches or exceeds the prescribed tolerance and a reducible verdict is not resolved at that tolerance.

The equal-read-out-direction null is used to calibrate $q_{95}$, whereas $\tau_1$ defines the largest noise-corrected directional spread regarded as acceptable for the intended reduction. A reducible verdict should therefore be interpreted as a tolerance-based decision on the sampled anchor set: no directional spread exceeding $\tau_1$ is resolved under the present evaluation-noise level and finite-sample protocol. It does not establish exact reducibility everywhere in $\Omega$.

\subsection*{Stage two: collapse residual and tolerance verdict}

\paragraph{Collapse residual.}
Stage two uses only the $M$ averaged anchor responses $\bar y_{i}^{\rm a}$ already evaluated in stage one and therefore requires no additional response evaluations. For a candidate direction $\theta$, the anchors are projected onto
\begin{align}
    u_i(\theta) = \cos\theta \cdot c_{i1} + \sin\theta \cdot c_{i2},
\end{align}
and the projected coordinate is standardised as
\begin{align}
    \tilde u_i(\theta) = \frac{u_i(\theta)-\bar u}{\mathrm{sd}(u)}.
\end{align}
The anchor responses are then fitted by a least-squares polynomial $P_d$ of degree $d$ in $\tilde u$. We define the normalised collapse residual as
\begin{align}
    \rho(\theta) = \frac{1}{s_y} \sqrt{ \frac{1}{M-d-1} \sum_{i=1}^{M} \left[ \bar y_{i}^{\rm a} - P_d \left(\tilde u_i\right) \right]^2 }, \quad s_y = \mathrm{sd} \bigl( \{ \bar y_{i}^{\rm a} \}_{i=1}^{M} \bigr).
    \label{eq:rho}
\end{align}
Here $s_y$ is the sample standard deviation of the averaged anchor responses, computed with denominator $M-1$.

The polynomial degree is selected from the fixed ladder $d\in\{2,3,5,8,12,16,20\}$, advancing to the next degree only when it reduces $\rho$ by more than $5\%$. The degrees-of-freedom correction in Eq.~\eqref{eq:rho}, together with this stopping rule, reduces the tendency to reward additional polynomial flexibility without a commensurate improvement in fit. Allowing $d$ to adapt is important because an insufficient fixed degree could confound non-linearity of the unknown univariate response $g$ with failure of one-dimensional collapse. Normalisation by $s_y$ makes $\rho$ dimensionless: it measures the root-mean-square departure from the fitted one-dimensional response relative to the overall variation in the anchor responses, and can therefore be interpreted as the relative thickness of the collapse plot. The collapse residual therefore provides a direct quantitative assessment of the selected one-dimensional representation on the sampled anchors.

\paragraph{Direction search.}
The collapse direction is allowed to vary locally around the stage-one estimate $\bar\alpha$ rather than being fixed to it. We evaluate $\rho(\theta)$ on an equally spaced grid of $N_{\rm W}$ candidate directions within
\begin{align}
    \theta \in \left[ \bar\alpha-\frac{3\sigma_\alpha}{\sqrt M}, \bar\alpha+\frac{3\sigma_\alpha}{\sqrt M} \right],
\end{align}
and report the minimum value as $\rho$. Here $N_{\rm W}$ is a numerical search-resolution parameter rather than a theoretically prescribed constant; unless otherwise stated, we use $N_{\rm W}=21$. For concentrated directions, $\sigma_\alpha/\sqrt M$ provides an approximate standard error of the axial mean~\citep{mardia2009directional}, so the search window allows modest uncertainty in the recovered direction while preventing the collapse fit from drifting far from the direction identified in stage one. At very low noise, the finite directional read-out can become comparable to the width of this search window; the resulting effect on the stage-two residual is examined in Supplementary Note~\ref{sec:sn2_stage2_low_noise}.

\paragraph{Analytical noise scale.}
Noise gives the collapse plot a non-zero thickness even when the underlying response is exactly one-dimensional. For a fixed projection direction and polynomial degree $d$, write the vector of averaged anchor responses as $\bar{\bm y}=\bm\mu+\bm e$, where $\bm\mu$ contains the noise-free responses and $\bm e$ contains the errors of the means of $n$ repeated evaluations. Under the evaluation-noise model specified above,
\begin{align}
    \operatorname{Cov}(\bm e) = \frac{\varepsilon^2}{n}\bm I.
\end{align}
Let $\bm H$ denote the least-squares projection onto the $(d+1)$-dimensional polynomial space in the standardised projected coordinate. The residual vector is $(\bm I-\bm H)(\bm\mu+\bm e)$, and the corresponding residual sum of squares satisfies (Supplementary Note~\ref{sec:sn4_tau_min})
\begin{align}
    \mathbb E \left[ \left\|(\bm I-\bm H)\bar{\bm y}\right\|^2 \right] = \left\|(\bm I-\bm H)\bm\mu\right\|^2 + (M-d-1)\frac{\varepsilon^2}{n}.
    \label{eq:rho_rss}
\end{align}
After the degrees-of-freedom normalisation in Eq.~\eqref{eq:rho}, the expected noise contribution to the residual mean square is therefore $\varepsilon^2/n$. Using the residual-pool estimate $\hat\varepsilon$ and the observed response scale $s_y$, we define the corresponding analytical noise scale of the normalised residual as
\begin{align}
    \tau_{\min} = \frac{\hat\varepsilon}{\sqrt n\,s_y}.
    \label{eq:tau_min}
\end{align}
For fixed direction and degree, the factor $M-d-1$ cancels exactly with the degrees-of-freedom normalisation, so the expected noise contribution does not depend explicitly on $d$. The calibration of $\tau_{\min}$ on exactly reducible synthetic fields is evaluated in Supplementary Fig.~\ref{fig:s3}a. The remaining noise-free fitting contribution may contain both genuine departure from one-dimensional collapse and finite polynomial-approximation error.

Equation~\eqref{eq:tau_min} provides an analytical noise scale rather than a formal detection threshold. In the deployed procedure, both the projection direction and polynomial degree are selected from the same noisy responses through the window search and degree ladder, so the fixed-direction, fixed-degree decomposition does not apply exactly to the selected fit. On exactly reducible synthetic fields, the pooled median ratio $\rho/\tau_{\min}$ is $0.97$--$1.04$ over the tested sixteenfold noise range (Supplementary Fig.~\ref{fig:s3}a). This ensemble-level agreement supports the use of \(\tau_{\min}\) as an analytical reference for the noise contribution to the collapse residual. The residual returned by the complete procedure can also contain direction-search and polynomial-approximation errors, whose effects become apparent at the lowest tested noise level (Supplementary Note~\ref{sec:sn2_stage2_low_noise}).

The additive decomposition in Eq.~\eqref{eq:rho_rss}, together with the analytical noise scale in Eq.~\eqref{eq:tau_min}, also motivates removal of the expected noise contribution in quadrature. For a fixed direction and degree, the noise-free fitting contribution and the noise contribution add at the level of the squared residual. We therefore define
\begin{align}
    \hat\rho_{\rm true} = \sqrt{ \max\bigl( \rho^2-\tau_{\min}^2, 0 \bigr) }.
    \label{eq:rho_true}
\end{align}
The truncation at zero accounts for finite-sample fluctuations that can yield $\rho<\tau_{\min}$. We interpret $\hat\rho_{\rm true}$ as the component of the observed collapse residual not accounted for by the analytical noise scale. Thus, $\hat\rho_{\rm true}=0$ indicates that the observed residual does not exceed the estimated noise contribution, analogous to $\hat\sigma_{\rm true}=0$ in stage one.

The accuracy of this quadrature correction after the complete direction-search and degree-selection procedure is verified on synthetic fields with known non-zero collapse residuals. For imposed departures with $\lambda\ge0.05$ and $\rho_{\rm clean}>\tau_{\min}$, the median ratio $\hat\rho_{\rm true}/\rho_{\rm clean}$ is $1.00$--$1.05$ (Supplementary Fig.~\ref{fig:s3}b).

\paragraph{Stage-two verdict.}
Given the user-defined tolerance $\tau$ on the relative collapse residual, the stage-two verdict is
\begin{align}
    \text{stage two}=
    \begin{cases}
        \text{not-red}   & \hat\rho_{\rm true}>\max(\tau,\tau_{\min}),\\[1pt]
        \text{reducible} & \hat\rho_{\rm true}\le\tau\ \text{and}\ \tau_{\min}<\tau, \\[1pt]
        \text{unresolved}& \text{otherwise},
    \end{cases}
    \label{eq:v2}
\end{align}

The two thresholds play distinct roles. The user-defined $\tau$ specifies the largest noise-corrected collapse residual regarded as acceptable for the intended use, whereas $\tau_{\min}$ is the analytical noise scale determined by $\hat\varepsilon$, $n$ and $s_y$. Thus, a not-red verdict requires the estimated structural residual to exceed both the admissible tolerance and the analytical noise scale. The comparison with $\tau_{\min}$ is an operational resolution criterion, not a statistical significance test: unlike the stage-one threshold $q_{95}$, $\tau_{\min}$ is not associated with a prescribed false-alarm probability.

A reducible verdict requires both that the estimated structural residual lie within the user-defined tolerance and that the analytical noise scale itself lie below that tolerance. With these inequalities, unresolved occurs only when
\begin{align}
    \tau_{\min}\ge\tau \quad\text{and} \quad \hat\rho_{\rm true}\le\tau_{\min}.
\end{align}
In this regime, the analytical noise scale reaches or exceeds the requested tolerance, while the estimated structural residual does not rise above that analytical noise scale. A reducible verdict is therefore not resolved at the specified tolerance. Whether the condition $\tau_{\min}<\tau$ is met follows directly from $\hat\varepsilon$, $n$, and $s_y$.

This rule is operationally analogous to Eq.~\eqref{eq:v1}, but the quantities controlling resolution in the two stages have different statistical meanings: $q_{95}$ is a percentile of the stage-one null distribution, whereas $\tau_{\min}$ is an analytically derived noise scale.

\subsection*{Joint verdict}

The two stage-wise verdicts are combined as follows (Fig.~\ref{fig:methodology}c, bottom row). The response is classified as reducible only when both stages return reducible, as not-red when either stage returns not-red, and as unresolved otherwise. The joint verdict therefore assesses whether the sampled response satisfies both the prescribed directional-consistency and collapse-accuracy criteria. A not-red verdict indicates failure of at least one of these operational criteria; in particular, rejection by stage one alone does not exclude a one-dimensional approximation whose response error lies below $\tau$. The verdict concerns the sampled regime and the implemented evaluation and fitting procedure, rather than establishing the existence or absence of every possible one-group approximation. The only user-specified quantities entering the decision rules are the two tolerances, $\tau_1$ in degrees and the dimensionless $\tau$; all other quantities are determined from the evaluations. Throughout this work, we use $\tau_1=4^\circ$ and $\tau=5\%$ as illustrative values. The effect of varying the stage-one tolerance $\tau_1$ on unresolved and incorrect verdicts is quantified in Supplementary Fig.~\ref{fig:s5}.

\subsection*{Benchmark implementation}

The eight benchmark regimes comprise three rough-pipe-flow regimes, three wound-closure regimes, one microfluidic-droplet regime and one column-buckling regime. Their physical definitions, parameter ranges and expected reducibility are given in Results and summarised in Supplementary Table~\ref{tab:cases}. Here we describe only the numerical evaluation of the benchmark responses.

For rough-pipe flow, the laminar relation is evaluated directly, whereas the Colebrook relation used in the other two regimes is solved by fixed-point iteration~\citep{moody1944friction,colebrook1939correspondence,nikuradse1950laws}. The constructed microfluidic power-law benchmark and the Rankine--Gordon column-buckling relation are evaluated directly from the expressions given in Results~\citep{thorsen2001dynamic,garstecki2006formation,ross1999strength,timoshenko2012theory}. The microfluidic power law is used only as a constructed test relation; its specific exponents do not constitute a published correlation.

For wound closure, each response evaluation is obtained by numerically solving the nondimensional Fisher--KPP equation
\begin{align}
    \partial_{\tilde t}\tilde n = \Pi_1\,\partial_{\tilde x}^2\tilde n + \tilde n(1-\tilde n)
\end{align}
on the half-domain $\tilde x\in[0,5]$~\citep{fisher1937wave,habbal2014assessing,johnston2015estimating}. No-flux boundary conditions are imposed, with initial condition $\tilde n=0$ for $\tilde x<1/2$ and $\tilde n=\Pi_2$ otherwise. The equation is discretised on $161$ spatial grid points with time step $0.03$, treating diffusion implicitly and reaction explicitly. The dimensionless closure time is defined as the first time at which $\tilde n(0,\tilde t)=1/2$, with the crossing time interpolated between adjacent time steps.

\paragraph{Noise added to the benchmarks.}
The underlying benchmark responses are noise-free. To emulate noisy evaluations, an independent zero-mean Gaussian error is added to every response evaluation. Its standard deviation is calibrated once for each regime and then held fixed throughout that regime,
\begin{align}
    \varepsilon = 0.05 \operatorname{med}_{j} R_j,
\end{align}
where
\begin{align}
    R_j = \max_{k=0,\ldots,K-1} \left| y\!\left( \bm c_j+(\eta L_1\cos t_k,\eta L_2\sin t_k) \right) - y(\bm c_j) \right|,
\end{align}
and the median is taken over $200$ independently sampled preliminary anchors. This calibration sets the noise level to $5\%$ of a characteristic local response variation within each regime. Once calibrated, $\varepsilon$ is fixed for all subsequent evaluations in that regime and does not vary with anchor location or response magnitude, consistent with the homoscedastic noise model.

Gaussian errors are used only to generate the benchmark observations. Neither the true value of $\varepsilon$ nor the generating distribution is supplied to the framework; instead, the noise scale $\hat\varepsilon$ is inferred from the repeated evaluations through the residual pool. The results in Tables~\ref{tab:pipe_verdicts}--\ref{tab:cross_domain} and Figs.~\ref{fig:flow_resistance}--\ref{fig:decision_map} correspond to one independent noise realisation for each regime.

\subsection*{Synthetic response fields for calibration}

The statistical properties of the two stages are evaluated using synthetic response fields whose noise-free structure is known. All fields are defined on $\Omega=[-2,2]^2$ in logarithmic coordinates and are processed with the same protocol used for the benchmarks, with $M=40$, $\eta=0.20$, $K=15$, $n=6$, $N_{\rm MC}=1200$ and $N_{\rm F}=6400$.

\paragraph{Field families.}
The first family consists of a single ridge with a linear transverse departure,
\begin{align}
    y(x_1,x_2) = c\,F(ax_1+bx_2) + \lambda c\,(-bx_1+ax_2),
    \label{eq:family_linear}
\end{align}
whereas the second combines two ridge components,
\begin{align}
    y(x_1,x_2) = c\,F(ax_1+bx_2) + \lambda c'\,G(a'x_1+b'x_2).
    \label{eq:family_tworidge}
\end{align}
For both families, $\lambda=0$ gives an exactly reducible field, whereas $\lambda>0$ introduces a controlled departure from one-group reducibility.

The primary profile $F$ is drawn with equal probability from
\begin{align}
    e^s,\quad \ln(1+e^s),\quad \operatorname{arcsinh}s,\quad \sinh s,\quad \tanh s,
\end{align}
with the corresponding response-scale constants
\begin{align}
    c=1,\quad 25,\quad 75,\quad 1,\quad 200,
\end{align}
respectively. The direction $(a,b)$ is obtained by drawing a point uniformly from $[-1,1]^2$ and normalising it to unit length; if the sampled point has norm below $0.2$, the direction is instead set to $(1,1)/\sqrt{2}$.

For the two-ridge family, the corresponding field index uses the same primary profile $F$, scale $c$ and direction $(a,b)$ as the single-ridge family. The secondary profile $G$ is drawn independently from the same five profiles, with $c'$ given by the corresponding scale above. For the secondary direction, a point is drawn independently and uniformly from $[-1,1]^2$ and redrawn until its norm is at least $0.2$. It is then normalised to give $(a',b')$ and accepted only if
\begin{align}
    |aa'+bb'|<0.9,
\end{align}
so that the two ridge directions differ axially by at least $25.8^\circ$.

\paragraph{Noise calibration.}
For each synthetic field, the reference noise scale $\varepsilon_0$ is calibrated using the same response-amplitude criterion as for the benchmarks,
\begin{align}
    \varepsilon_0=0.05\,\operatorname{med}_j R_j.
\end{align}
Here $R_j$ is the response amplitude spanned by a probe with semi-axes equal to the minimum deployed size, $\eta L_1$ and $\eta L_2$, centred at preliminary anchor $j$. The median is evaluated over $60$ preliminary anchors sampled independently of the anchors used in the run. The calibration is performed on the corresponding field with $\lambda=0$, so that $\varepsilon_0$ is independent of the imposed departure $\lambda$.

Synthetic runs use
\begin{align}
    \varepsilon=0.25\varepsilon_0, \quad \varepsilon=\varepsilon_0,\quad \text{or}\quad \varepsilon=4\varepsilon_0,
\end{align}
as specified for each test. Independent zero-mean Gaussian errors with the selected standard deviation are added to every evaluation. As for the physical benchmarks, neither the generating distribution nor the value of $\varepsilon$ is supplied to the framework.

\paragraph{Noise-free reference quantities.}
The stage-one reference is the noise-free gradient-direction spread $\sigma_{\rm grad}$ evaluated at the $40$ anchors of the run. Gradient components are computed from the noise-free field by central differences with step $10^{-4}$, and their axial spread is evaluated using the same doubled-angle statistic as for the directional read-out.

The stage-two reference is the noise-free collapse residual $\rho_{\rm clean}$. It is computed from the noise-free anchor responses using the same residual definition and adaptive polynomial-degree ladder as in stage two, with normalisation by the sample standard deviation of the noise-free anchor responses. The residual is minimised over $181$ equally spaced projection directions on $[0,\pi]$. This provides a noise-free reference over the full axial direction range for assessing the residual returned by the deployed procedure.

Within each synthetic design, a field retains the same anchors when $\lambda$ or the evaluation-noise level is varied. The field families, imposed values of $\lambda$, noise levels and numbers of runs used for each Supplementary Figure are summarised in Supplementary Table~\ref{tab:synthetic}.

\subsection*{Known limitations and numerical approximations}

Several practical features of the evaluation protocol introduce small deviations from the ideal continuous-field quantities, including finite anchor sampling, finite probe size, discrete query angles and finite-sample estimation of the directional statistics. First, the $M$ anchors provide only a finite sampling of $\Omega$. Stage one therefore assesses directional consistency at the sampled locations rather than establishing it everywhere in the regime.

Second, the finite-scale direction read-out differs from the local differential gradient direction. A finite probe identifies the chord joining the anchor to an equal-response point rather than the tangent to the contour at the anchor; for curved contours, the resulting chord error decreases as the probe size is reduced (Supplementary Fig.~\ref{fig:s6}a). In addition, the equal-response crossing is located by interpolation between the $K=15$ queried angles, introducing an angular-interpolation error that is reduced by increasing the angular resolution. Supplementary Fig.~\ref{fig:s6}b separates these two contributions at the deployed probe. The chord contribution vanishes for benchmarks whose responses depend exactly on a single linear combination of the logarithmic inputs, whereas angular interpolation sets their remaining finite read-out floor. For the two non-reducible benchmark regimes, curved response contours produce a larger chord contribution that cannot be removed simply by increasing the number of queried angles.

Third, the axial resultant is estimated from a finite number of anchors. Even when the underlying read-out perturbations are homogeneous, the finite-$M$ resultant has a small upward bias in magnitude and therefore a corresponding downward bias in the inferred spread. In the homogeneous small-spread limit,
\begin{align}
    \mathbb E[\sigma_\alpha^2] \simeq \left(1-\frac{1}{M}\right) \sigma_{\rm noise}^2,
\end{align}
giving a $2.5\%$ effect for $M=40$ (Supplementary Note~\ref{sec:sn3_axial_quadrature}). Because the same finite-$M$ construction is used for the observed statistic and its null calibration, no explicit correction is applied.

Finally, the noise treatment assumes additive evaluation errors that are independent and identically distributed, have a common mean that is constant across evaluated points, and have finite variance. Temporal drift or other correlations between repeated evaluations violate this assumption: their contribution need not decrease as $1/\sqrt n$, and independent resampling from the pooled residuals does not preserve such dependence. Spatially varying bias, temporal drift, correlated errors and heteroscedasticity therefore lie outside the present noise model.

\section*{Data availability}

All data generated in this study, including the benchmark realisations underlying Tables~\ref{tab:pipe_verdicts}--\ref{tab:cross_domain} and Figs.~\ref{fig:flow_resistance}--\ref{fig:decision_map}, and the synthetic-field runs underlying the Supplementary Figures and Tables, will be released together with the code in a public GitHub repository upon publication (\url{https://github.com/yandingbang/reducibility-two-input}).

\section*{Code availability}

All code used in this study, including the implementation of the framework, the benchmark and synthetic-field generators, and the scripts used to produce every figure and table, will be released in the same GitHub repository upon publication (\url{https://github.com/yandingbang/reducibility-two-input}).

\section*{Acknowledgements}

This study was funded by the European Commission through the Marie Skłodowska-Curie Actions Doctoral Network project ``Coupled Problems for Decarbonisation in Industry and Power Generation'', (COMBINE, Grant ID: 101227547), and by the Swedish Transport Administration in the project ``Digital aerodynamics prediction and Stability safety for enhancing Wingsail integrity under dynamic ship Motions'' (SafeWinds, Grant ID: TRV~2024/98491).

\section*{Author contributions}

Conceptualization, H.Y.; Methodology, H.Y. and D.Y.; Software, D.Y. and Z.X.; Validation, D.Y.; Formal analysis, D.Y.; Investigation, D.Y.; Resources, H.Y. and Z.X.; Data curation, D.Y. and Z.X.; Writing – original draft, D.Y.; Writing – review \& editing, H.Y. and Z.X.; Visualization, D.Y. and Z.X.; Supervision, H.Y.; Project administration, H.Y.; Funding acquisition, H.Y.

\section*{Competing interests}

The authors declare no competing interests.

\bibliographystyle{naturemag} 
\bibliography{
references/bakarji2022dimensionally,
references/Barenblatt1996ScalingSA,
references/Bridgman1922,
references/Buckingham1914,
references/colebrook1939correspondence,
references/constantine2014active,
references/constantine2015active,
references/constantine2016many,
references/constantine2017data,
references/davison1997bootstrap,
references/efron1992bootstrap,
references/fisher1937wave,
references/FriedmanTukey1974,
references/garstecki2006formation,
references/gautier2022fully,
references/habbal2014assessing,
references/hang2023novel,
references/jofre2020data,
references/johnston2015estimating,
references/kline2012similitude,
references/li1991sliced,
references/liang2007vitro,
references/mardia2009directional,
references/mendez2005scaling,
references/moody1944friction,
references/murari2023combining,
references/nikuradse1950laws,
references/pinkus2015ridge,
references/romor2024local,
references/ross1999strength,
references/sedov1960similarity,
references/seshadri2019dimension,
references/shi2025hybrid,
references/thorsen2001dynamic,
references/timoshenko2012theory,
references/trepat2009physical,
references/tripathy2016gaussian,
references/villar2023dimensionless,
references/White2016,
references/wycoff2021sequential,
references/Xie2022,
references/XU2022111145,
references/Yuan2025,
references/zhang2025mutual,
references/ZHANG2024116728
}

\clearpage
\setcounter{page}{1}
\renewcommand{\thepage}{S\arabic{page}}
% ===================== Supplementary Information =====================

\begin{center}
    {\Large\bfseries Supplementary Information}\\[1.5em]

    {\large\bfseries
    Equal-response probing with replicate calibration identifies scaling relations from noisy observations of complex systems
    }
\end{center}

\vspace{1.5em}

% Reset counters
\setcounter{equation}{0}
\setcounter{figure}{0}
\setcounter{table}{0}
\setcounter{algorithm}{0}

% Supplementary numbering
\renewcommand{\theequation}{S\arabic{equation}}

\renewcommand{\figurename}{Supplementary Figure}
\renewcommand{\thefigure}{\arabic{figure}}

\renewcommand{\tablename}{Supplementary Table}
\renewcommand{\thetable}{\arabic{table}}

\newcounter{suppmethod}
\renewcommand{\thesuppmethod}{SM\arabic{suppmethod}}

\newcounter{suppnote}
\renewcommand{\thesuppnote}{\arabic{suppnote}}

% =====================================================================
\section*{Supplementary Methods}

% ---------------------------------------------------------------------
\refstepcounter{suppmethod}
\subsection*{\thesuppmethod\quad Null-space representation of dimensionless groups}
\label{sec:sm1_nullspace}

Let $q_1,\ldots,q_n$ denote the dimensional variables of a physical problem. Each variable is represented by its exponent vector with respect to the chosen fundamental dimensions. For example, with mass, length and time as the basis dimensions, velocity has dimensions
\begin{align}
    [M]^0[L]^1[T]^{-1},
\end{align}
and is represented by the vector
\begin{align}
    \bm v=[0,1,-1]^{\mathsf T}.
\end{align}

Stacking the dimensional exponent vectors columnwise gives the dimensional matrix
\begin{align}
    \bm D=
    \begin{bmatrix}
        \bm v_1 & \bm v_2 & \cdots & \bm v_n
    \end{bmatrix}
    \in\mathbb R^{k\times n}.
\end{align}
A monomial
\begin{align}
    \Pi=\prod_{j=1}^{n}q_j^{w_j}
\end{align}
is dimensionless if and only if its exponent vector
\begin{align}
    \bm w=[w_1,\ldots,w_n]^{\mathsf T}
\end{align}
satisfies
\begin{align}
    \bm D\bm w=\bm 0.
\end{align}
Hence the admissible dimensionless exponent vectors form the null space $\mathcal N(\bm D)$.

If $r=\operatorname{rank}(\bm D)$, the rank--nullity theorem gives
\begin{align}
    \dim\mathcal N(\bm D)=n-r.
\end{align}
A basis
\begin{align}
    \bm w^{(1)},\ldots,\bm w^{(m)}, \quad m=n-r,
\end{align}
therefore defines $m$ independent dimensionless groups,
\begin{align}
    \Pi_\ell = \prod_{j=1}^{n} q_j^{\,w_j^{(\ell)}}, \quad \ell=1,\ldots,m.
\end{align}

Any linear combination of the null-space basis vectors,
\begin{align}
    \bm w = \sum_{\ell=1}^{m} a_\ell \bm w^{(\ell)},
\end{align}
also belongs to $\mathcal N(\bm D)$. The corresponding dimensionless group can therefore be written as
\begin{align}
    \Pi = \prod_{\ell=1}^{m} \Pi_\ell^{\,a_\ell}.
\end{align}
For the two-input case considered in the main text, this gives
\begin{align}
    \hat u=\Pi_1^{a}\Pi_2^{b},
\end{align}
showing that the reduced coordinate remains dimensionless for any exponent pair $(a,b)$. The additional constraint $a^2+b^2=1$ used in the main text fixes only the scale of the exponent vector and does not affect dimensional consistency.

% ============================================================
\clearpage
\section*{Supplementary Notes}

\refstepcounter{suppnote}
\subsection*{Supplementary Note~\thesuppnote: Direction statistics do not determine collapse accuracy}
\label{sec:sn1_direction_collapse}

Stage one characterises the consistency of local response directions, whereas stage two evaluates the accuracy of the resulting one-dimensional collapse. The latter cannot in general be inferred from statistics of the gradient vectors alone, because such statistics need not retain information about where those gradients occur in the response domain. The following construction provides an explicit example.

Consider
\begin{align}
    y_n(w_1,w_2) = w_2 + \frac{A}{2\pi n} \sin(2\pi n w_1), \quad A=1.481, \quad n\in\{1,4,16\},
    \label{eq:sn1_fields}
\end{align}
with gradient
\begin{align}
    \nabla y_n = \left( A\cos(2\pi n w_1),\,1 \right).
    \label{eq:sn1_gradients}
\end{align}
The factor $1/n$ in the response amplitude cancels the factor $n$ introduced by differentiation. The three fields therefore contain the same set of gradient vectors when sampled on the lattice of $49\times8$ anchors
\begin{align}
    w_1^{(i)} = \frac{i}{49}, \quad i=0,\ldots,48, \quad w_2^{(j)} = \frac{j+\tfrac12}{8}, \quad j=0,\ldots,7.
    \label{eq:sn1_lattice}
\end{align}
The gradient depends only on $w_1$, and because $\gcd(n,49)=1$ for $n=1,4,16$, multiplication by $n$ permutes the residues modulo $49$. Hence the values
\begin{align}
    \cos\left(\frac{2\pi n i}{49}\right)
\end{align}
form the same multiset for all three fields, differing only in the anchor to which each gradient vector is assigned.

Consequently, any permutation-invariant statistic constructed solely from the sampled gradient vectors is identical for the three fields. In particular, their axial directional spreads, computed from the analytic gradient directions, are identical,
\begin{align}
    \sigma_{\rm grad} = 59.09^\circ,
\end{align}
up to floating-point precision, and their axial means are all
\begin{align}
    \bar\alpha=90^\circ.
\end{align}
Their active-subspace matrices are also identical,
\begin{align}
    \bm C = \frac{1}{49} \sum_{i=0}^{48} \nabla y_n^{(i)} \nabla y_n^{(i)\mathsf T} =
    \begin{pmatrix}
        A^2/2 & 0\\
        0 & 1
    \end{pmatrix},
    \label{eq:sn1_active}
\end{align}
where the average over the full lattice reduces to the average over the $49$ values of $w_1$ because the gradient is independent of $w_2$. The three fields therefore have the same eigenvalues, $A^2/2=1.097$ and $1$, the same spectral gap and the same leading eigenvector of the gradient-covariance matrix. The eigenvalue ratio is only $1.097$, so this matrix provides little evidence for a strongly dominant one-dimensional direction. At the same time, the leading eigenvector points along $w_1$, whereas the $n=16$ response is best collapsed along a direction close to $w_2$. This difference arises because the oscillation amplitude, $A/(2\pi n)$, decreases as $1/n$, while the amplitude of its gradient, $A$, remains unchanged.

The response fields nevertheless have markedly different one-dimensional collapse errors. The residual was evaluated without noise on the lattice of Eq.~\eqref{eq:sn1_lattice} using the collapse residual defined in Methods, Eq.~\eqref{eq:rho}, including the same adaptive degree ladder. To give each field the most favourable collapse available under the implemented fitting procedure, the residual was minimised over $360$ equally spaced projection directions on $[0,\pi)$ rather than over the stage-two search window. The resulting residuals are
\begin{align}
    \rho = 0.316,\quad 0.142,\quad 0.036
\end{align}
for $n=1$, $4$ and $16$, respectively, attained at projection directions of $114^\circ$, $91.5^\circ$ and $90^\circ$. The first and third fields therefore differ in collapse residual by a factor of approximately $8.8$, despite having identical gradient-vector statistics.

The same conclusion holds under more restrictive choices of projection direction. Restricting the search to the window
\begin{align}
    \bar\alpha \pm \frac{3\sigma_\alpha}{\sqrt M},
\end{align}
constructed from the common analytic directional statistics, gives residuals of $0.404$, $0.142$ and $0.036$ for $n=1$, $4$ and $16$. Projecting directly onto $w_2$ gives $0.504$, $0.144$ and $0.036$. Thus the separation between the fields is not produced by the unrestricted directional search. Increasing $n$ reduces the amplitude of the oscillatory term in Eq.~\eqref{eq:sn1_fields} without changing the sampled gradient vectors, so the response becomes progressively closer to a function of $w_2$ even though its gradient statistics remain unchanged.

The same distinction holds in the continuous limit. For $w_1$ uniformly distributed on $[0,1]$,
\begin{align}
    \mathbb E[\cos(2\pi n w_1)] &=0,\\
    \mathbb E[\cos^2(2\pi n w_1)] &=\frac12
\end{align}
for every integer $n\ge1$, so
\begin{align}
    \bm C = \operatorname{diag}(A^2/2,1)
\end{align}
is independent of $n$. By contrast, the amplitude of the departure from the $w_2$ dependence decreases as $1/n$, so the accuracy of a one-dimensional collapse improves as $n$ increases.

This example shows that directional information and collapse accuracy are distinct properties of a response field. Statistics constructed from the gradient vectors alone can characterise directional structure, but they do not in general determine the residual error of a one-dimensional representation. This does not contradict active-subspace analysis, which consistently identifies the direction associated with the largest mean squared directional derivative on these fields. The point is that this information alone does not determine the accuracy of a one-dimensional collapse, because the spatial arrangement of the gradients has been discarded. Stage two therefore provides a complementary test by evaluating the collapse directly, rather than inferring its accuracy from the stage-one directional statistic.

\refstepcounter{suppnote}
\subsection*{Supplementary Note~\thesuppnote: Stage-two residuals at very low noise}
\label{sec:sn2_stage2_low_noise}

The analytical noise scale $\tau_{\min}$ quantifies the contribution of evaluation noise to the collapse residual for a fixed projection direction and polynomial fit. When the evaluation noise becomes very small, however, the residual produced by the complete stage-two procedure can contain small numerical and fitting contributions that do not decrease with the noise. As the evaluation noise decreases, $\tau_{\min}$ decreases, whereas these contributions need not. They can therefore become comparable to, or larger than, the noise contribution at sufficiently low noise. This behaviour is visible for a subset of the exactly reducible synthetic fields at $0.25\times$ the calibration noise (Supplementary Fig.~\ref{fig:s3}a). The discrepancy can be traced to two implementation effects: the narrowing of the direction-search window and the adaptive polynomial fit. Neither effect changes any stage-two verdict at the tolerances used in the main text.

\paragraph{Direction-search effect.}
Stage two searches for the collapse direction within
\begin{align}
    \bar\alpha \pm \frac{3\sigma_\alpha}{\sqrt M}.
\end{align}
As the evaluation noise decreases, $\sigma_\alpha$ decreases and this search window becomes narrower. The finite-$K$ directional read-out, however, can retain a small deterministic discrepancy from the analytic ridge direction because the equal-response crossing is located from discretely sampled probe angles. This discrepancy does not decrease with the evaluation noise.

The effect is most visible for the exponential ridge profile. At $0.25\times$ the calibration noise, the analytic ridge direction lies outside the stage-two search window in $88\%$ of the exponential-profile runs. The collapse is then evaluated at a slightly displaced direction, giving a median
\begin{align}
    \rho/\tau_{\min}=1.89.
\end{align}
Evaluating the same responses at the analytic ridge direction reduces the median ratio to $0.99$, and imposing a minimum search-window half-width of $0.5^\circ$ reduces it to $1.01$. The excess residual at very low noise therefore arises when the search-window half-width becomes smaller than the deterministic offset between the recovered mean direction $\bar{\alpha}$ and the analytic ridge direction. The analytic direction can then fall outside the search window, so the collapse is evaluated at a slightly displaced direction. This effect is distinct from the evaluation-noise contribution quantified by $\tau_{\min}$. The finite-scale directional read-out introduces small deterministic errors through finite probe size and angular interpolation. These contributions remain small for the benchmark settings used here; the chord contribution decreases as the probe size is reduced, whereas the interpolation contribution is controlled by the angular discretisation (Supplementary Fig.~\ref{fig:s6}).

\paragraph{Noise-independent polynomial approximation.}
A second contribution that does not decrease with the evaluation noise comes from the adaptive polynomial approximation. Stage two considers the degree sequence
\begin{align}
    d\in\{2,3,5,8,12,16,20\},
\end{align}
and advances to the next degree only when the resulting residual decreases by more than $5\%$. For the softplus profile, after removal of its linear component the remaining variation is even about its centre. The additional cubic term can therefore reduce the residual only slightly, even though a higher-order polynomial containing additional even terms would improve the approximation substantially.

At $0.25\times$ the calibration noise, the transition from degree $2$ to degree $3$ reduces $\rho$ by less than $5\%$ in $81$ of the $159$ softplus runs, so the degree ladder stops at $d=2$. Nearly the same stopping occurs for the corresponding noise-free responses ($79$ of $159$), showing that the residual is produced by the fitting rule rather than by evaluation noise. Allowing the fit to reach degree $5$ removes $96\%$ of the residual left by the quadratic fit.

Because this polynomial-approximation error is approximately independent of the evaluation noise while $\tau_{\min}$ continues to decrease, their ratio becomes most visible at very low noise. For the affected softplus fields, the median ratio at the analytic ridge direction is
\begin{align}
    \rho/\tau_{\min}=2.87,
\end{align}
whereas fixing the polynomial degree at $20$ reduces the median ratio to $0.98$. At the calibration noise level, the effect is small: the median $\rho/\tau_{\min}$ lies between $0.94$ and $0.98$ for the other ridge profiles and is $1.08$ for softplus.

\paragraph{Implications for the stage-two test.}
These results distinguish the analytical noise scale from the total residual returned by the implemented stage-two procedure. The scale $\tau_{\min}$ describes the contribution expected from evaluation noise, whereas the final $\rho$ can additionally contain small direction-search and polynomial-approximation errors. At the calibration noise level and above, these contributions are minor, but they become visible as the evaluation noise is reduced and $\tau_{\min}$ approaches their magnitude.

Both effects increase rather than decrease the collapse residual and therefore make a reducible verdict less likely. At $\tau=5\%$, none of the $2400$ exactly reducible runs in Supplementary Fig.~\ref{fig:s3}a receives a stage-two not-red verdict. At sufficiently small prescribed tolerances, these implementation residuals may exceed both $\tau$ and $\tau_{\min}$, causing an exactly reducible field to receive a stage-two not-red verdict.

The low-noise effects could be reduced, for example, by imposing a minimum search-window width or by allowing the degree ladder to look beyond a single unsuccessful step. We retain the same fitting and search rules throughout this work rather than introducing additional flexibility solely to improve the exactly reducible cases. Any such modification would require revalidation across both reducible and non-reducible fields, because greater fitting flexibility could reduce numerical approximation error in the former while also overfitting the finite-sample manifestation of a genuine two-dimensional departure in the latter. Supplementary Fig.~\ref{fig:s3} therefore reports the end-to-end behaviour of the procedure used throughout the study.

\refstepcounter{suppnote}
\subsection*{Supplementary Note~\thesuppnote: Quadrature decomposition of the axial spread}
\label{sec:sn3_axial_quadrature}

Stage one estimates the structural component of the observed directional spread by subtracting the estimated noise contribution in quadrature,
\begin{align}
    \hat\sigma_{\rm true} = \sqrt{ \max\left( \sigma_\alpha^2-\sigma_{\rm noise}^2, 0 \right) }
    \label{eq:sn3_estimator}
\end{align}
(Methods, Eq.~\eqref{eq:quadrature}). This note establishes the decomposition that motivates Eq.~\eqref{eq:sn3_estimator}, quantifies the effect of anchor-to-anchor variation in the read-out uncertainty, and evaluates the accuracy and finite-$M$ bias of the resulting estimator.

\paragraph{Setting.}
For a fixed response field and anchor set, let $\alpha_i^{0}$ denote the axial direction returned at anchor $i$ by the finite-scale direction read-out in the absence of evaluation noise. The observed direction is written as
\begin{align}
    \alpha_i = \alpha_i^{0} + \delta_i \pmod{\pi},
\end{align}
where $\delta_i$ is the random perturbation produced by evaluation noise after propagation through the complete direction read-out. Its distribution may differ between anchors because the local response profile and probe geometry affect the sensitivity of the read-out to noise.

Define
\begin{align}
    z &= \frac1M \sum_{i=1}^{M} e^{2\mathrm i\alpha_i}, \\
    z^{0} &= \frac1M \sum_{i=1}^{M} e^{2\mathrm i\alpha_i^{0}}, \\
    \varphi_i &= \mathbb E \left[ e^{2\mathrm i\delta_i} \right], \quad \bar\varphi = \frac1M \sum_{i=1}^{M} \varphi_i ,
    \label{eq:sn3_defs_z}
\end{align}
where the expectations are conditional on the fixed response field and anchor set. We define
\begin{align}
    \sigma_{\rm true}^2 &= -\frac12\ln|z^{0}|, \\
    \sigma_{\rm noise}^2 &= -\frac12\ln|\bar\varphi|.
    \label{eq:sn3_defs}
\end{align}
Here $\sigma_{\rm true}$ denotes the structural spread of the \emph{noise-free finite-scale read-out directions}. It is therefore distinct from the spread of the analytic gradient directions; the deterministic discrepancy between the two is examined separately in Supplementary Fig.~\ref{fig:s6}.

For $\bar\varphi\neq0$, write
\begin{align}
    \varphi_i = \bar\varphi(1+\epsilon_i), \quad \sum_{i=1}^{M}\epsilon_i=0,
\end{align}
and define
\begin{align}
    \epsilon_{\rm rms} &= \left( \frac1M \sum_{i=1}^{M} |\epsilon_i|^2 \right)^{1/2}, \\
    \epsilon_{\max} &= \max_i |\epsilon_i|.
\end{align}

\paragraph{Expected-resultant decomposition.}
By linearity of expectation,
\begin{align}
    \mathbb E[z] &= \frac1M \sum_{i=1}^{M} e^{2\mathrm i\alpha_i^{0}} \varphi_i \\
    &= \bar\varphi \left( z^{0}+r \right),
    \label{eq:sn3_factor}
\end{align}
where
\begin{align}
    r = \frac1M \sum_{i=1}^{M} \epsilon_i e^{2\mathrm i\alpha_i^{0}}.
    \label{eq:sn3_r}
\end{align}

If the read-out perturbation distributions are identical at all anchors, then $\varphi_i\equiv\varphi$, so $\epsilon_i=0$ and $r=0$. Equation~\eqref{eq:sn3_factor} then gives
\begin{align}
    -\frac12\ln|\mathbb E[z]| = \sigma_{\rm true}^2 + \sigma_{\rm noise}^2.
    \label{eq:sn3_exact_additivity}
\end{align}
Thus the structural and noise-induced spreads add exactly in quadrature at the level of the expected axial resultant when the read-out perturbations share a common distribution.

A common perturbation distribution is not required under the equal-read-out-direction null used to calibrate $q_{95}$. Under this null, all noise-free finite-scale read-out directions coincide modulo $\pi$, so
\begin{align}
    \alpha_i^{0} = \alpha^{0} \pmod{\pi} \quad \text{for all }i.
\end{align}
It then follows that
\begin{align}
    r = e^{2\mathrm i\alpha^{0}} \frac1M\sum_{i=1}^{M}\epsilon_i = 0,
\end{align}
regardless of the differences among the $\varphi_i$. Therefore,
\begin{align}
    \mathbb E[z] = e^{2\mathrm i\alpha^{0}}\bar\varphi,
\end{align}
and hence
\begin{align}
    -\frac12\ln|\mathbb E[z]| = \sigma_{\rm noise}^2.
\end{align}
Thus, under this null, differences in noise sensitivity between anchors do not generate a structural directional spread.

This statement concerns the population quantity $\mathbb E[z]$. The finite-sample threshold $q_{95}$ is obtained separately by resampling the estimated anchor-specific perturbation distributions, and its empirical false-alarm behaviour is evaluated in Supplementary Fig.~\ref{fig:s2}.

\paragraph{Heterogeneous read-out uncertainty away from the null.}
When the noise-free directions vary across anchors, $r$ in Eq.~\eqref{eq:sn3_factor} need not vanish. Using Eqs.~\eqref{eq:sn3_defs} and \eqref{eq:sn3_factor},
\begin{align}
    -\frac12\ln|\mathbb E[z]| = \sigma_{\rm true}^2 + \sigma_{\rm noise}^2 + \Delta,
    \label{eq:sn3_delta_decomp}
\end{align}
with
\begin{align}
    \Delta = -\frac12 \ln \left| 1 + \frac{r}{z^{0}} \right|.
    \label{eq:sn3_delta}
\end{align}
The term $\Delta$ therefore measures the coupling between structural variation in the noise-free directions and anchor-to-anchor variation in their noise sensitivity.

The size of $r$ can be bounded directly. Because $\sum_i\epsilon_i=0$, for any real $\psi$,
\begin{align}
    r = \frac1M \sum_{i=1}^{M} \epsilon_i \left( e^{2\mathrm i\alpha_i^{0}} - e^{\mathrm i\psi} \right).
\end{align}
Choosing $\psi=\arg z^{0}$ and applying the Cauchy--Schwarz inequality gives
\begin{align}
    |r| &\le \epsilon_{\rm rms} \sqrt{ 2\left(1-|z^{0}|\right) } \\
    &\le 2\epsilon_{\rm rms}\sigma_{\rm true},
    \label{eq:sn3_r_bound}
\end{align}
where the second inequality follows from
\begin{align}
    |z^{0}| = e^{-2\sigma_{\rm true}^2}
\end{align}
and $1-e^{-x}\le x$ for $x\ge0$. Thus the heterogeneous correction vanishes as the structural spread tends to zero.

Let
\begin{align}
    \kappa = \frac{|r|}{|z^{0}|}.
\end{align}
For $\kappa<1$,
\begin{align}
    |\Delta| &\le -\frac12\ln(1-\kappa) \\
    &\le \frac{\kappa}{2(1-\kappa)},
    \label{eq:sn3_delta_bound}
\end{align}
and Eq.~\eqref{eq:sn3_r_bound} gives
\begin{align}
    \kappa \le 2\epsilon_{\rm rms} \sigma_{\rm true} e^{2\sigma_{\rm true}^2}.
    \label{eq:sn3_kappa}
\end{align}

These expressions show that departure from exact quadrature additivity is controlled by two quantities: the structural directional spread and the heterogeneity of the anchor-specific read-out uncertainty. Because $\sigma_{\rm true}$ is not known in a noisy run, Supplementary Table~\ref{tab:kappa} evaluates the right-hand side of Eq.~\eqref{eq:sn3_kappa} using $\hat\sigma_{\rm true}$ as a plug-in estimate. The resulting values are below $10^{-3}$ for all eight benchmark cases. Over the first $100$ synthetic fields from Supplementary Fig.~\ref{fig:s5}, the maximum plug-in values are $0.005$ at the calibration noise and $0.071$ at four times that noise. These values provide numerical diagnostics of the heterogeneous correction, suggesting a small contribution in the benchmark cases and a larger potential contribution at higher noise. Because they use estimated quantities, they should be interpreted as plug-in diagnostics rather than certified upper bounds. The finite-run accuracy of the quadrature correction is evaluated directly below. At the level of the expected resultant, define the quadrature estimate
\begin{align}
    \sigma_{\rm quad}^2 = -\frac12\ln|\mathbb E[z]| - \sigma_{\rm noise}^2 = \sigma_{\rm true}^2+\Delta .
    \label{eq:sn3_quad_population}
\end{align}
When the right-hand side is non-negative,
\begin{align}
    \frac{|\sigma_{\rm quad}-\sigma_{\rm true}|}{\sigma_{\rm true}}
    &= \frac{|\Delta|}{\sigma_{\rm true}(\sigma_{\rm quad}+\sigma_{\rm true})} \\
    &\le \frac{|\Delta|}{\sigma_{\rm true}^2} \\
    &\le \frac{\epsilon_{\rm rms}e^{2\sigma_{\rm true}^2}}{\sigma_{\rm true}(1-\kappa)}.
    \label{eq:sn3_relerr}
\end{align}
This upper bound can become loose, particularly when the structural spread approaches zero, because it is expressed relative to $\sigma_{\rm true}$ and does not account for cancellation among the anchor-specific contributions to $r$. We therefore use it only to characterise the theoretical effect of heterogeneous read-out uncertainty; the actual finite-run accuracy of the quadrature correction is evaluated directly below.

\paragraph{Finite-run accuracy of the quadrature correction.}
The preceding results concern the expected resultant, whereas Eq.~\eqref{eq:sn3_estimator} is applied to one finite noisy run. We therefore test the complete correction directly on the $902$ synthetic fields with non-zero departures used in Supplementary Fig.~\ref{fig:s5}. For each field, $\hat\sigma_{\rm true}$ from the noisy run is compared with
\begin{align}
    \sigma_{\alpha,\mathrm{clean}},
\end{align}
the directional spread obtained from the same finite-scale read-out, anchors and probes with evaluation noise removed. This comparison directly tests the quadrature correction because both quantities are based on the same finite-scale read-out, with evaluation noise being the only difference. The deterministic discrepancy between the finite-scale read-out and the analytic gradient direction is assessed separately in Supplementary Fig.~\ref{fig:s6}.

When
\begin{align}
    \sigma_{\alpha,\mathrm{clean}} > 2\sigma_{\rm noise},
\end{align}
the median ratio
\begin{align}
    \hat\sigma_{\rm true}/\sigma_{\alpha,\mathrm{clean}}
\end{align}
is $1.00$ at the calibration noise and $0.98$ at four times that noise, with interquartile ranges within approximately $\pm6\%$ at both noise levels. These subsets contain $295$ and $70$ runs, respectively. For $\sigma_{\alpha,\mathrm{clean}}>5\sigma_{\rm noise}$, the median ratio is $1.00$ at the calibration noise, with the interquartile range within approximately $\pm2.5\%$ across $137$ runs. Only three runs at four times the calibration noise reach this range, so no corresponding distributional summary is inferred there.

As $\sigma_{\alpha,\mathrm{clean}}$ approaches $\sigma_{\rm noise}$, the correction tends to read low. For
\begin{align}
    \sigma_{\rm noise} < \sigma_{\alpha,\mathrm{clean}} \le 2\sigma_{\rm noise},
\end{align}
the median ratios are $0.96$ at the calibration noise and $0.90$ at four times that noise, with interquartile ranges of $0.75$--$1.07$ and $0.72$--$1.03$, respectively. These subsets contain $104$ and $97$ runs. Below $\sigma_{\rm noise}$, the positive-part correction in Eq.~\eqref{eq:sn3_estimator} returns zero in $50\%$ and $65\%$ of the runs at the two noise levels, respectively. Thus the estimator is most accurate when the structural spread is clearly separated from $\sigma_{\rm noise}$, whereas it tends to underestimate the structural spread as the two scales approach one another. This downward tendency makes Stage one less likely to declare not-red for departures close to $\sigma_{\rm noise}$.

\paragraph{Finite number of anchors.}
The observed statistic uses the resultant of one finite set of $M$ directions rather than its expectation. Under the independent-evaluation noise model used in this work,
\begin{align}
    \mathbb E|z|^2 = |\mathbb E[z]|^2 + \frac1{M^2} \sum_{i=1}^{M} \left(1-|\varphi_i|^2\right).
    \label{eq:sn3_finiteM}
\end{align}
Under the equal-read-out-direction null, if the directional perturbations are identically distributed across anchors and the resulting spread is small, Eq.~\eqref{eq:sn3_finiteM} gives
\begin{align}
    \mathbb E[\sigma_\alpha^2] \simeq \left(1-\frac1M\right) \sigma_{\rm noise}^2.
    \label{eq:sn3_finiteM_bias}
\end{align}
Finite $M$ therefore introduces a small downward bias in the noise-induced squared spread under the null. For $M=40$, the factor $1-1/M=0.975$ corresponds to a $2.5\%$ reduction relative to $\sigma_{\rm noise}^2$. Because both the observed statistic and the synthetic null distribution used to determine $q_{95}$ are based on the same finite-$M$ construction, no separate finite-$M$ correction is applied.

The finite-$M$ effect accounts for only part of the downward tendency in $\sigma_\alpha/\sigma_{\rm noise}$ under the null. In the exactly reducible synthetic runs, the median ratio $\sigma_\alpha/\sigma_{\rm noise}$ is $0.93$ at the calibration noise and $0.85$ at four times that noise. The remaining difference arises in part because the perturbation distributions used for the null simulation are estimated around the noisy averaged profiles rather than the inaccessible noise-free profiles. This tendency lowers the corrected spread over much of the tested range, making Stage one less likely to declare not-red. This does not guarantee an exact finite-sample false-alarm rate when the deterministic spread introduced by the finite-scale read-out becomes comparable to $q_{95}$; that regime is examined directly in Supplementary Fig.~\ref{fig:s2}.

\paragraph{Scope.}
The expected-resultant decomposition in Eqs.~\eqref{eq:sn3_factor}--\eqref{eq:sn3_delta_bound} is conditional on a fixed response field and does not require a parametric distribution for the directional perturbations, symmetry of those perturbations, or identical perturbation distributions across anchors. The finite-$M$ expression in Eq.~\eqref{eq:sn3_finiteM} and the synthetic construction of $q_{95}$ additionally use the independent-evaluation noise assumption adopted in Methods. A common point-independent mean in the additive evaluation errors does not alter the stage-one directional perturbations considered here because it cancels from the probe--anchor response differences and is removed by the within-repeat centring used to construct the residual pool.

This note concerns the random perturbation of the finite-scale direction read-out by evaluation noise. Deterministic differences between the noise-free read-out direction $\alpha_i^{0}$ and the analytic gradient direction are separate from this decomposition and are quantified in Supplementary Fig.~\ref{fig:s6}.

\refstepcounter{suppnote}
\subsection*{Supplementary Note~\thesuppnote: Derivation of the stage-two noise scale $\tau_{\min}$}
\label{sec:sn4_tau_min}

This note derives the analytical noise scale $\tau_{\min}$ used in stage two. The derivation applies to a fixed projection direction and a fixed polynomial degree $d$. It first quantifies the contribution of evaluation noise to the residual sum of squares and then shows how this leads to the quadrature correction used for $\hat\rho_{\rm true}$. The derivation allows the single-evaluation error to have an unknown common mean that is constant across evaluated points; such a constant offset is absorbed by the intercept of the polynomial fit and therefore does not contribute to the collapse residual.

\paragraph{Noise in the averaged responses.}
Let
\begin{align}
    \bar{\bm y} = \bm\mu + \bm e
\end{align}
denote the vector of the $M$ averaged anchor responses, where $\bm\mu\in\mathbb R^M$ contains the corresponding noise-free responses and $\bm e\in\mathbb R^M$ contains the errors in the averages. At anchor $i$,
\begin{align}
    e_i = \frac{1}{n} \sum_{j=1}^{n}e_{ij},
\end{align}
where the single-evaluation errors $e_{ij}$ are independent and identically distributed, have a common mean $m_e$ that does not depend on the evaluated point, and have finite common variance $\varepsilon^2$. Therefore,
\begin{align}
    \mathbb E[e_i] &= m_e,\\
    \operatorname{Var}(e_i) &= \frac{\varepsilon^2}{n}.
\end{align}
Because evaluations at different anchors are also independent,
\begin{align}
    \mathbb E[\bm e] &= m_e\bm 1,\\
    \operatorname{Cov}(\bm e) &= \frac{\varepsilon^2}{n} \bm I,
    \label{eq:sn4_cov}
\end{align}
where $\bm 1$ denotes the vector of ones. The common mean $m_e$ is not assumed known and need not vanish; as shown below, it is removed by the fitted intercept.

\paragraph{Polynomial fitting as an orthogonal projection.}
For a fixed projection direction, let $\tilde u_i$ denote the standardised projected coordinate of anchor $i$. A polynomial of degree $d$ has the design matrix
\begin{align}
    \bm X=
    \begin{bmatrix}
        1 & \tilde u_1 & \tilde u_1^2 & \cdots & \tilde u_1^d \\
        1 & \tilde u_2 & \tilde u_2^2 & \cdots & \tilde u_2^d \\
        \vdots & \vdots & \vdots & & \vdots \\
        1 & \tilde u_M & \tilde u_M^2 & \cdots & \tilde u_M^d
    \end{bmatrix}.
\end{align}
For the non-degenerate projected coordinates considered here, $\bm X$ has full column rank. The least-squares fitted response is
\begin{align}
    \hat{\bm y}=\bm H \bar{\bm y},
\end{align}
where
\begin{align}
    \bm H=\bm X(\bm X^{\mathsf T}\bm X)^{-1}\bm X^{\mathsf T}
\end{align}
is the orthogonal projection matrix onto the column space of $\bm X$.

Define
\begin{align}
    \bm A=\bm I - \bm H.
\end{align}
Because $\bm H$ is an orthogonal projection,
\begin{align}
    \bm A^{\mathsf T} = \bm A,\quad \bm A^2 = \bm A.
\end{align}
Its rank and trace are
\begin{align}
    \operatorname{rank}(\bm A) = \operatorname{tr}(\bm A) = M-d-1.
    \label{eq:sn4_rank}
\end{align}
The residual vector of the polynomial fit is therefore
\begin{align}
    \bm r = \bm A\bar{\bm y} = \bm A(\bm\mu+\bm e).
\end{align}

Because the constant vector $\bm 1$ is the first column of $\bm X$, it lies in the column space of $\bm X$. Hence
\begin{align}
    \bm H\bm 1=\bm 1, \quad \bm A\bm 1=\bm 0.
    \label{eq:sn4_intercept}
\end{align}
The fitted intercept therefore removes any error component that is common to all evaluated points.

\paragraph{Expected residual sum of squares.}
The residual sum of squares is
\begin{align}
    \|\bm r\|^2 &= \|\bm A(\bm\mu+\bm e)\|^2\\
    &= \|\bm A\bm\mu\|^2 + 2\bm\mu^{\mathsf T}\bm A \bm e + \bm e^{\mathsf T} \bm A \bm e,
\end{align}
where symmetry and idempotence of $\bm A$ have been used.

For the cross term,
\begin{align}
    \mathbb E \left[ 2\bm\mu^{\mathsf T}\bm A\bm e \right] &= 2\bm\mu^{\mathsf T}\bm A\,\mathbb E[\bm e]\\
    &= 2 m_e\,\bm\mu^{\mathsf T}\bm A\bm 1\\
    &= 0,
\end{align}
where Eq.~\eqref{eq:sn4_intercept} has been used.

For the final term, because $\bm A\bm 1=\bm 0$,
\begin{align}
    \bm e^{\mathsf T}\bm A\bm e = (\bm e-m_e\bm 1)^{\mathsf T} \bm A (\bm e-m_e\bm 1).
\end{align}
The centred vector $\bm e-m_e\bm 1$ has zero mean and covariance $\varepsilon^2\bm I/n$, so
\begin{align}
    \mathbb E \left[ \bm e^{\mathsf T}\bm A\bm e \right] &= \operatorname{tr} \left[ \bm A\operatorname{Cov}(\bm e) \right]\\
    &= \frac{\varepsilon^2}{n}\operatorname{tr}(\bm A)\\
    &= (M-d-1)\frac{\varepsilon^2}{n}.
\end{align}
Hence,
\begin{align}
    \mathbb E \left[ \|(\bm I-\bm H)\bar{\bm y}\|^2 \right] = \|(\bm I-\bm H)\bm\mu\|^2 + (M-d-1)\frac{\varepsilon^2}{n}.
    \label{eq:sn4_rss}
\end{align}

Equation~\eqref{eq:sn4_rss} separates the expected residual sum of squares into a noise-free fitting contribution and a contribution from evaluation noise. The latter is exact for a fixed projection direction and polynomial degree under the noise model in Eq.~\eqref{eq:sn4_cov}. Importantly, the common error mean $m_e$ makes no contribution because it lies entirely in the intercept direction removed by the residual projection.

\paragraph{Stage-two noise scale.}
The collapse residual used in Methods is
\begin{align}
    \rho = \frac{1}{s_y} \sqrt{ \frac{ \|(\bm I-\bm H)\bar{\bm y}\|^2 }{M-d-1} }.
\end{align}
Dividing the noise term in Eq.~\eqref{eq:sn4_rss} by $M-d-1$ gives
\begin{align}
    \frac{\varepsilon^2}{n}.
\end{align}
Thus, before normalisation by the response scale, the root-mean-square contribution of evaluation noise to the collapse residual is
\begin{align}
    \frac{\varepsilon}{\sqrt n}.
\end{align}

Replacing $\varepsilon$ by its residual-pool estimate $\hat\varepsilon$ and normalising by the observed response scale $s_y$ gives
\begin{align}
    \tau_{\min} = \frac{\hat\varepsilon}{\sqrt n\,s_y}.
    \label{eq:sn4_tau}
\end{align}
The scale $s_y$ is a sample standard deviation and is therefore invariant to a common additive shift of all anchor responses. Consequently, neither the residual numerator nor its normalising response scale depends on the common error mean $m_e$. The estimate $\hat\varepsilon$ is likewise unaffected, since the residual pool consists of within-repeat residuals, from which a common mean cancels.

For a fixed direction and degree, the factor $M-d-1$ therefore cancels exactly between the expected noise contribution to the residual sum of squares and the degrees-of-freedom normalisation. The analytical noise contribution consequently has no explicit dependence on the polynomial degree $d$.

\paragraph{Quadrature correction.}
Equation~\eqref{eq:sn4_rss} motivates subtracting the noise contribution at the level of the squared residual. For a fixed projection direction $\theta$ and polynomial degree $d$, define the noise-free fitting contribution, normalised by the observed response scale, as
\begin{align}
    \rho_0(\theta,d;s_y) = \frac{1}{s_y} \sqrt{\frac{\|(\bm I-\bm H)\bm\mu\|^2}{M-d-1}}.
\end{align}
Here $\bm H$ is the projection matrix for that fixed direction and degree, and $s_y$ is the standard deviation of the observed averaged anchor responses. The expected-residual decomposition then motivates the run-level approximation
\begin{align}
    \rho^2 \simeq \rho_0^2(\theta,d;s_y)+\tau_{\min}^2,
    \label{eq:sn4_quadrature}
\end{align}
which uses the same normalising scale for both contributions. The quantity $\rho_0$ describes a fixed fit and is distinct from the optimised noise-free reference residual $\rho_{\rm clean}$ used in the synthetic tests.

We therefore define
\begin{align}
    \hat\rho_{\rm true} = \sqrt{ \max \left( \rho^2-\tau_{\min}^2, 0 \right) }.
    \label{eq:sn4_rho_true}
\end{align}
The truncation at zero accounts for finite-sample fluctuations for which $\rho<\tau_{\min}$. We interpret $\hat\rho_{\rm true}$ as the part of the observed collapse residual that remains after subtracting the estimated noise contribution in quadrature. It is not an exact unbiased estimator of the structural collapse residual.

\paragraph{Relation to the deployed stage-two procedure.}
The exact decomposition in Eq.~\eqref{eq:sn4_rss} assumes that the projection direction and polynomial degree are fixed independently of the evaluation noise. In the deployed procedure, however, the direction is selected by the local direction search and the polynomial degree by the adaptive degree ladder, both using the noisy averaged anchor responses. Because the selected direction and polynomial degree can change with the noise realisation, the final residual need not follow the fixed-direction, fixed-degree decomposition exactly.

In addition, the normalising scale $s_y$ is computed from the same noisy anchor responses. Equation~\eqref{eq:sn4_tau} should therefore be understood as an analytical noise scale for the deployed normalised residual rather than as an exact expression for its full finite-sample distribution.

The complete procedure is tested directly in Supplementary Fig.~\ref{fig:s3}. For exactly reducible synthetic fields, the median $\rho/\tau_{\min}$ is $1.04$, $0.98$ and $0.97$ at $0.25$, $1$ and $4$ times the calibration noise, respectively. The departures observed at the lowest noise are traced to the direction-search window and adaptive degree ladder rather than to Eq.~\eqref{eq:sn4_tau} (Supplementary Note~\ref{sec:sn2_stage2_low_noise}).

For synthetic fields with non-zero imposed departures from one-group reducibility, $\hat\rho_{\rm true}$ is compared directly with the corresponding noise-free collapse residual in Supplementary Fig.~\ref{fig:s3}b. For imposed departures with $\lambda\ge0.05$ and $\rho_{\rm clean}>\tau_{\min}$, the median ratio $\hat\rho_{\rm true}/\rho_{\rm clean}$ is $1.00$--$1.05$. These tests show that Eq.~\eqref{eq:sn4_tau} provides an accurate noise scale over the tested range and that, for the tested imposed departures satisfying $\lambda\ge0.05$ and $\rho_{\rm clean}>\tau_{\min}$, the quadrature correction closely recovers the corresponding noise-free residual.

% ============================================================
\clearpage
\section*{Supplementary Figures}

\begin{figure}[htpb]
    \centering
    \includegraphics[width=88mm]{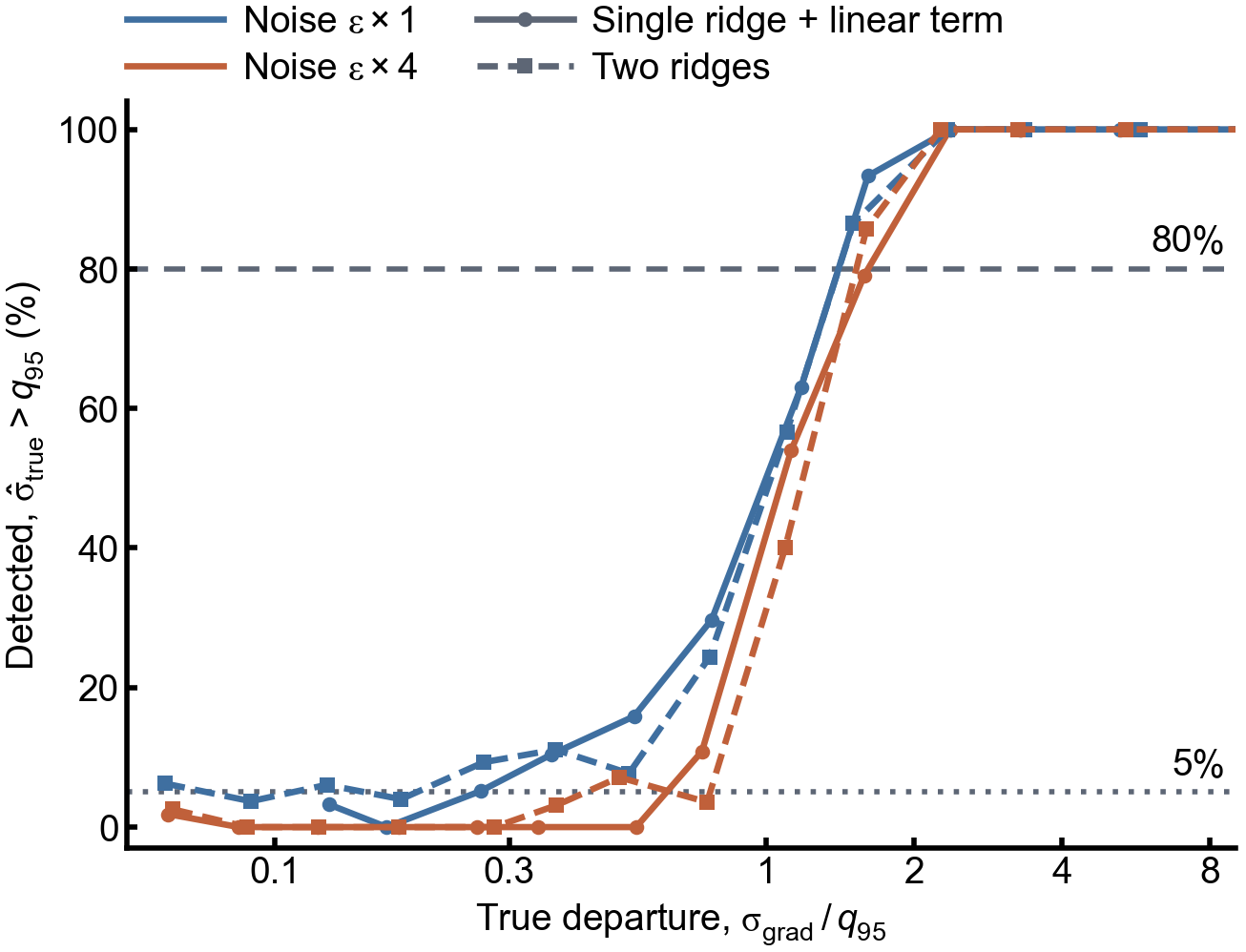}
    \caption{\small \textbf{Detection power of the stage-one $q_{95}$ threshold.} The fraction of runs satisfying $\hat\sigma_{\rm true}>q_{95}$ is plotted against $\sigma_{\rm grad}/q_{95}$. Colours denote the two noise levels and line styles the two synthetic-field families. The dashed horizontal line marks $80\%$ detection and the dotted line the nominal $5\%$ level associated with the $q_{95}$ threshold. Detection reaches $80\%$ at $\sigma_{\rm grad}/q_{95}=1.41$--$1.62$, and all runs with $\sigma_{\rm grad}\ge2q_{95}$ are detected.}
    \label{fig:s1}
\end{figure}

Supplementary Fig.~\ref{fig:s1} evaluates the detection power of the stage-one $q_{95}$ threshold using the two synthetic-field families defined in Methods. In both families, $\lambda=0$ gives an exactly reducible field, whereas $\lambda>0$ introduces a departure from one-group reducibility. For each field, the noise-free gradient-direction spread $\sigma_{\rm grad}$ is computed at the sampled anchors and compared with the run-specific null threshold $q_{95}$. A run is counted as detected when $\hat\sigma_{\rm true}>q_{95}$; $\tau_1$ is not involved in this test.

When the departure is expressed relative to the run-specific $q_{95}$, the two field families and two noise levels show broadly similar detection behaviour. This indicates that $q_{95}$ provides a useful noise-dependent reference scale for the stage-one directional departure over the tested cases. Detection increases rapidly once $\sigma_{\rm grad}$ becomes comparable to and then exceeds $q_{95}$, reaching $80\%$ at $\sigma_{\rm grad}/q_{95}=1.41$--$1.62$ and $100\%$ for all tested runs with $\sigma_{\rm grad}\ge2q_{95}$. Thus, departures well below $q_{95}$ are difficult to distinguish from noise, whereas departures exceeding the null threshold by roughly a factor of two are consistently detected in these synthetic tests.

\begin{figure}[htpb]
    \centering
    \begin{minipage}[t]{0.48\textwidth}
        \textbf{a}\par
        \centering
        \includegraphics[width=\linewidth]{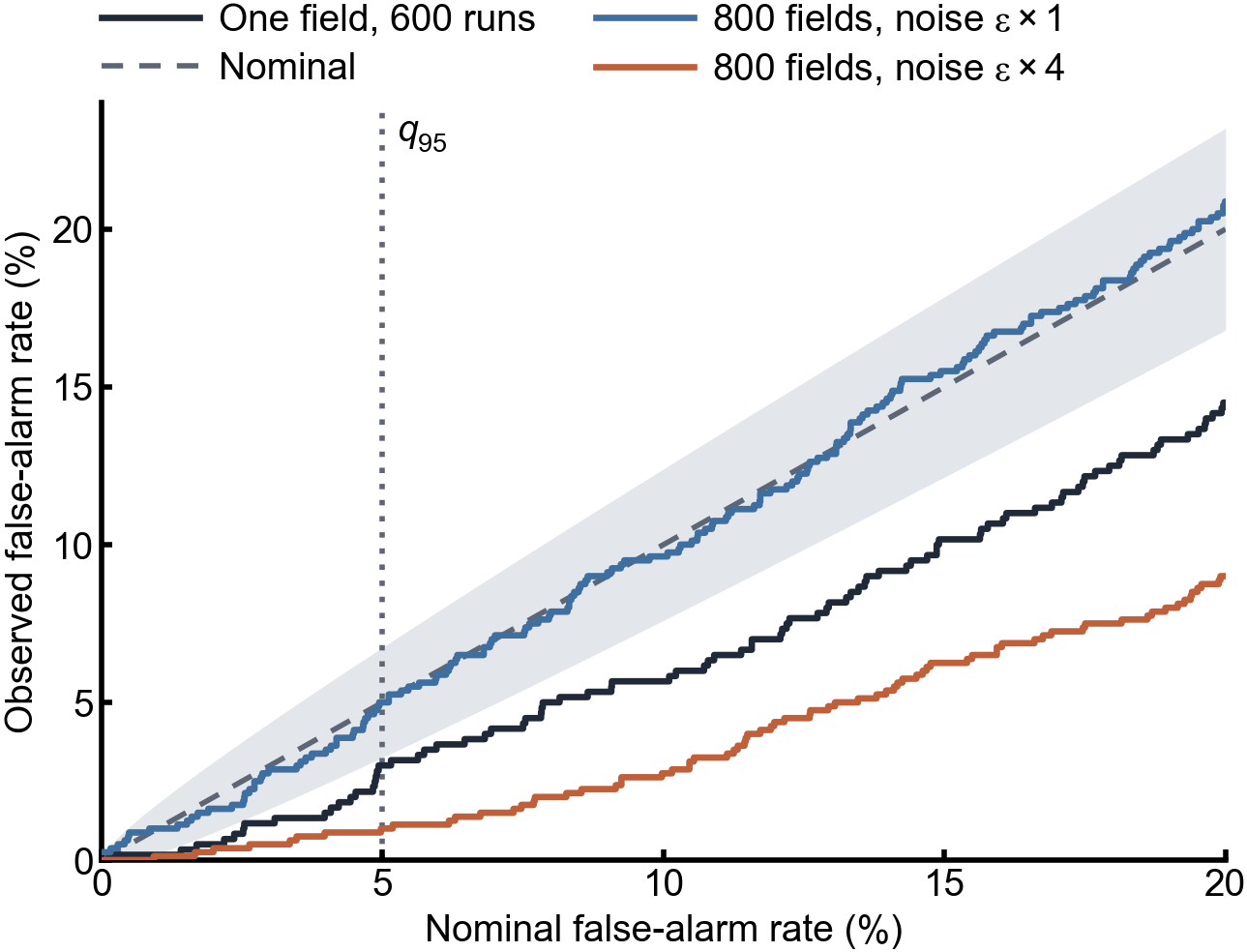}
    \end{minipage}
    \hfill
    \begin{minipage}[t]{0.48\textwidth}
        \textbf{b}\par
        \centering
        \includegraphics[width=\linewidth]{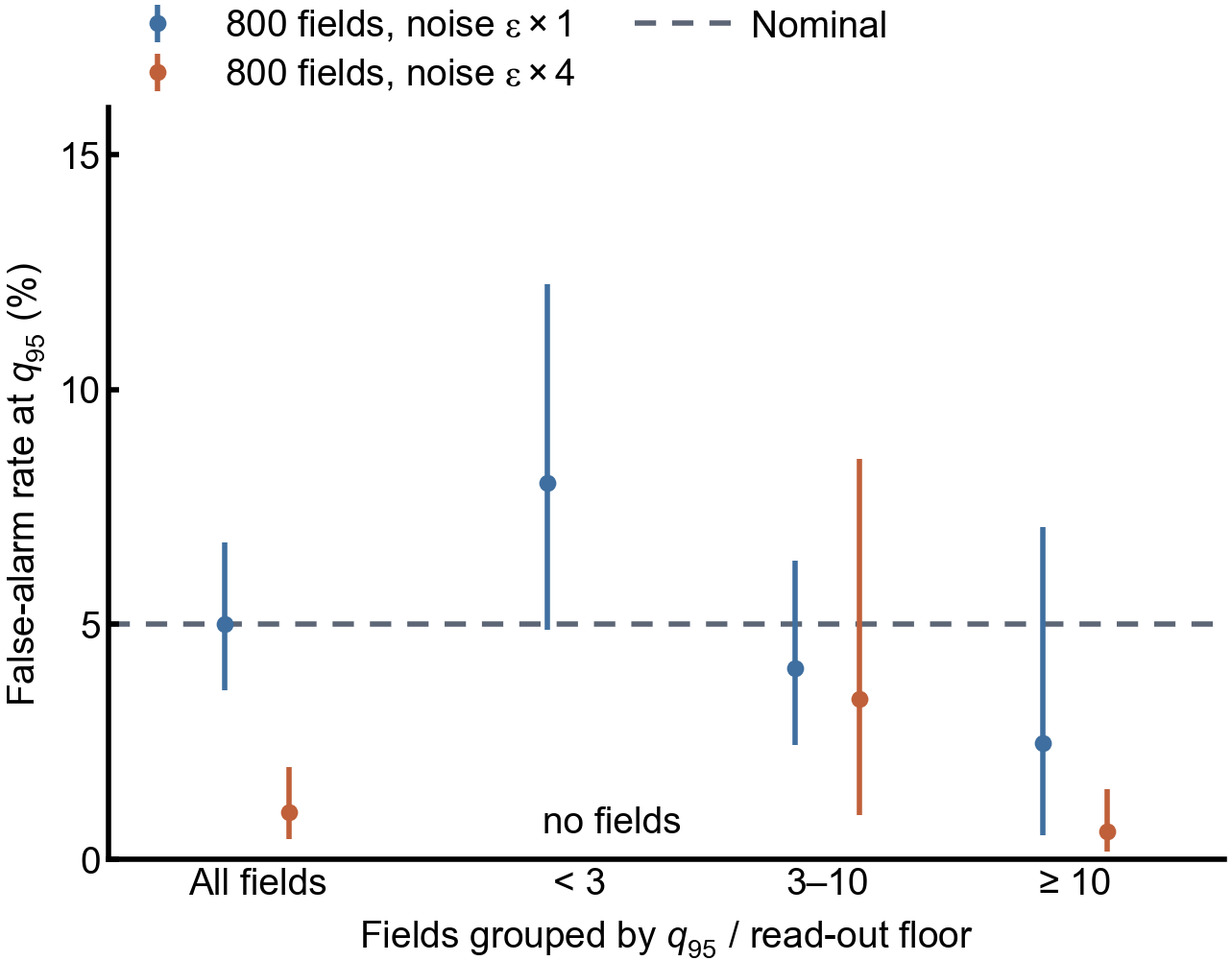}
    \end{minipage}
    \caption{\small
    \textbf{False-alarm calibration of the stage-one $q_{95}$ threshold.}
    \textbf{a}, Observed versus nominal false-alarm rate. Black denotes $600$ repeated runs of one fixed ridge field; blue and terracotta denote $800$ randomly generated ridge fields at $1$ and $4$ times the calibration noise, respectively. At the nominal $5\%$ level corresponding to $q_{95}$, the observed rates are $3.00\%$, $5.00\%$ and $1.00\%$. The grey band shows the central $95\%$ reference range for $600$ runs of an exactly calibrated test.
    \textbf{b}, False-alarm rate at $q_{95}$ grouped by $q_{95}/\sigma_{\alpha,\mathrm{clean}}$, where $\sigma_{\alpha,\mathrm{clean}}$ is the noise-free finite-scale read-out spread. Error bars show Clopper--Pearson $95\%$ intervals; the dashed line marks the nominal $5\%$ rate.}
    \label{fig:s2}
\end{figure}

Supplementary Fig.~\ref{fig:s2} evaluates the false-alarm calibration of the stage-one $q_{95}$ threshold on exactly reducible synthetic fields. These fields have a constant noise-free gradient direction, so $\sigma_{\rm grad}=0$, although the finite-scale direction read-out can retain a small non-zero spread even without evaluation noise. For the calibration of $q_{95}$, a run is counted as a false alarm when $\hat\sigma_{\rm true}>q_{95}$. Panel~a compares the observed and nominal false-alarm rates, whereas panel~b examines how the false-alarm rate depends on the ratio of $q_{95}$ to the noise-free finite-scale read-out floor.

At the deployed $5\%$ level, the overall false-alarm rates are at or below the nominal rate (Supplementary Fig.~\ref{fig:s2}a). Panel~b shows that the largest departure occurs when $q_{95}$ approaches the finite-scale read-out floor. At $1\times$ noise, the false-alarm rate is $8.0\%$ ($19$ of $237$ runs) when $q_{95}/\sigma_{\alpha,\mathrm{clean}}<3$, but decreases to $4.1\%$ ($18$ of $442$) for ratios between $3$ and $10$ and to $2.5\%$ ($3$ of $121$) above $10$. At $4\times$ noise, no field lies in the lowest-ratio group, and the rates are $3.4\%$ ($4$ of $117$) and $0.6\%$ ($4$ of $683$) in the remaining two groups. Thus, departures from the nominal calibration become most apparent when $q_{95}$ is only a few times larger than the deterministic finite-scale read-out floor.

\begin{figure}[htpb]
    \centering
    \begin{minipage}[t]{0.48\textwidth}
        \textbf{a}\par\centering
        \includegraphics[width=\linewidth]{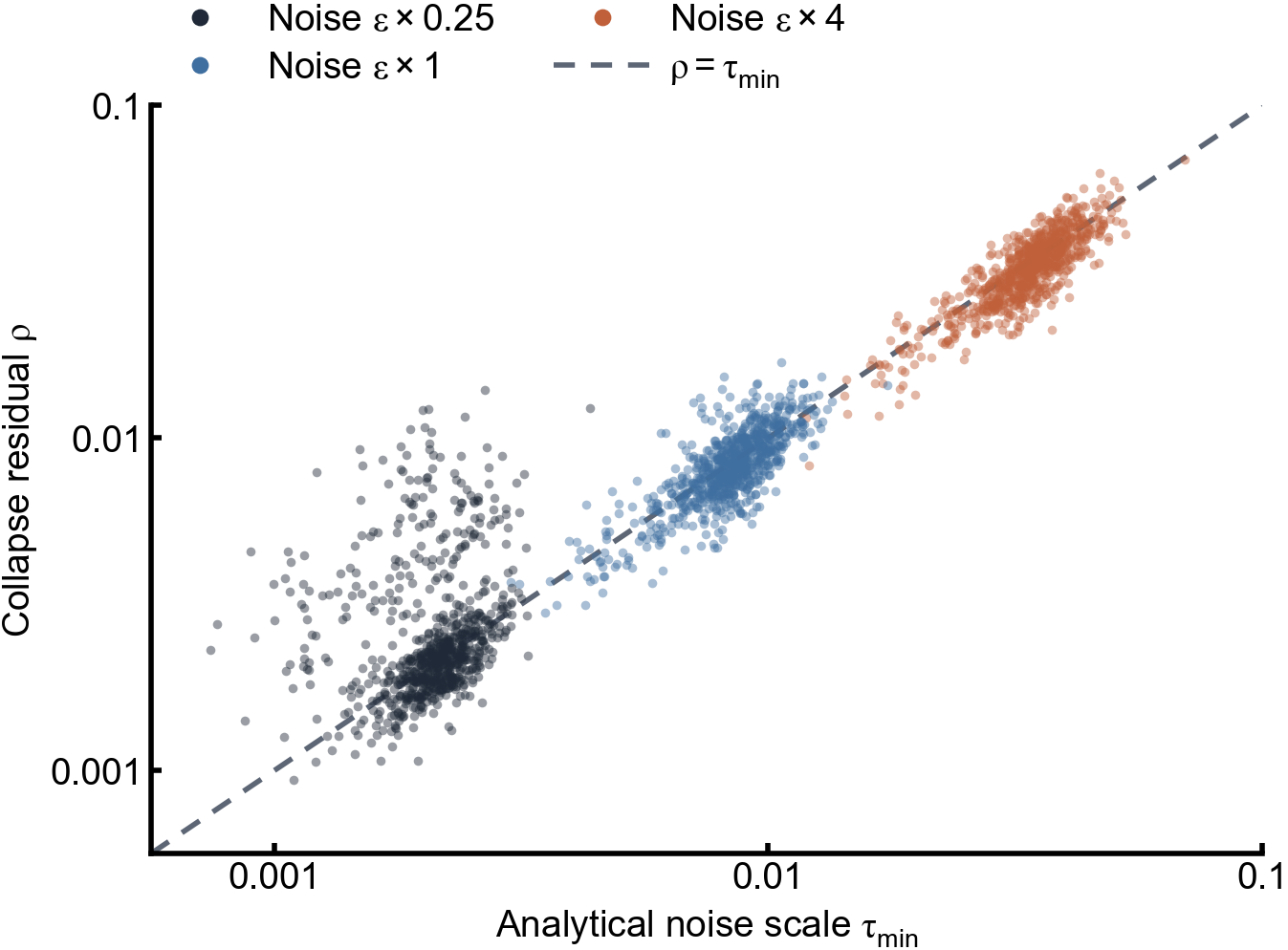}
    \end{minipage}
    \hfill
    \begin{minipage}[t]{0.48\textwidth}
        \textbf{b}\par\centering
        \includegraphics[width=\linewidth]{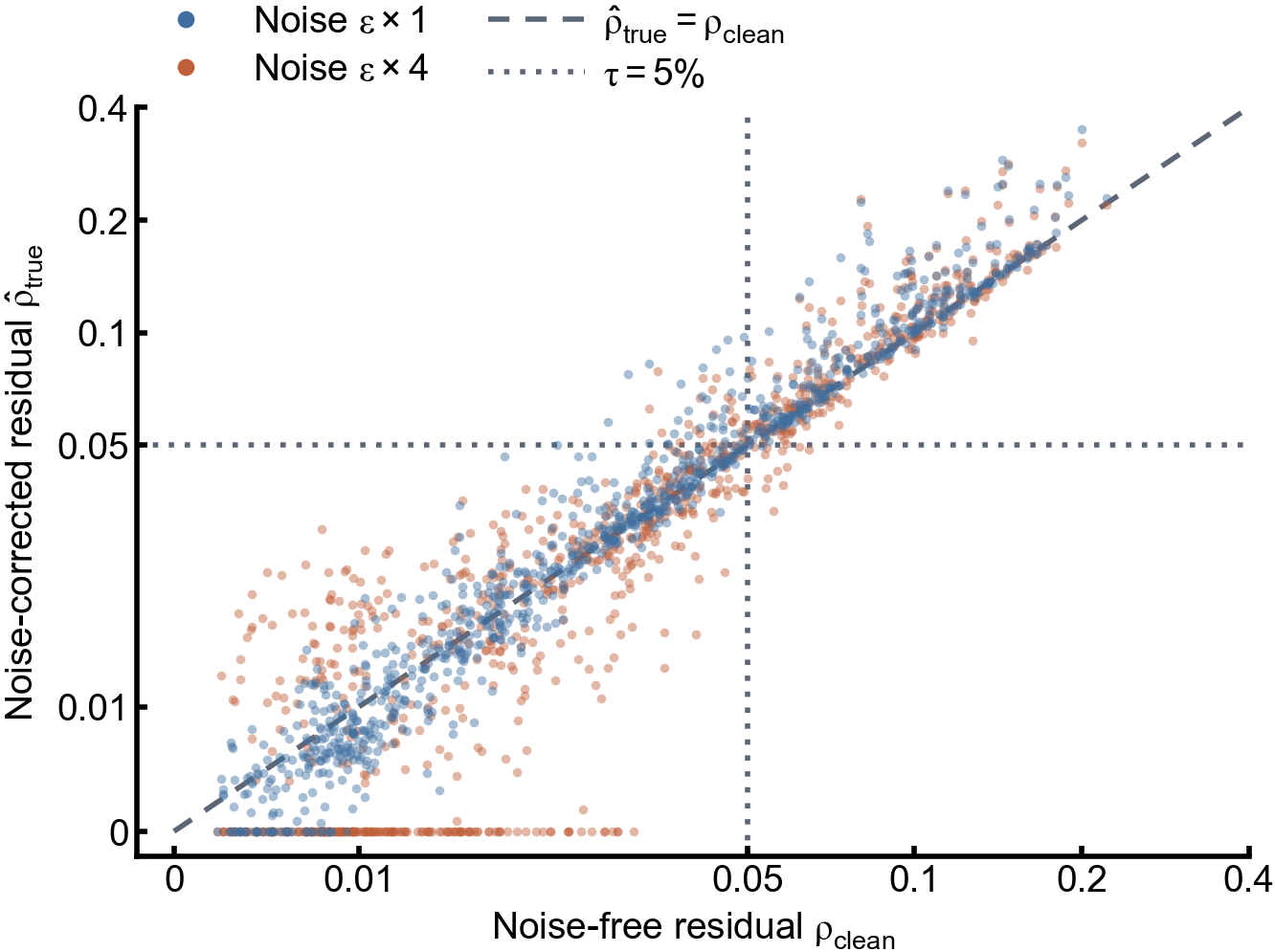}
    \end{minipage}
    \caption{\small \textbf{Calibration of the stage-two noise scale and quadrature correction.} 
    \textbf{a}, Collapse residual $\rho$ versus the analytical noise scale $\tau_{\min}$ for exactly reducible ridge fields at $0.25$, $1$ and $4$ times the calibration noise. The dashed line denotes $\rho=\tau_{\min}$. 
    \textbf{b}, Noise-corrected residual $\hat\rho_{\rm true}$ versus the noise-free residual $\rho_{\rm clean}$ for fields with controlled departures from one-group reducibility at $1$ and $4$ times the calibration noise. The dashed line denotes $\hat\rho_{\rm true}=\rho_{\rm clean}$ and the dotted lines mark the prescribed tolerance $\tau=5\%$. Points on the horizontal axis correspond to runs with $\rho<\tau_{\min}$, for which the positive-part correction gives $\hat\rho_{\rm true}=0$; this occurs in $36$ runs at $1\times$ noise and $197$ at $4\times$ noise, all with $\rho_{\rm clean}<5\%$.}
    \label{fig:s3}
\end{figure}

Supplementary Fig.~\ref{fig:s3} evaluates the stage-two noise scale $\tau_{\min}$ and the quadrature correction for the collapse residual. Panel~a uses exactly reducible ridge fields to compare the observed residual $\rho$ with $\tau_{\min}$, for which agreement with $\rho=\tau_{\min}$ is expected when evaluation noise dominates the residual. Panel~b introduces controlled departures from one-group reducibility and compares the corrected residual $\hat\rho_{\rm true}$ with the corresponding noise-free residual $\rho_{\rm clean}$.

For the exactly reducible fields, $\rho$ remains close to $\tau_{\min}$ over the tested noise range, with median $\rho/\tau_{\min}$ of $1.04$, $0.98$ and $0.97$ at $0.25$, $1$ and $4$ times the calibration noise, respectively. The larger scatter at the lowest noise arises when small noise-independent contributions from the direction-search window and adaptive degree ladder become comparable to the shrinking noise scale. These effects are examined in Supplementary Note~\ref{sec:sn2_stage2_low_noise}. None of the $2400$ exactly reducible runs receives a stage-two not-red verdict at the prescribed tolerance $\tau=5\%$.

For fields with imposed departures from one-group reducibility, the corrected residual closely follows the corresponding noise-free residual once the latter exceeds the noise scale. For $\lambda\ge0.05$ and $\rho_{\rm clean}>\tau_{\min}$, the median ratio $\hat\rho_{\rm true}/\rho_{\rm clean}$ is $1.00$--$1.05$. Misclassifications occur mainly close to the prescribed tolerance. Among fields with $\rho_{\rm clean}>5\%$, stage two returns reducible for $1$ of $309$ runs at $1\times$ noise and $12$ of $309$ runs at $4\times$ noise. Among runs with $\rho_{\rm clean}\le5\%$, stage two returns not-red for $34$ of $691$ runs at $1\times$ noise and $25$ of $691$ runs at $4\times$ noise. Eight runs at $4\times$ noise are unresolved because $\tau_{\min}>5\%$.

\begin{figure}[htpb]
    \centering
    \includegraphics[width=\textwidth]{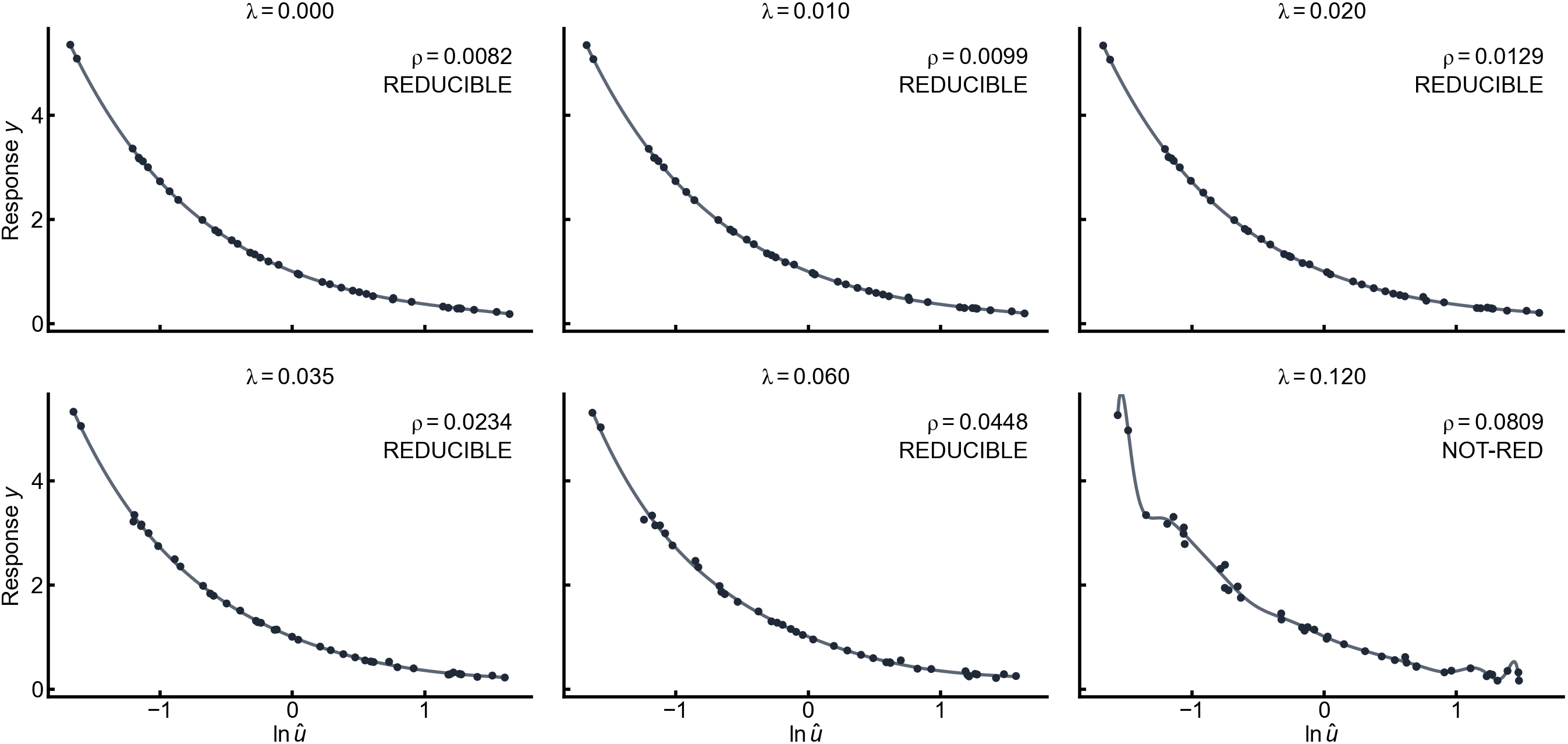}
    \caption{\small
    \textbf{Visual similarity of collapse plots can mask changes in the stage-two residual.} Collapse plots are shown for the synthetic family $y=c[\exp(u)+\lambda v]$, with $u=ax_1+bx_2$ and $v=-bx_1+ax_2$, using the same anchors and noise realisation while increasing $\lambda$ from $0$ to $0.12$. Points show the averaged anchor responses against the stage-two group coordinate $\ln\hat u$, and curves show the selected polynomial fits. Each panel reports the collapse residual $\rho$ and the resulting stage-two verdict; the verdict uses $\hat\rho_{\rm true}$ with $\tau=5\%$ and $\tau_{\min}=0.0085$--$0.0086$.}
    \label{fig:s4}
\end{figure}

Supplementary Fig.~\ref{fig:s4} illustrates why stage-two collapse is assessed quantitatively rather than from visual inspection alone. The same synthetic field family, anchor set and noise realisation are used in all six panels, while the imposed transverse departure $\lambda$ is increased from $0$ to $0.12$. For each value of $\lambda$, stage two selects the projection direction and polynomial fit and evaluates the collapse residual.

Although the first five collapse plots appear very similar, the quantitative residual increases steadily with the imposed departure. As $\lambda$ increases from $0$ to $0.060$, $\rho$ rises from $0.0082$ to $0.0448$ and $\hat\rho_{\rm true}$ from $0$ (the residual at $\lambda=0$ lies below $\tau_{\min}$) to $0.044$. The corrected residual remains below the prescribed tolerance $\tau=0.05$ in all five cases, so stage two returns reducible. At $\lambda=0.120$, the corrected residual exceeds $\tau=5\%$, and stage two returns not-red. The figure therefore shows that visual inspection alone can miss substantial changes in collapse accuracy; the stage-two verdict is determined by the quantitative residual relative to the prescribed tolerance and analytical noise scale.

\begin{figure}[htpb]
    \centering
    \begin{minipage}[t]{0.48\textwidth}
        \textbf{a}\par\centering
        \includegraphics[width=\linewidth]{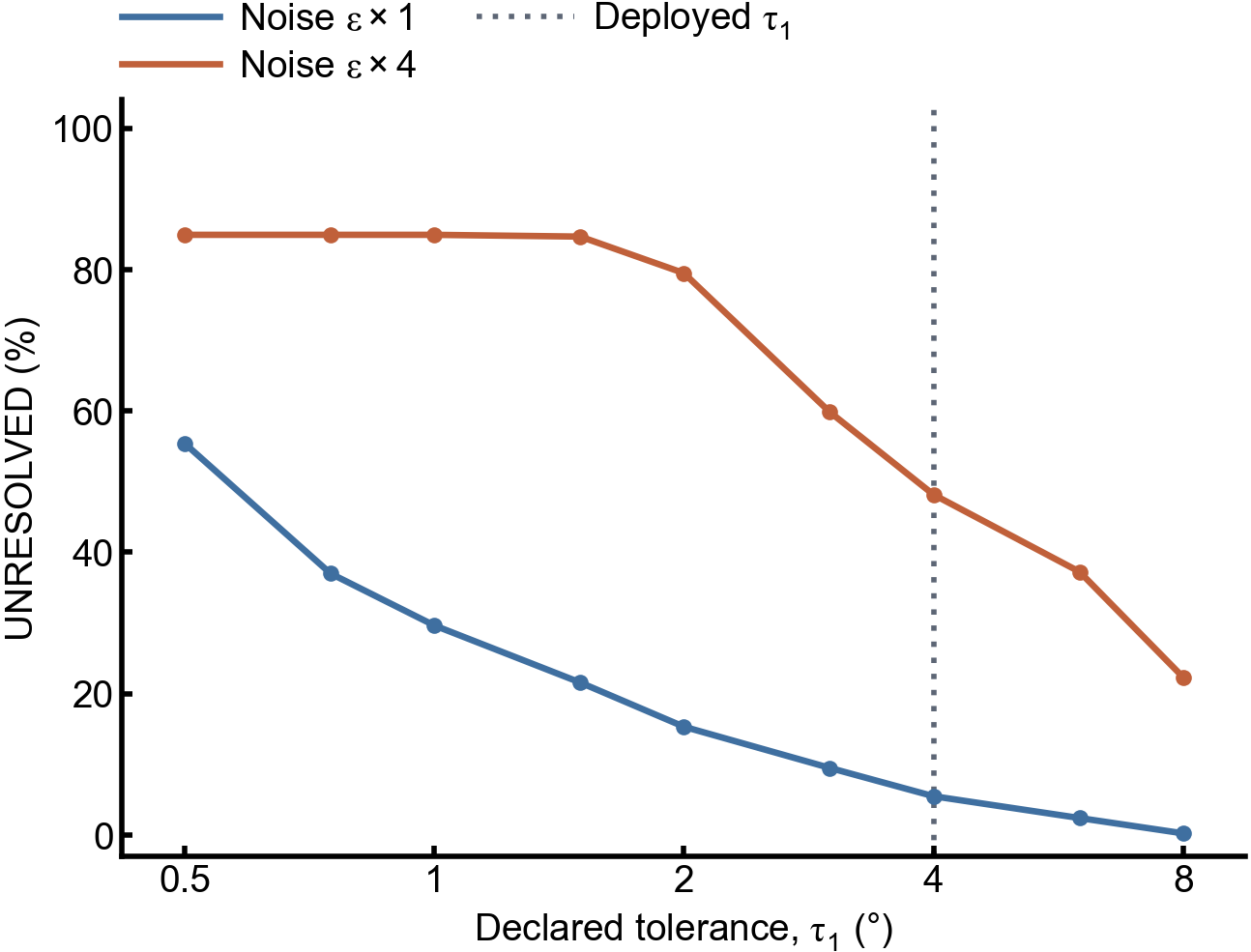}
    \end{minipage}
    \hfill
    \begin{minipage}[t]{0.48\textwidth}
        \textbf{b}\par\centering
        \includegraphics[width=\linewidth]{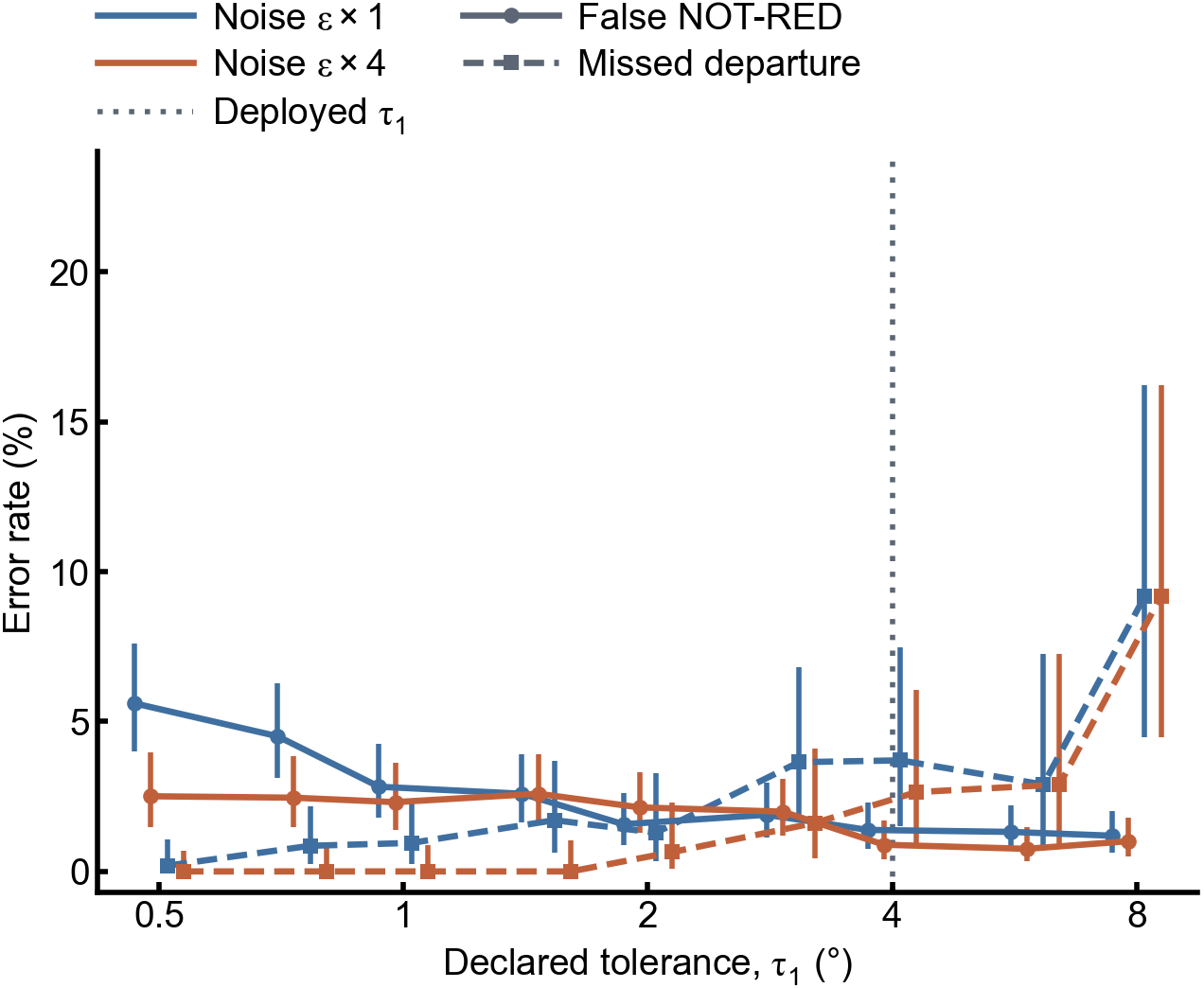}
    \end{minipage}
    \caption{\small
    \textbf{Effect of the declared stage-one tolerance $\tau_1$ on unresolved and incorrect verdicts.}
    \textbf{a}, Fraction of runs classified as unresolved at $1$ and $4$ times the calibration noise.
    \textbf{b}, Incorrect resolved verdicts: not-red for fields with $\sigma_{\rm grad}\le\tau_1$ (solid circles) and reducible for fields with $\sigma_{\rm grad}>\tau_1$ (dashed squares). Error bars show Clopper--Pearson $95\%$ intervals. The dotted vertical line marks the deployed value $\tau_1=4^\circ$.}
    \label{fig:s5}
\end{figure}

Supplementary Fig.~\ref{fig:s5} examines how the declared stage-one tolerance $\tau_1$ affects the balance between unresolved and incorrect verdicts. The same $1200$ synthetic fields at each noise level are reclassified at each value of $\tau_1$; the directional read-out, $\hat\sigma_{\rm true}$ and $q_{95}$ are unchanged. For evaluating the verdicts, the noise-free gradient-direction spread $\sigma_{\rm grad}$ defines whether a field lies within the declared tolerance ($\sigma_{\rm grad}\le\tau_1$) or beyond it ($\sigma_{\rm grad}>\tau_1$). Panel~a shows the fraction of unresolved runs, whereas panel~b shows the two types of incorrect resolved verdict. The corresponding numerical rates and confidence intervals are reported in Supplementary Table~\ref{tab:tau1}.

Increasing $\tau_1$ reduces the fraction of unresolved runs because a larger tolerance is more likely to exceed the noise-based threshold $q_{95}$ (Supplementary Fig.~\ref{fig:s5}a). At $1\times$ noise, the unresolved fraction decreases from $55.4\%$ at $\tau_1=0.5^\circ$ to $0.25\%$ at $8^\circ$; at $4\times$ noise it decreases from $84.9\%$ to $22.3\%$. At the deployed $\tau_1=4^\circ$, the corresponding fractions are $5.5\%$ and $48.2\%$. The larger unresolved fraction at $4\times$ noise reflects its larger $q_{95}$, whose median is approximately $4.6^\circ$ compared with $0.88^\circ$ at $1\times$ noise.

The two resolved error modes respond differently to $\tau_1$ (Supplementary Fig.~\ref{fig:s5}b). False not-red verdicts generally decrease as $\tau_1$ increases, whereas missed departures become more frequent when the declared tolerance approaches the upper end of the tested departure distribution. At $4\times$ noise, missed departures remain uncommon at small and moderate values of $\tau_1$, because many runs near the decision boundary are returned as unresolved rather than being assigned an incorrect resolved verdict. The increase in missed departures at large $\tau_1$ occurs because many of the sampled departures lie only slightly above the declared tolerance.

\begin{figure}[htpb]
    \centering
    \begin{minipage}[t]{0.48\textwidth}
        \textbf{a}\par\centering
        \includegraphics[width=\linewidth]{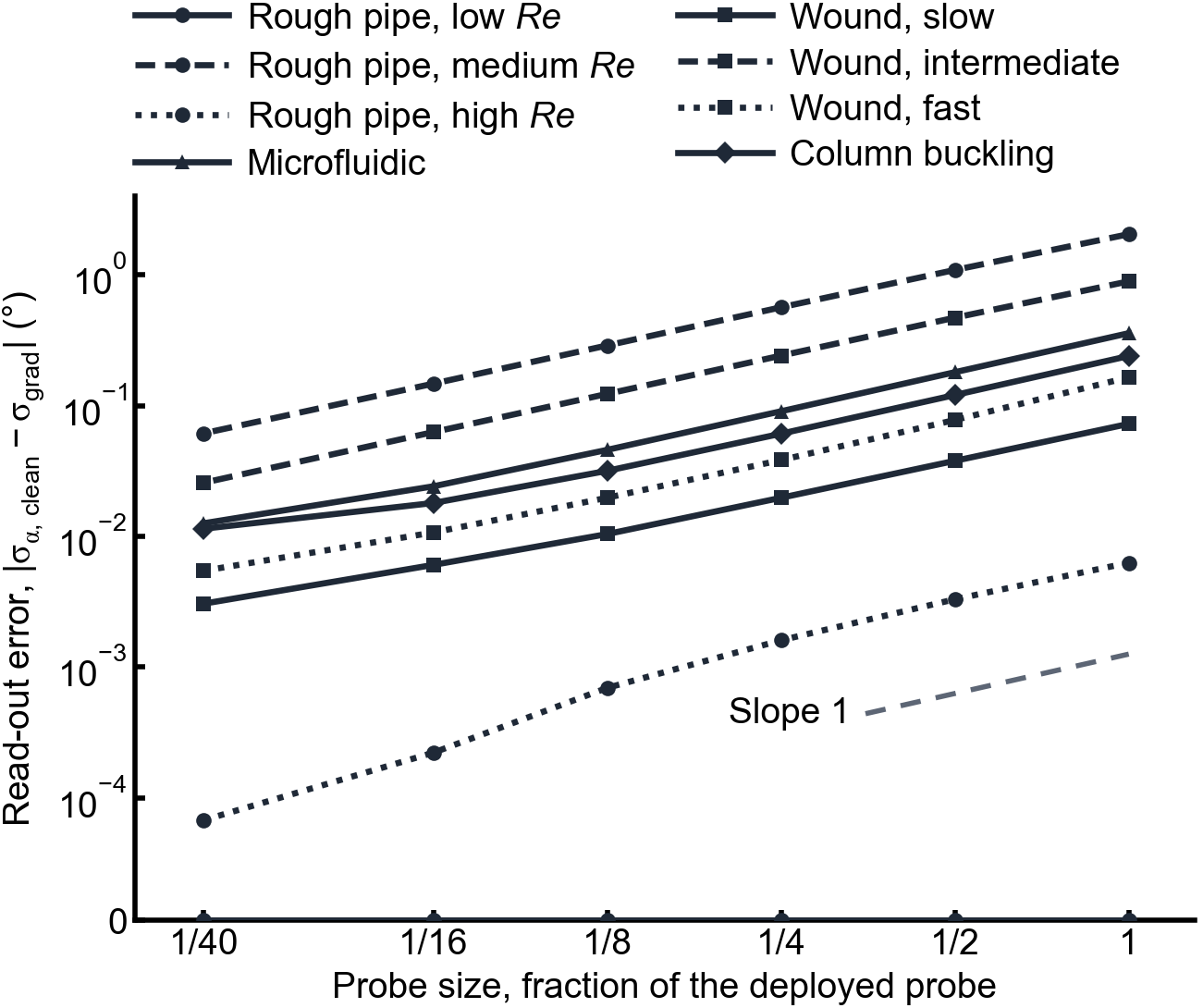}
    \end{minipage}
    \hfill
    \begin{minipage}[t]{0.48\textwidth}
        \textbf{b}\par\centering
        \includegraphics[width=\linewidth]{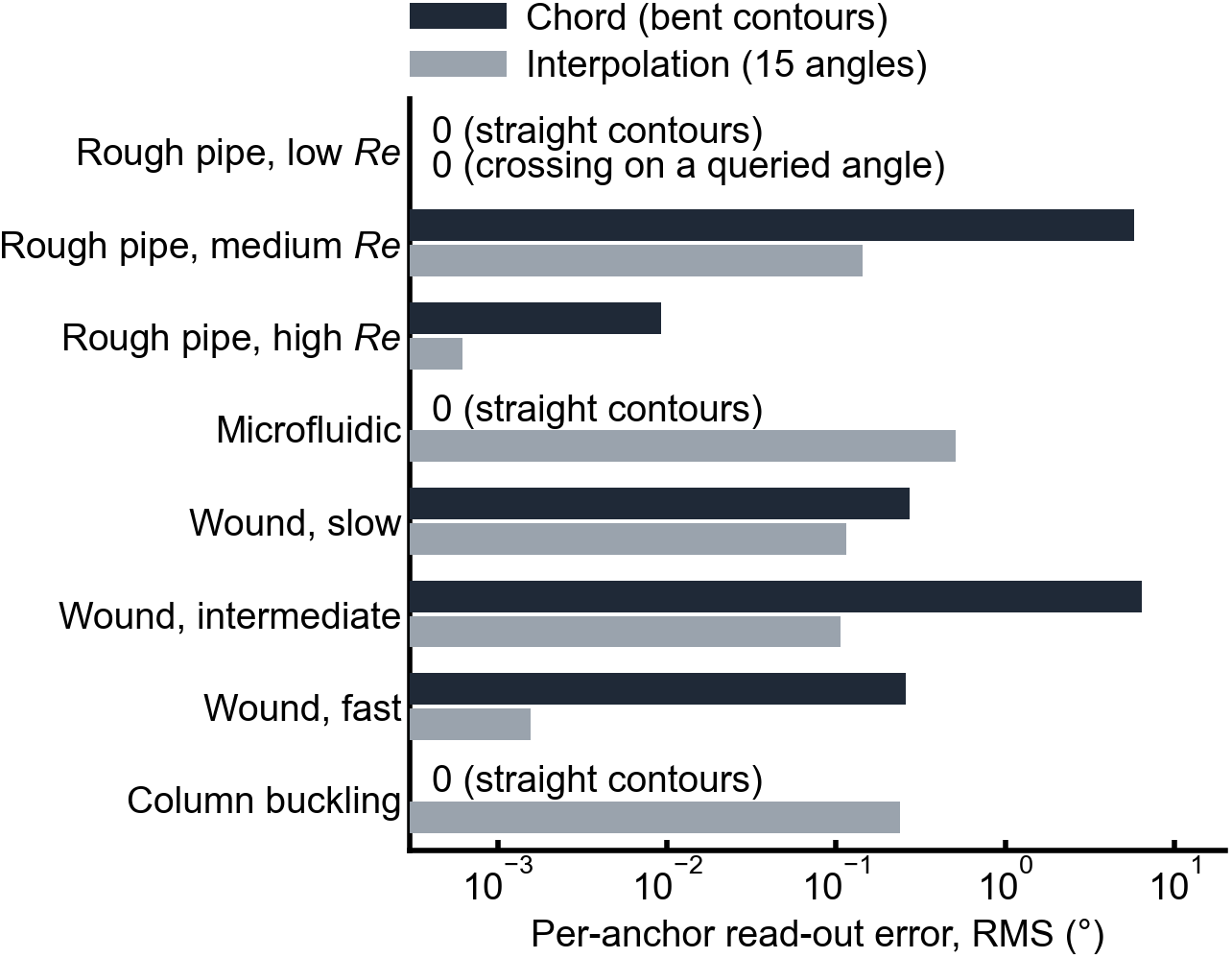}
    \end{minipage}
    \caption{\small \textbf{Finite-scale and angular-interpolation contributions to the stage-one read-out error.} 
    \textbf{a}, Absolute difference between the noise-free read-out spread $\sigma_{\alpha,\mathrm{clean}}$ and the analytic gradient-direction spread $\sigma_{\rm grad}$ as the probe semi-axes are reduced from the deployed size to $1/40$ of that size. The dashed reference segment has slope one. 
    \textbf{b}, Per-anchor root-mean-square read-out error at the deployed probe, separated into the finite-probe chord contribution (dark) and the angular-interpolation contribution from the deployed $15$ probe angles (grey). Exact zero chord contributions are annotated explicitly; the horizontal axis is logarithmic.}
    \label{fig:s6}
\end{figure}

Supplementary Fig.~\ref{fig:s6} quantifies the deterministic error introduced by the finite-scale stage-one direction read-out in the absence of evaluation noise. Panel~a varies the probe size while keeping the benchmark anchors fixed and compares the resulting noise-free directional spread $\sigma_{\alpha,\mathrm{clean}}$ with the analytic gradient-direction spread $\sigma_{\rm grad}$. Panel~b examines the deployed probe and separates the per-anchor read-out error into a finite-probe chord contribution and an angular-interpolation contribution.

For seven of the eight benchmarks, the non-zero read-out error decreases by a factor of approximately $21$--$35$ over the fortyfold reduction in probe size, close to the first-order trend indicated by the slope-one reference in Supplementary Fig.~\ref{fig:s6}a. The low-$Re$ rough-pipe case has zero error at every probe size because its contours are straight and the equal-response crossing occurs at one of the queried angles.

The decomposition at the deployed probe clarifies the two sources of read-out error (Supplementary Fig.~\ref{fig:s6}b). For the low-$Re$ rough-pipe, microfluidic and column-buckling cases, the response depends on a single linear combination of the logarithmic inputs, so the contours are straight and the finite-probe chord contribution vanishes. The remaining error in the microfluidic and column-buckling cases is therefore due to angular interpolation. By contrast, curved contours produce a non-zero chord contribution that cannot be removed simply by increasing the number of queried angles. This contribution is largest for the medium-$Re$ rough-pipe and intermediate wound-closure regimes, where the response is genuinely non-reducible over the sampled domain.

% ============================================================
\clearpage
\section*{Supplementary Tables}

\begin{table}[htpb]
    \centering
    \caption{\small Benchmark regimes, dimensionless responses, parameter ranges and expected reduction structure. Ranges are given in the original dimensionless variables; the framework operates on their logarithms.}
    \label{tab:cases}
    \footnotesize
    \setlength{\tabcolsep}{3pt}
    \begin{tabular}{@{}l c l l l@{}}
        \toprule
        Regime & $\Pi_o$ & $\Pi_1$ range & $\Pi_2$ range & \shortstack{Expected\\structure} \\
        \midrule
        Pipe, low $Re$ & $\lambda$ & $Re\in[10^2,\,2\times10^3]$ & $\epsilon/D\in[10^{-3},\,5\times10^{-2}]$ & Exact: $Re$ \\
        Pipe, medium $Re$ & $\lambda$ & $Re\in[10^4,\,10^5]$ & $\epsilon/D\in[2\times10^{-3},\,2\times10^{-2}]$ & Non-reducible \\
        Pipe, high $Re$ & $\lambda$ & $Re\in[5\times10^6,\,5\times10^7]$ & $\epsilon/D\in[2\times10^{-3},\,5\times10^{-2}]$ & Asymptotic: $\epsilon/D$ \\
        Wound, slow & $T_c r$ & $\Pi_1\in[6\times10^{-4},\,5\times10^{-3}]$ & $\Pi_2\in[0.40,\,0.90]$ & Asymptotic: $\Pi_1$ \\
        Wound, intermediate & $T_c r$ & $\Pi_1\in[0.03,\,3]$ & $\Pi_2\in[0.12,\,0.90]$ & Non-reducible \\
        Wound, fast & $T_c r$ & $\Pi_1\in[8,\,50]$ & $\Pi_2\in[0.10,\,0.45]$ & Asymptotic: $\Pi_2$ \\
        Microfluidic & $L_d/w$ & $Ca\in[10^{-2},\,3\times10^{-1}]$ & $\phi\in[0.1,\,2]$ & Exact: $Ca^{-1}\phi^{1/3}$ \\
        Column buckling & $\sigma_{cr}/\sigma_y$ & $\kappa L/r\in[20,\,200]$ & $\sigma_y/E\in[10^{-3},\,10^{-2}]$ & Exact: $(\kappa L/r)^2\sigma_y/E$ \\
        \bottomrule
    \end{tabular}
\end{table}

\begin{table}[htpb]
    \centering
    \caption{\small Protocol constants and fixed analysis settings. Unless otherwise noted, the same settings are used across all benchmarks and synthetic fields.}
    \label{tab:protocol}
    \footnotesize
    \renewcommand{\arraystretch}{1.10}
    \begin{tabular}{@{}l l p{0.50\textwidth}@{}}
        \toprule
        Quantity & Value & Role \\
        \midrule
        $M$ & $40$ & Independently sampled anchors per run \\
        $\eta$ & $0.20$ & Anchor margin and minimum probe semi-axis \\
        Probe & As large as fits & Semi-axes set by the distances from the anchor to the regime boundaries \\
        $K$ & $15$ & Equally spaced queried angles, $t_j=j\pi/14$, $j=0,\dots,14$ \\
        $n$ & $6$ & Repeats at every probe point and anchor \\
        Evaluations per run & $M(K+1)n=3840$ & Total deployed response evaluations \\
        Residual pool & $M(K+1)n=3840$ residuals & Within-repeat residuals of the run, scaled by $\sqrt{n/(n-1)}$; used to estimate $\hat\varepsilon$ and to generate the Monte Carlo directional perturbations \\
        \addlinespace[2pt]

        $N_{\rm MC}$ & $1200$ & Perturbation profiles per anchor for estimating directional noise \\
        $N_{\rm F}$ & $6400$ & Synthetic null realisations used to determine $q_{95}$ \\
        \addlinespace[2pt]

        Degree ladder & $2,3,5,8,12,16,20$ & Advance only when $\rho$ decreases by more than $5\%$ \\
        Stage-two window & $\bar\alpha\pm3\sigma_\alpha/\sqrt M$, $21$ candidates & Local direction search for collapse fitting \\
        $\tau_1$, $\tau$ & $4^\circ$, $5\%$ & Prescribed tolerances used in the main text \\
        \addlinespace[2pt]

        Benchmark noise & $\varepsilon=0.05\,\mathrm{med}_j R_j$ & Independent zero-mean Gaussian evaluation noise; $R_j$ is the response amplitude over a probe with semi-axes $\eta L_1$ and $\eta L_2$ at each of $200$ preliminary anchors \\
        Fine grid & $99$ angles, spacing $\pi/98$ & Used only in Supplementary Fig.~\ref{fig:s6}; not part of the deployed protocol \\
        \bottomrule
    \end{tabular}
\end{table}

\begin{table}[htpb]
    \centering
    \caption{\small Stage-one verdicts as a function of the declared tolerance $\tau_1$ for the $1200$ synthetic fields in Supplementary Fig.~\ref{fig:s5}. Unresolved is reported over all runs. False not-red is the fraction of runs with $\sigma_{\rm grad}\le\tau_1$ that are classified as not-red, whereas missed departure is the fraction with $\sigma_{\rm grad}>\tau_1$ that are classified as reducible. Brackets show Clopper--Pearson $95\%$ confidence intervals.}
    \label{tab:tau1}
    \footnotesize
    \setlength{\tabcolsep}{2.5pt}
    \renewcommand{\arraystretch}{1.12}
    \begin{tabular}{@{}r rcc rcc@{}}
        \toprule
        & \multicolumn{3}{c}{$1\times$ calibration noise}
        & \multicolumn{3}{c}{$4\times$ calibration noise} \\
        \cmidrule(lr){2-4}\cmidrule(lr){5-7}
        $\tau_1$ ($^\circ$)
        & Unresolved
        & \shortstack{False\\not-red}
        & \shortstack{Missed\\departure}
        & Unresolved
        & \shortstack{False\\not-red}
        & \shortstack{Missed\\departure} \\
        \midrule

        0.5
        & $55.4\%$
        & \shortstack{$5.60\%$\\[-1pt]{\scriptsize [4.0--7.6]}}
        & \shortstack{$0.19\%$\\[-1pt]{\scriptsize [0.0--1.1]}}
        & $84.9\%$
        & \shortstack{$2.51\%$\\[-1pt]{\scriptsize [1.5--4.0]}}
        & \shortstack{$0\%$\\[-1pt]{\scriptsize [0.0--0.7]}} \\

        0.75
        & $37.0\%$
        & \shortstack{$4.51\%$\\[-1pt]{\scriptsize [3.1--6.3]}}
        & \shortstack{$0.85\%$\\[-1pt]{\scriptsize [0.2--2.2]}}
        & $84.9\%$
        & \shortstack{$2.46\%$\\[-1pt]{\scriptsize [1.5--3.9]}}
        & \shortstack{$0\%$\\[-1pt]{\scriptsize [0.0--0.8]}} \\

        1
        & $29.7\%$
        & \shortstack{$2.82\%$\\[-1pt]{\scriptsize [1.8--4.2]}}
        & \shortstack{$0.95\%$\\[-1pt]{\scriptsize [0.3--2.4]}}
        & $84.9\%$
        & \shortstack{$2.31\%$\\[-1pt]{\scriptsize [1.4--3.6]}}
        & \shortstack{$0\%$\\[-1pt]{\scriptsize [0.0--0.9]}} \\

        1.5
        & $21.6\%$
        & \shortstack{$2.59\%$\\[-1pt]{\scriptsize [1.6--3.9]}}
        & \shortstack{$1.71\%$\\[-1pt]{\scriptsize [0.6--3.7]}}
        & $84.7\%$
        & \shortstack{$2.59\%$\\[-1pt]{\scriptsize [1.6--3.9]}}
        & \shortstack{$0\%$\\[-1pt]{\scriptsize [0.0--1.0]}} \\

        2
        & $15.3\%$
        & \shortstack{$1.57\%$\\[-1pt]{\scriptsize [0.9--2.6]}}
        & \shortstack{$1.29\%$\\[-1pt]{\scriptsize [0.4--3.3]}}
        & $79.5\%$
        & \shortstack{$2.13\%$\\[-1pt]{\scriptsize [1.3--3.3]}}
        & \shortstack{$0.65\%$\\[-1pt]{\scriptsize [0.1--2.3]}} \\

        3
        & $9.5\%$
        & \shortstack{$1.89\%$\\[-1pt]{\scriptsize [1.1--3.0]}}
        & \shortstack{$3.64\%$\\[-1pt]{\scriptsize [1.7--6.8]}}
        & $59.8\%$
        & \shortstack{$1.99\%$\\[-1pt]{\scriptsize [1.2--3.1]}}
        & \shortstack{$1.62\%$\\[-1pt]{\scriptsize [0.4--4.1]}} \\

        \addlinespace[2pt]
        \textbf{4}
        & $\mathbf{5.5\%}$
        & \shortstack{\textbf{$1.38\%$}\\[-1pt]{\scriptsize [0.8--2.3]}}
        & \shortstack{\textbf{$3.70\%$}\\[-1pt]{\scriptsize [1.5--7.5]}}
        & $\mathbf{48.2\%}$
        & \shortstack{\textbf{$0.89\%$}\\[-1pt]{\scriptsize [0.4--1.7]}}
        & \shortstack{\textbf{$2.65\%$}\\[-1pt]{\scriptsize [0.9--6.1]}} \\
        \addlinespace[2pt]

        6
        & $2.4\%$
        & \shortstack{$1.32\%$\\[-1pt]{\scriptsize [0.7--2.2]}}
        & \shortstack{$2.90\%$\\[-1pt]{\scriptsize [0.8--7.3]}}
        & $37.3\%$
        & \shortstack{$0.75\%$\\[-1pt]{\scriptsize [0.3--1.5]}}
        & \shortstack{$2.90\%$\\[-1pt]{\scriptsize [0.8--7.3]}} \\

        8
        & $0.25\%$
        & \shortstack{$1.19\%$\\[-1pt]{\scriptsize [0.6--2.0]}}
        & \shortstack{$9.17\%$\\[-1pt]{\scriptsize [4.5--16.2]}}
        & $22.3\%$
        & \shortstack{$1.01\%$\\[-1pt]{\scriptsize [0.5--1.8]}}
        & \shortstack{$9.17\%$\\[-1pt]{\scriptsize [4.5--16.2]}} \\

        \bottomrule
    \end{tabular}
\end{table}

\begin{table}[htpb]
    \centering
    \caption{\small Anchor-to-anchor variation in $\varphi_i$ and plug-in diagnostics of the heterogeneous correction in the quadrature decomposition (Supplementary Note~\ref{sec:sn3_axial_quadrature}). $\epsilon_{\rm rms}$ and $\epsilon_{\max}$ quantify the variation of the anchor-specific $\varphi_i$ around their mean. The diagnostic for $\kappa$ is obtained by evaluating the right-hand side of Eq.~\eqref{eq:sn3_kappa} with $\hat\sigma_{\rm true}$ substituted for the unknown noise-free spread. For the synthetic rows, $\epsilon_{\rm rms}$ and $\epsilon_{\max}$ are reported as median / maximum values over the first $100$ fields of Supplementary Fig.~\ref{fig:s5}, whereas the final column reports the maximum field-wise diagnostic value.}
    \label{tab:kappa}
    \footnotesize
    \setlength{\tabcolsep}{5pt}
    \begin{tabular}{@{}l r r r r@{}}
        \toprule
        Run
        & \shortstack{$\hat\sigma_{\rm true}$\\($^\circ$)}
        & $\epsilon_{\rm rms}$
        & $\epsilon_{\max}$
        & \shortstack{$\kappa$\\plug-in diagnostic} \\
        \midrule
        Pipe, low $Re$            & 0     & 0.0020 & 0.0089 & $0$        \\
        Pipe, medium $Re$         & 10.10 & 0.0009 & 0.0026 & $<10^{-3}$ \\
        Pipe, high $Re$           & 0     & 0.0265 & 0.1075 & $0$        \\
        Wound, slow               & 1.55  & 0.0038 & 0.0107 & $<10^{-3}$ \\
        Wound, intermediate       & 11.10 & 0.0006 & 0.0021 & $<10^{-3}$ \\
        Wound, fast               & 0.17  & 0.0068 & 0.0187 & $<10^{-3}$ \\
        Microfluidic              & 0     & 0.0038 & 0.0173 & $0$        \\
        Column buckling           & 0     & 0.0017 & 0.0060 & $0$        \\
        \addlinespace[2pt]
        Synthetic, $1\times$ noise & -- & 0.0022 / 0.039 & 0.010 / 0.24 & $\le0.005$ \\
        Synthetic, $4\times$ noise & -- & 0.049 / 0.27   & 0.20 / 1.21  & $\le0.071$  \\
        \bottomrule
    \end{tabular}
\end{table}

\begin{table}[htpb]
    \centering
    \caption{\small Synthetic-field designs used in Supplementary Figs.~\ref{fig:s1}--\ref{fig:s5}. Field families are defined in Methods, Eqs.~\eqref{eq:family_linear} and \eqref{eq:family_tworidge}; noise levels are given relative to the field-specific calibration scale $\varepsilon_0$. One run denotes one application of the complete evaluation and inference protocol to one field at one noise level. For designs with a listed base seed $s_0$, field $j$ uses seed $s_j=s_0+j$, with $j=0,\ldots,N-1$.}
    \label{tab:synthetic}

    \scriptsize
    \setlength{\tabcolsep}{2pt}
    \renewcommand{\arraystretch}{1.10}

    \begin{tabular}{@{}
        p{0.045\textwidth}
        p{0.155\textwidth}
        p{0.190\textwidth}
        p{0.315\textwidth}
        p{0.085\textwidth}
        p{0.055\textwidth}
    @{}}
        \toprule
        Figure & Family & Base fields (seed) & $\lambda$ & Noise & Runs \\
        \midrule

        S1
        & Single ridge + linear
        & $100$ ($51300000$)
        & $0$, $0.005$, $0.01$, $0.02$, $0.03$, $0.05$, $0.10$, $0.20$
        & $1$, $4$
        & $1600$ \\

        S1
        & Two ridges
        & Same $100$ primary ridges ($51300000$)
        & $0$, $0.002$, $0.005$, $0.01$, $0.02$, $0.05$, $0.10$, $0.30$
        & $1$, $4$
        & $1600$ \\

        S2a
        & Fixed ridge$^{\ast}$
        & $1$
        & $0$
        & $1$
        & $600$ \\

        S2a, b
        & Single ridge
        & $800$ ($61300000$)
        & $0$
        & $1$, $4$
        & $1600$ \\

        S3a
        & Single ridge
        & Same $800$ fields as S2b$^{\dagger}$
        & $0$
        & $0.25$, $1$, $4$
        & $2400$ \\

        S3b
        & Single ridge + linear
        & $200$ ($62300000$)
        & $0.02$, $0.05$, $0.10$, $0.20$, $0.40$
        & $1$, $4$
        & $2000$ \\

        S4
        & Single ridge + linear
        & $1$ ($51300003$; $F=e^s$)
        & $0$, $0.010$, $0.020$, $0.035$, $0.060$, $0.120$
        & $1$
        & $6$ \\

        S5
        & Single ridge + linear
        & $1200$ ($63300000$)
        & $0$ with probability $1/4$ ($298$ fields); otherwise log-uniform on $[0.001,0.3]$
        & $1$, $4$
        & $2400$ \\

        \bottomrule
    \end{tabular}

    \vspace{2pt}
    \raggedright\scriptsize
    $^{\ast}$ The fixed field is $y=\exp[0.7(0.8x_1-0.6x_2)]$, evaluated using the anchors and probes of the low-$Re$ benchmark in its logarithmic coordinates. The $600$ runs use independent noise realisations with $\varepsilon$ equal to $5\%$ of the median response amplitude over the deployed probes.

    $^{\dagger}$ At $1$ and $4$ times $\varepsilon_0$, the noise realisations are the same as those used in S2b. The $\lambda=0$ runs in S1 have $\sigma_{\rm grad}=0$ and are therefore omitted from the logarithmic horizontal axis of Supplementary Fig.~\ref{fig:s1}. The finite-run quadrature test in Supplementary Note~\ref{sec:sn3_axial_quadrature} uses the $902$ S5 fields with $\lambda>0$, and the synthetic rows of Supplementary Table~\ref{tab:kappa} use the first $100$ S5 fields.
\end{table}

\end{document}